\documentclass[12pt]{article}
\usepackage{comment}
\usepackage{fancybox}
\usepackage{color}
\usepackage{titlesec}
\titlespacing*{\section}{0pt}{0.6ex}{0.6ex}
\titlespacing*{\subsection}{0pt}{0.6ex}{0.6ex}
\titlespacing*{\subsubsection}{0pt}{0.4ex}{0.4ex}

\titleformat{\section}
{\normalfont\fontsize{14}{24}\bfseries}{\thesection}{1em}{}
\titleformat{\subsection}
{\normalfont\fontsize{12}{24}\bfseries}{\thesubsection}{1em}{}

\titleformat{\subsubsection}
{\normalfont\fontsize{12}{24}\bfseries}{\thesubsubsection}{1em}{}
\definecolor{light-gray}{gray}{0.9}

\newcommand{\rev}[1]{\textcolor{red}{#1}}

\usepackage[margin=1in]{geometry}
\usepackage{amsmath,bm}
\usepackage{amssymb}
\usepackage{hyperref}
\usepackage{subcaption}
\usepackage{tikz}
\usepackage{pgfplots}

\usepackage{amsthm}

\date{}

\usepackage{graphicx}
\usepackage{epstopdf,epsfig}
\usepackage{newtxtext}
\usepackage{newtxmath}
\usepackage{natbib}
\usepackage{hyperref}
\usepackage{caption}
\usepackage{subcaption}

\hypersetup{
    colorlinks = true,
    urlcolor   = blue,
    citecolor  = black,
}

\title{The Minimal Attached Eddy in Wall Turbulence: Statistical Foundations, Inverse Identification and  Influence Kernels}

\author{Karthik Duraisamy \\
	Department of Aerospace Engineering,  \\ University of Michigan, Ann Arbor, MI 48109
  }

\begin{document}

\maketitle
 
\begin{abstract}
Townsend's attached eddy hypothesis models the logarithmic region of high Reynolds number wall turbulence as a random superposition of wall-attached, self-similar eddies whose sizes obey a scale-invariant population law. Building on the statistical framework of \cite{woodcock2015statistical}, the present work (i) poses an inverse problem to infer the ideal single-eddy contribution (influence) functions for the mean velocity and Reynolds stresses from DNS moments, (ii) uses these inferred kernels to guide a minimal Biot--Savart-consistent hairpin-type eddy built from Rankine vortex rods with an inviscid image system, and (iii) introduces and infers a spectral Influence kernel  that maps a self-similar eddy footprint to its one-dimensional energy spectrum. The Influence-kernel viewpoint yields an explanation for the emergence of the linear part of the energy spectrum, provides a clear scale-by-scale decomposition  and helps rationalize why simple eddy templates can reproduce a broad set of log-layer statistics once the mean-flow anchoring is fixed. Closed-form expressions for the mean influence function and the Fourier-space streamwise velocity of a general straight-segment hairpin family with image are derived, revealing a mean-variance duality: For a rectangular hairpin, the horizontal head determines the entire mean kernel while the inclined legs dominate the spectral energy. It is shown that the rectangular hairpin yields a precise plateau in the influence function, resulting in a logarithmic mean velocity profile and desirable properties for the streamwise fluctuations. The model is further extended by calibrating the eddy population density as a function of  scale, yielding accurate descriptions of mean velocity and streamwise variance across $Re_{\tau}=6000-20000$.
\end{abstract}

Characterizing and predicting the turbulent flow in the vicinity of a wall continues to not only be a problem of scientific interest, but also one of  great practical relevance. Significant knowledge and insight has been gained from the development of experimental diagnostics in the 1960s to the emergence of direct numerical simulations in the 1980s, sophisticated measurements in the 1990s,  an emphasis on coherent structures in the 2000s and operator-based dynamical models over the past decade. Against this backdrop of increasingly sophisticated experimentation, computations and analyses, the attached eddy hypothesis - originally proposed more than 50 years ago by ~\cite{townsend1976structure} - continues to serve as a simple, yet effective theory to qualitatively and quantitatively describe  structural and statistical aspects of turbulent boundary layers, particularly focusing on the logarithmic region. 

The attached eddy model (AEM) was given a firm mathematical footing by ~\cite{perry1982mechanism}, and further developed by Marusic and co-workers over the past 25 years ~\citep{perry1995wall, marusic2013logarithmic,woodcock2015statistical,de2016influence}. Among other useful features, the AEM is able to explain scaling behaviors of velocity moments, provide an explanation for uniform momentum zones~\citep{de2016uniform}, and serve as a predictive model for the {von K\'arm\'an} constant as a function of Reynolds number.  A review of the theory and developments of the AEM of wall turbulence can be found in ~\cite{marusic2019attached}. Over the past few years, theoretical and numerical  analyses (e.g. ~\cite{mckeon2019self,lozano2019characteristic}) have added further credibility   to this theory. Despite the success of the AEM, it is pertinent to remember that it is fundamentally a statistical theory, and other hypotheses (e.g. ~\cite{davidson2006logarithmic,davidson2009simple})  can also be used to explain the statistical characteristics of the log layer. 

The attached eddy hypothesis is based on the principle that the physics and statistical properties of the logarithmic layer can be explained by considering geometrically self-similar eddies that extend from the wall. A foundational assumption is that the length scale of each individual eddy follows a probability distribution that is a function of the distance from the wall. Hence, the term `attached' alludes to the fact that every eddy can be assumed to be  randomly placed on the wall. The AEM is effectively an inviscid theory, yet the range of scales is set by the Reynolds number.  

A key advance was made by~\cite{woodcock2015statistical} (henceforth W\&M) who established a statistical foundation for  AEM. They provide a complete derivation for {\em all} the velocity moments and demonstrate logarithmic scaling relationships therein. They were also able to provide expressions for the skewness and flatness of the wall-normal and spanwise fluctuations as a function of the Reynolds number. While variants of the AEM continue to be developed in the literature (e.g.~\cite{hwang2020attached}), we consider W\&M as the starting point of our exploration and exclusively consider zero pressure gradient boundary layers. The main contributions and guiding questions in this work are:

\begin{enumerate}
	
	\item \textbf{Inverse identification of attached-eddy influence functions.} Townsend introduced the notion of an ``eddy intensity'' (or contribution) function: a wall-parallel average of the single-eddy induced velocity and its products.  While Townsend gave remarkably insightful descriptions of the nature of this function (Figure 5.7,  page 155 of~\cite{townsend1976structure}), his and most other treatments are qualitative. Here we pose a concrete inverse problem that infers the \emph{ideal} influence functions $I_1(y/h)$ and $I_{ij}(y/h)$ implied by reference one-point moments (DNS/experiment). The inferred kernels expose the wall-normal support, sign structure, and plateau/decay behavior required for log-layer mean and Reynolds stresses, and motivate a simplified kernel model.
	
	\item \textbf{A minimal Biot--Savart-consistent attached eddy.} W\&M utilize an eddy that has a complex shape  (Figure 1 in W\&M), presumably configured using insight from  DNS and/or PIV fields.  Guided by the inferred kernels, we construct an attached-eddy template from a small number of Rankine vortex rods (a ``rectangular hairpin'') together with an inviscid image system. The objective is not an exact Navier--Stokes solution, but a kinematically consistent building block that (i) respects wall-impermeability, (ii) has well-posed planar influence integrals, and (iii) admits a transparent connection between morphology and the influence functions.
	
	\item \textbf{What aspects of morphology control log-layer statistics?} We quantify the sensitivity of $I_1$, $I_{ij}$ and the resulting moments to eddy geometry (e.g.\ triangular vs.\ square hairpins, inclination angle, and simple packet/nesting configurations treated as composite marks). This clarifies which structural features are essential (e.g.\ realizing the near-constant portion of $I_1$ for $y/h\lesssim 1$) and which details have weak impact once similarity and attachedness are enforced.
	
	\item \textbf{A spectral Influence kernel and interpretation of attached-eddy spectra.} We introduce an explicit spectral \emph{Influence kernel} $I_\phi(\kappa_x,\eta)$ that maps a self-similar wall-parallel footprint to its one-dimensional streamwise energy spectrum as a function of relative height $\eta=y/h$. Combining $I_\phi$ with the scale-invariant population law $p(h)\propto h^{-3}$ yields a compact explanation for the emergence (and limitations) of the $k_x^{-1}$ range, the location of the premultiplied hump, and the origin of low-$k_x$ discrepancies under the independent-mark (Poisson) hypothesis.  The ideal influence kernel is extracted by defining a Fredholm-type inverse problem.
	
	\item \textbf{Why the rectangular hairpin is unusually predictive.} Exact closed-form expressions for the mean kernel and Fourier-space velocity of a general three-segment hairpin family  reveal a clean mean-variance duality: the horizontal head determines the entire mean kernel $I_1$, while the inclined legs dominate the spectral energy $I_\phi$. Within the straight segment family, only the rectangular hairpin  produces an exact plateau in $I_1$ and hence a purely logarithmic mean.

		\end{enumerate}

	Forward and inverse modeling tools used here are open sourced\footnote{\url{https://github.com/CaslabUM/AttachedEddy}}.

\section{Statistics of attached eddies }

We begin with a short description of the attached eddy hypothesis and modeling. This presentation generally follows W\&M with a slightly different pedagogy. As noted above, we adopt the standard coordinate system: $x$ streamwise, $y$ wall-normal, and $z$ spanwise. Planar homogenization/influence integrals are taken over the wall-parallel $(x,z)$ plane, so one-point statistics depend only on $y$.

\begin{figure}
    \centering
    \includegraphics[width=0.8\textwidth]{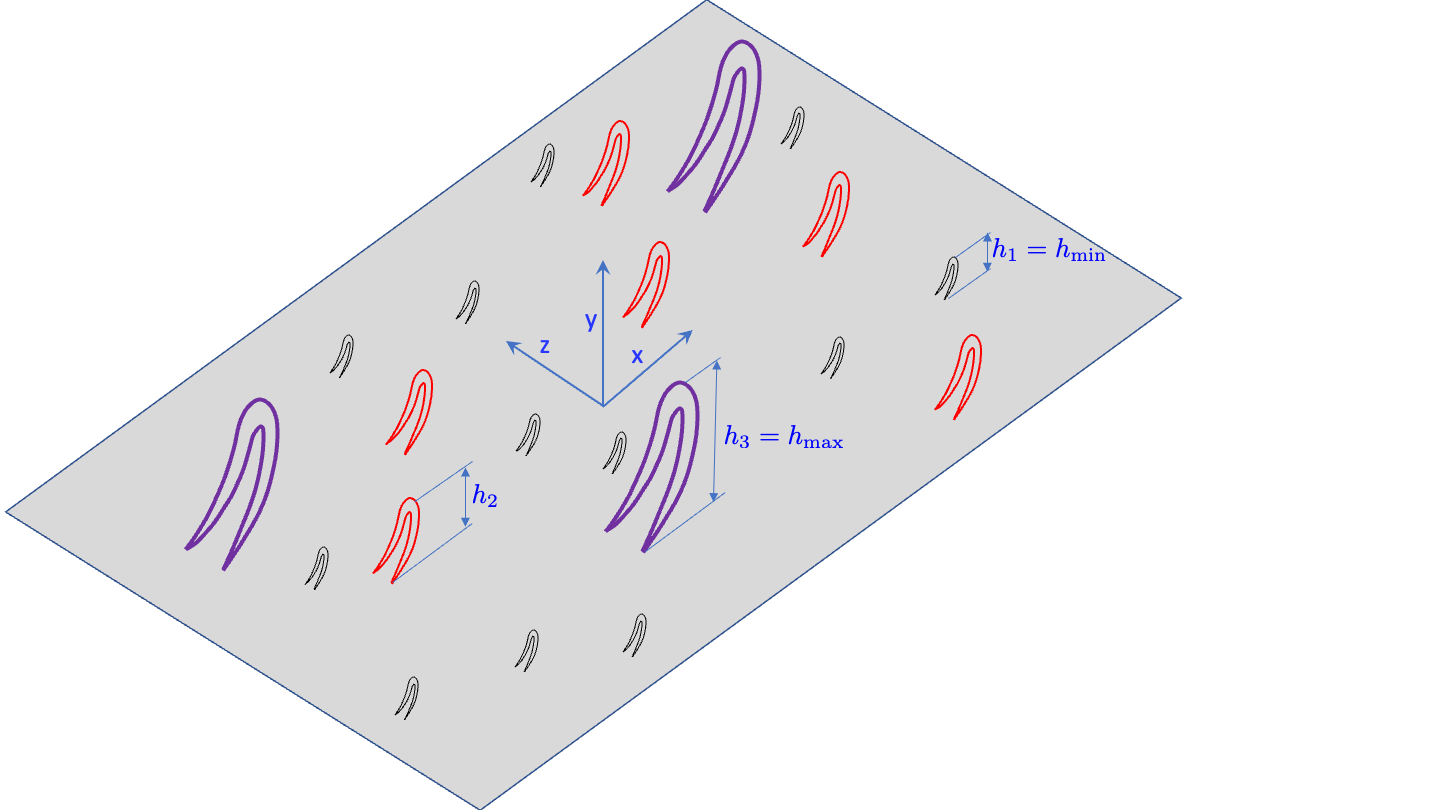}
    \caption{Schematic of discrete representation of attached eddies with $n=21$ and $m=3$.}
    \label{fig:attachededdy}
\end{figure}

Consider a collection of $n$ eddies (Figure~\ref{fig:attachededdy}) of length scale $h_{e,i}$ that are placed on the wall at locations $\mathbf{x}_{e,i}=(x_{e,i},z_{e,i})$. Since these are attached eddies, it is implicit that only the streamwise and spanwise components of $\mathbf{x}_{e,i}$ are variable. Define $\mathbf{h}_e \triangleq \{h_{e,1},h_{e,2},..,h_{e,n}\}$ and $\mathbf{X}_e \triangleq \{\mathbf{x}_{e,1},\mathbf{x}_{e,2},..,\mathbf{x}_{e,n}\}$. Assuming self-similarity and linearity, a quantity $q$ (e.g. a velocity component) evaluated at a location $\mathbf{x}$ can be determined using the superposition
\[
q(\mathbf{x},\mathbf{X}_e,\mathbf{h}_e,n) \triangleq \sum_{i=1}^n q\!\left(\frac{\mathbf{x}-\mathbf{x}_{e,i}}{h_{e,i}}\right).
\]

Restricting our attention to a streamwise--spanwise square wall patch of side $2L$ in which the eddies are assumed to be independently and uniformly distributed, and assuming that the range of eddy length scales follows a probability density function $p(h)$, the expectation of $q$ in this region is given by
\begin{equation}
	q(\mathbf{x},n) \triangleq \mathbb{E}_{\mathbf{X}_e,\mathbf{h}_e}\!\left[q(\mathbf{x},\mathbf{X}_e,\mathbf{h}_e,n)\right]
	=
	\frac{1}{4L^2}\int_{h_{\min}}^{h_{\max}}\int_{-L}^L\int_{-L}^L
	\sum_{i=1}^n q\!\left(\frac{\mathbf{x}-\mathbf{x}_{e,i}}{h}\right)\,p(h)\,dx_{e,i}\,dz_{e,i}\,dh.
	\label{eq:mean_shotnoise}
\end{equation}

Homogenizing in the $(x,z)$ directions, and assuming $L$ is large enough that each eddy centered at the origin has a negligible induced contribution outside an area $4L^2$ (\rev{cf.}\ Campbell's theorem~\cite{rice1944mathematical} and the appendix of W\&M), it can be shown that
\[
q(y,n)\triangleq \mathbb{E}_{x,z}\!\left[q(\mathbf{x},n)\right]
\approx
\frac{n}{4L^2}\int_{h_{\min}}^{h_{\max}}\int_{-L}^L\int_{-L}^L q\!\left(\frac{\mathbf{x}}{h}\right)\,p(h)\,dx\,dz\,dh.
\]
Note that in the above equation, the entire field is written as a function of \emph{one} prototypical eddy, scaled by the probability density function of the eddy sizes $p(h)$ and the eddy density $n/(4L^2)$. If the mean eddy density is $\beta$, then using Poisson's law, the expected value (over all numbers of eddies) is
\begin{equation}
Q(y) \triangleq \mathbb{E}_{n}\!\left[q(y,n)\right]
=
\beta \int_{h_{\min}}^{h_{\max}}\int_{-L}^L\int_{-L}^L q\!\left(\frac{\mathbf{x}}{h}\right)\,p(h)\,dx\,dz\,dh.
\label{eq:mean_Qy}
\end{equation}
Note that all the $q$'s defined above are random variables, yet $Q(y)$ is a deterministic quantity. Now we are in a position to define the mean streamwise velocity $U(y)$ as a superposition of eddies of various sizes $h$. This can be written in terms of the induced velocity field $u_1(\cdot)$ of one prototypical eddy:
\begin{align}
U(y)
&\triangleq
\beta \int_{h_{\min}}^{h_{\max}}\int_{-L}^L\int_{-L}^L u_1\!\left(\frac{\mathbf{x}}{h}\right)\,p(h)\,dx\,dz\,dh
+U_{ref} \nonumber\\
&=
\beta \int_{h_{\min}}^{h_{\max}} p(h)\,h^2\,I_1\!\left(\frac{y}{h}\right)\,dh
+U_{ref},
\label{eq:U_of_y}
\end{align}
where the mean-flow eddy contribution (influence) function is
\begin{align}
I_1\!\left(\frac{y}{h}\right)
&\triangleq
\int_{-L/h}^{L/h}\int_{-L/h}^{L/h}
u_1\!\left(\frac{\mathbf{x}}{h}\right)\,d\!\left(\frac{x}{h}\right)\,d\!\left(\frac{z}{h}\right).
\label{eq:zz}
\end{align}
Similarly, we can define the Reynolds stress tensor as
\begin{align*}
R_{ij}(y)
&\triangleq
\beta \int_{h_{\min}}^{h_{\max}} p(h)\,h^2\,I_{ij}\!\left(\frac{y}{h}\right)\,dh,\\
\textrm{where }\;
I_{ij}\!\left(\frac{y}{h}\right)
&\triangleq
\int_{-L/h}^{L/h}\int_{-L/h}^{L/h}
u_i\!\left(\frac{\mathbf{x}}{h}\right)\,u_j\!\left(\frac{\mathbf{x}}{h}\right)\,d\!\left(\frac{x}{h}\right)\,d\!\left(\frac{z}{h}\right).
\end{align*}
 Note the presence of an additional freestream velocity in the definition of $U(y)$. This is required because we are working with induced velocity fluctuations.

The final piece we need is the probability distribution of the eddy sizes. Using insight from~\cite{townsend1976structure} and~\cite{perry1982mechanism}, W\&M propose that $p(h)\propto 1/h^3$. In particular, for $h\in[h_{\min},h_{\max}]$,
\[
p(h) \triangleq \frac{C}{h^3},
\qquad
C \triangleq \frac{2}{1/h_{\min}^2-1/h_{\max}^2}.
\]
Here $h_{\max}\approx \delta^+$ is the outer length scale (boundary layer thickness in wall units) and $h_{\min}$ is set by the friction Reynolds number in the form $2.6 \sqrt{Re_\tau}$. 

\subsection{Statistical assumptions and derivation of the Reynolds-stress formula}
\label{sec:stress_assumptions}

The attached-eddy framework above is a \emph{shot-noise} model: the instantaneous velocity
perturbation at a point is written as a superposition of contributions from a random collection of eddies.
While the mean follows immediately from linearity of expectation, the Reynolds stresses are quadratic
and require an explicit statement of the underlying point-process assumptions.

We idealize the set of attached eddies as a stationary \emph{marked Poisson point process}
\(\Phi\) on the wall-parallel plane \((x,z)\). Each eddy is characterized by a wall-parallel center location
\(\mathbf{x}_e=(x_e,z_e)\) and a mark \(\theta\) that includes its size \(h\) (and possibly additional parameters,
e.g. orientation, circulation distribution, etc.). The following are assumed:
\begin{enumerate}
	\item[\textbf{A1.}] \textbf{Homogeneity:} the process is stationary in \((x,z)\) with constant areal intensity \(\beta\)
	(so the expected number of eddies in area \(A\) is \(\beta A\)).
	\item[\textbf{A2.}] \textbf{Independent marking:} conditional on the locations, the marks \(\{\theta_k\}\) (collectively, all eddy properties beyond the wall-parallel center location) are i.i.d.
	and independent of the point locations; in particular the size distribution is \(p(h)\).
	\item[\textbf{A3.}] \textbf{Moment existence:} the required single-eddy planar integrals defining \(I_i\) and \(I_{ij}\)
	exist in the \(L/h\to\infty\) limit (see Appendix~A for a convergence discussion).
	\item[\textbf{A4.}] \textbf{Ergodic/homogenization interpretation:} planar averaging over a sufficiently large patch
	is identified with the ensemble mean for the stationary process.
\end{enumerate}
Assumptions \textbf{A1--A2} are the essential statistical hypotheses behind the Reynolds-stress closure.

\paragraph{Velocity superposition.}
Let a single prototypical eddy with mark \(\theta\) induce a velocity field
\(\,u_i^{(e)}(\xi,\zeta,\eta;\theta)\,\) in dimensionless coordinates
\(
(\xi,\zeta,\eta)=(x/h,\;z/h,\;y/h).
\)
Define the single-eddy planar influence integral
\begin{equation}
	I_i(\eta;\theta)
	\;\triangleq\;
	\int_{\mathbb{R}^2} u_i^{(e)}(\xi,\zeta,\eta;\theta)\,d\xi\,d\zeta,
	\qquad
	(\eta=y/h).
	\label{eq:mean_kernel_def}
\end{equation}
Under \textbf{A1--A3}, first-order Campbell/Mecke yields
\begin{equation}
	U_i(y)\;\triangleq \;\mathbb{E}[u_i(y)]
	\;=\;
	\beta \int p(h)\,h^2\,
	\mathbb{E}_{\theta|h}\!\left[I_i\!\left(\frac{y}{h};\theta\right)\right]\,dh.
	\label{eq:mean_campbell}
\end{equation}

\paragraph{Second moment and the cross-term issue.}
The central subtlety is that
\(
u_i u_j = (\sum_k u_i^{(k)})(\sum_\ell u_j^{(\ell)})
\)
contains \emph{cross terms} with \(k\neq \ell\), so one cannot in general write the second moment as a
sum of ``individual Reynolds stresses''. To make this explicit, expand
\begin{equation}
	\mathbb{E}[u_i(y)u_j(y)]
	=
	\mathbb{E}\!\left[\sum_k u_i^{(k)}u_j^{(k)}\right]
	+
	\mathbb{E}\!\left[\sum_{k\neq \ell} u_i^{(k)}u_j^{(\ell)}\right].
	\label{eq:second_moment_decomp}
\end{equation}
Define the single-eddy quadratic kernel
\begin{equation}
	I_{ij}(\eta;\theta)
	\;\triangleq \;
	\int_{\mathbb{R}^2} u_i^{(e)}(\xi,\zeta,\eta;\theta)\,
	u_j^{(e)}(\xi,\zeta,\eta;\theta)\,d\xi\,d\zeta.
	\label{eq:stress_kernel_def}
\end{equation}
For a \emph{Poisson} process with \emph{independent} marks (\textbf{A1--A2}), the second-order Campbell theorem
gives the closed form
\begin{equation}
	\mathbb{E}[u_i(y)u_j(y)]
	=
	\beta \int p(h)\,h^2\,\mathbb{E}_{\theta|h}\!\left[I_{ij}\!\left(\frac{y}{h};\theta\right)\right]\,dh
	\;+\;
	U_i(y)\,U_j(y),
	\label{eq:raw_second_moment}
\end{equation}
where the \(U_iU_j\) term is precisely the contribution of the \(k\neq \ell\) cross terms in
\eqref{eq:second_moment_decomp}. In other words, under \textbf{A1--A2}, the cross terms do \emph{not}
vanish individually; rather, they factorize into the product of means.

\paragraph{Reynolds stresses as covariances (cancellation of the cross term).}
The one-point Reynolds stress tensor is the covariance of the total field:
\begin{equation}
	R_{ij}(y)
	\;\triangleq\;
	\mathbb{E}\!\left[(u_i(y)-U_i(y))(u_j(y)-U_j(y))\right]
	=
	\mathbb{E}[u_i(y)u_j(y)]-U_i(y)U_j(y).
	\label{eq:rij_cov_def}
\end{equation}
Substituting \eqref{eq:raw_second_moment} into \eqref{eq:rij_cov_def} yields the 
single-eddy representation
\begin{equation}
		R_{ij}(y)
		=
		\beta \int p(h)\,h^2\,\mathbb{E}_{\theta|h}\!\left[I_{ij}\!\left(\frac{y}{h};\theta\right)\right]\,dh.
	\label{eq:rij_final}
\end{equation}
Equation~\eqref{eq:rij_final} is therefore justified \emph{only} when the eddy field is modeled as a Poisson
(or otherwise ``factorizable'') superposition with independent marks. 

If the eddy process is \emph{not} Poisson (e.g. clustering, exclusion, or explicit packet/nesting correlations),
then the cross term in \eqref{eq:second_moment_decomp} no longer factorizes into \(U_iU_j\). In that case,
\eqref{eq:raw_second_moment} acquires an additional correction involving the pair (second factorial) moment
measure of the process (equivalently a pair-correlation function), and the covariance is no longer determined
solely by the single-eddy kernel \(I_{ij}\). A statistically consistent way to incorporate a correlated internal
structure (e.g. a hairpin packet) while retaining \eqref{eq:rij_final} is to treat the \emph{packet as the Poisson object}:
internal correlations are absorbed into the mark \(\theta\), while packet centers remain a stationary Poisson field.

\subsection{Convergence of planar influence integrals}
\label{sec:influence_convergence}

Equation~\eqref{eq:zz} defines the mean-flow eddy contribution (influence) function as a planar integral of the
streamwise velocity induced by a single prototypical eddy. Writing the dimensionless coordinates
\(
(\xi,\zeta,\eta) \triangleq (x/h,\;z/h,\;y/h)
\)
and \(R \triangleq L/h\), we can make the dependence on the truncation radius explicit:
\begin{equation}
I_1(\eta;R)
\;\triangleq\;
\int_{-R}^{R}\int_{-R}^{R} u_1(\xi,\zeta,\eta)\,d\xi\,d\zeta,
\qquad
I_1(\eta)\;\triangleq\;\lim_{R\to\infty} I_1(\eta;R),
\label{eq:I1_def_R}
\end{equation}
whenever the limit exists.
In particular, we do \emph{not} assume compact support of the induced velocity field. Rather, the
required condition for Eq.~\eqref{eq:zz} to be well-defined is that the planar integral converges as \(R\to\infty\),
or equivalently that the tail contribution beyond \(|(\xi,\zeta)|\gtrsim R\) becomes negligible for the
chosen truncation. A detailed analysis of convergence is presented in Appendix A.


\section{The ideal eddy contribution function}
\label{sec:inverse}

The functions $I_1(y/h)$ and $I_{ij}(y/h)$ introduced in Eq.~\eqref{eq:zz} are key for the moments produced by the attached-eddy model: once these single-eddy planar integrals are specified (together with $\beta$ and $p(h)$), the attached-eddy integrals yield the one-point mean and Reynolds stresses, and (through similar constructions) higher-order moments. Rather than prescribing a particular eddy geometry \emph{a priori}, we first ask a more direct question:

\begin{quote}
\emph{Given  one-point statistics $U_{\rm data}(y)$ and $R_{ij,{\rm data}}(y)$ (from DNS or experiment), what shape of $I_1(\eta)$ and $I_{ij}(\eta)$, with $\eta=y/h$, is implied by the attached-eddy integral relations?}
\end{quote}

This is an inverse problem for a Fredholm integral equation of the first kind. In its continuous form,
\begin{align}
U_{\rm data}(y)-U_{ref,\rm{data}} &= \beta \int_{h_{\min}}^{h_{\max}} p(h)\,h^2\,I_1\!\left(\frac{y}{h}\right)\,dh, \label{eq:inv_mean_cont} \\
R_{ij,\rm{data}}(y) &= \beta \int_{h_{\min}}^{h_{\max}} p(h)\,h^2\,I_{ij}\!\left(\frac{y}{h}\right)\,dh. \label{eq:inv_stress_cont}
\end{align}
where $U_{ref,\rm{data}}$ is the outer velocity (e.g. center line or freestream velocity). For the classical attached-eddy size distribution $p(h)=C/h^3$, these become
\begin{equation}
U_{\rm data}(y)-U_{ref,\rm{data}} = \beta\,C \int_{h_{\min}}^{h_{\max}} \frac{I_1(y/h)}{h}\,dh,
\qquad
R_{ij,\rm{data}}(y) = \beta\,C \int_{h_{\min}}^{h_{\max}} \frac{I_{ij}(y/h)}{h}\,dh,
\label{eq:inv_logform}
\end{equation}
and, after the change of variables $\eta=y/h$ (so that $dh/h=-d\eta/\eta$),
\begin{equation}
U_{\rm data}(y)-U_{ref,\rm{data}}= \beta\,C \int_{y/h_{\max}}^{y/h_{\min}} \frac{I_1(\eta)}{\eta}\,d\eta,
\qquad
R_{ij,\rm{data}}(y) = \beta\,C \int_{y/h_{\max}}^{y/h_{\min}} \frac{I_{ij}(\eta)}{\eta}\,d\eta.
\label{eq:inv_eta_form}
\end{equation}
Equation~\eqref{eq:inv_eta_form} makes two important points transparent: (i) the mapping $I(\eta)\mapsto U(y)$ is \emph{smoothing} (ill-conditioned to invert without regularization), and (ii) a plateau of $I_1(\eta)$ over $\eta<1$ immediately yields logarithmic dependence of $U(y)$ on $y$.

\subsection{Discrete formulation and identifiability}

In practice, we discretize the eddy-size range using $\tilde{\mathbf{h}}=\{h_1,\ldots,h_m\}$ (log-spaced between $h_{\min}$ and $h_{\max}$) and sample the wall-normal coordinate at $\tilde{\mathbf{y}}=\{y_1,\ldots,y_{m_y}\}$. The forward map for the mean can be written compactly as
\begin{equation}
\mathbf{U}_{\rm data}(\tilde{\mathbf y})-U_{ref,\rm{data}}\mathbf{1}=\mathbf{K}(\mathbf{c})\mathbf{b},
\label{eq:inv_discrete_basic}
\end{equation}
where $\mathbf{c}$ contains the unknown nodal values of $I_1(\eta)$ (in a chosen basis) and $\mathbf{K}$ is the corresponding discretized integral operator (assembled from the weights $p(h)\,h^2$ and the ratios $\eta=y/h$). An analogous system holds for each component $I_{ij}$.

 The discretized  velocity moments are 
\begin{align*}
	U_{\rm data}(y_r; \mathbf{d}) &= \beta \sum_{l=1}^{m}  p(h_l) h_l^2 \left[\sum_q \sum_p u_1\left(\frac{{x_p,z_q,y_r}}{h_l}; \mathbf{d}\right)   \left(\frac{\Delta x}{h_l} \right) \left(\frac{\Delta z}{h_l} \right) \right] \Delta h + U_{ref,\rm{data}}\\
	R_{ij,\rm{data}}(y_r; \mathbf{d}) &= \beta \sum_{l=1}^{m}  p(h_l) h_l^2 \left[\sum_q \sum_p u_i\left(\frac{{x_p,z_q,y_r}}{h_l}; \mathbf{d}\right)  u_j\left(\frac{{x_p,z_q,y_r}}{h_l}; \mathbf{d}\right)   \left(\frac{\Delta x}{h_l} \right) \left(\frac{\Delta z}{h_l} \right) \right] \Delta h.
\end{align*}
Explicitly $K_{ij} = I_1({y}_i/h_j)$ and $b_i = \beta p(h_i) h_i^2 \Delta h. $
We would now like to extract these coefficients. Towards this end, we define the eddy contribution function at selected locations, and interpolate for the values in other locations in a piecewise linear fashion. In other words
$$ \mathbf{K}=
\begin{bmatrix}
	I_1(1) & I_1(h_1/h_2) & I_1(h_1/h_3) & .. & I_1(h_1/h_{m-1}) & I_1(h_1/h_m)\\
	I_1(h_2/h_1) & I_1(1) & I_1(h_2/h_3) & .. & I_1(h_2/h_{m-1}) & I_1(h_2/h_m)\\
	.. & .. & .. & .. & .. & ..\\
	I_1(h_m/h_1) & I_1(h_m/h_2) & I_1(h_m/h_3) & .. & I_1(h_m/h_{m-1}) & I_1(1)
\end{bmatrix} $$ $$\triangleq \begin{bmatrix}
	c_1 & c_{m+1} & c_{m+2} & .. & c_{2m-2} & c_{2m-1}\\
	c_2 & c_1& c_{m+1} & .. & c_{2m-3} & c_{2m-2}\\
	.. & .. & .. & .. & .. & ..\\
	c_m &  c_{m-1} & c_{m-2} & .. & c_2 & c_1\\
\end{bmatrix}.
$$
The unknown $I_1(\cdot)$ values are then interpolated~\footnote{For instance, if $h_{\min}=186$ and $h_{\max}=5186$ and $m=11$, then $I_1(h_2/h_3)=0.42106 c_1 + 0.57894 c_{12}$} from the  nodal locations $\mathbf{c}$.
Note that there are more unknowns ($2m$) than equations ($m$), and so while it is possible to determine the unknowns that lead to perfect match of a reference (DNS or experiment) velocity profiles, a choice has to be made on the problem formulation. We choose to solve the following minimum-norm exact interpolation problem:
$$
\mathbf{c}_\textrm{opt} = \min \mathbf{c}^T \mathbf{c} \ \ 
\textrm{such that} \ \  \mathbf{U}_{\rm{data}}(\tilde{\mathbf y})-U_{ref,\rm{data}} = \mathbf{K} \mathbf{b}.
$$

\begin{figure}
    \centering
    \includegraphics[width=0.4\textwidth]{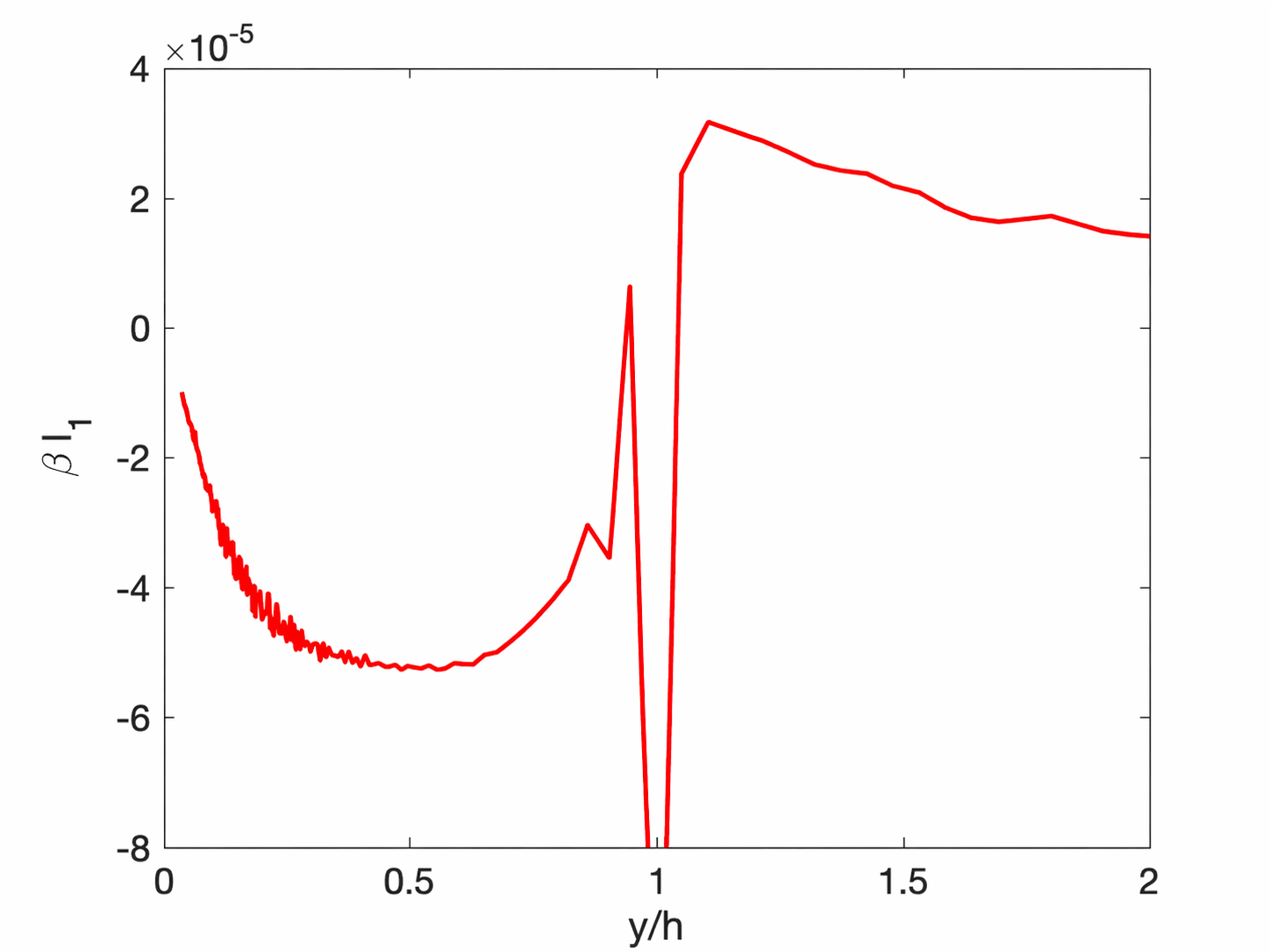}
        \includegraphics[width=0.44\textwidth]{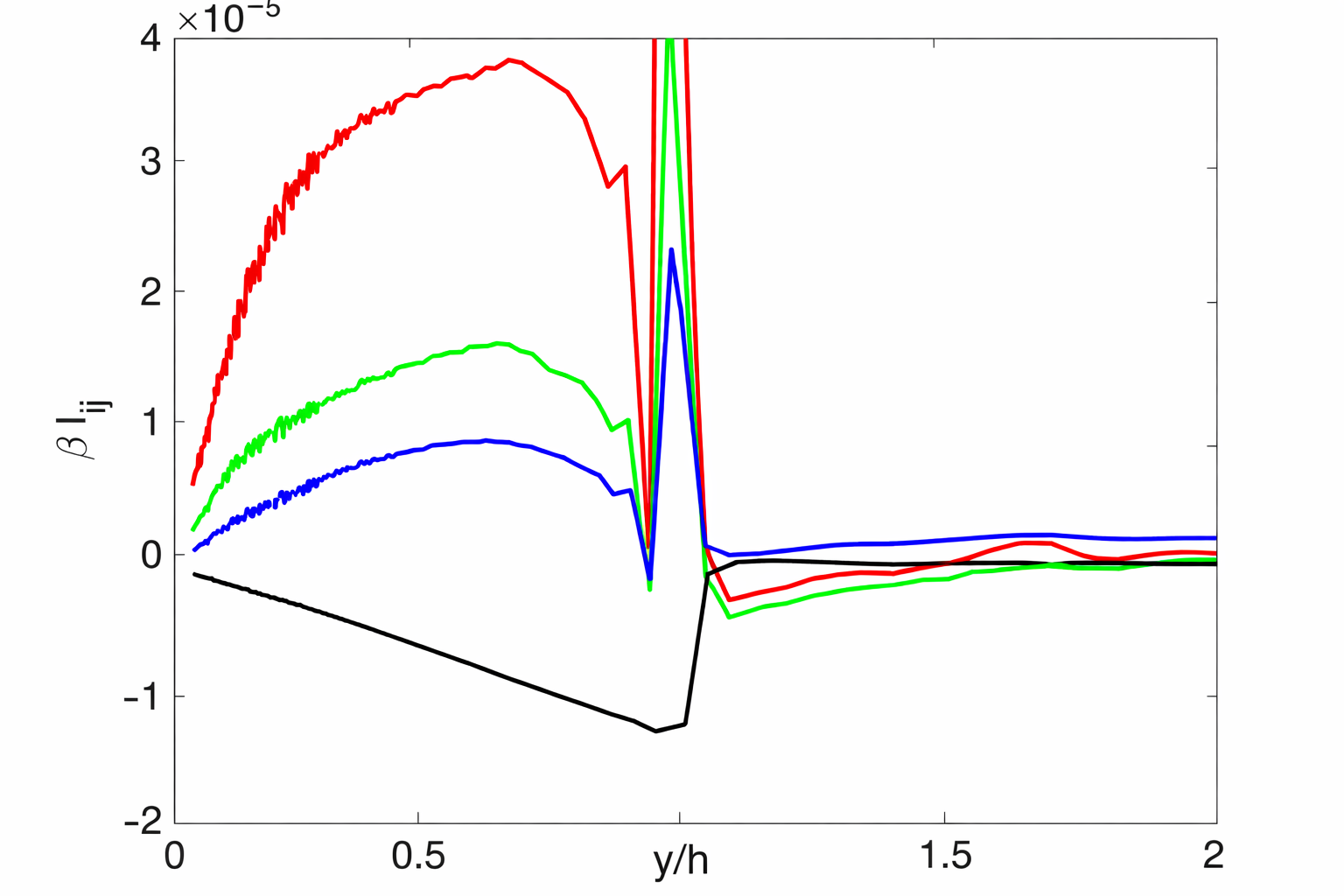}
    \caption{Optimal influence functions for $Re_\tau \approx 5200$ for the mean flow (left) and Reynolds stresses (right, with red=streamwise; green=spanwise; blue=wall-normal, and black=shear).}
    \label{fig:Optimal1}
\end{figure}

\begin{figure}
    \centering
    \includegraphics[width=0.4\textwidth]{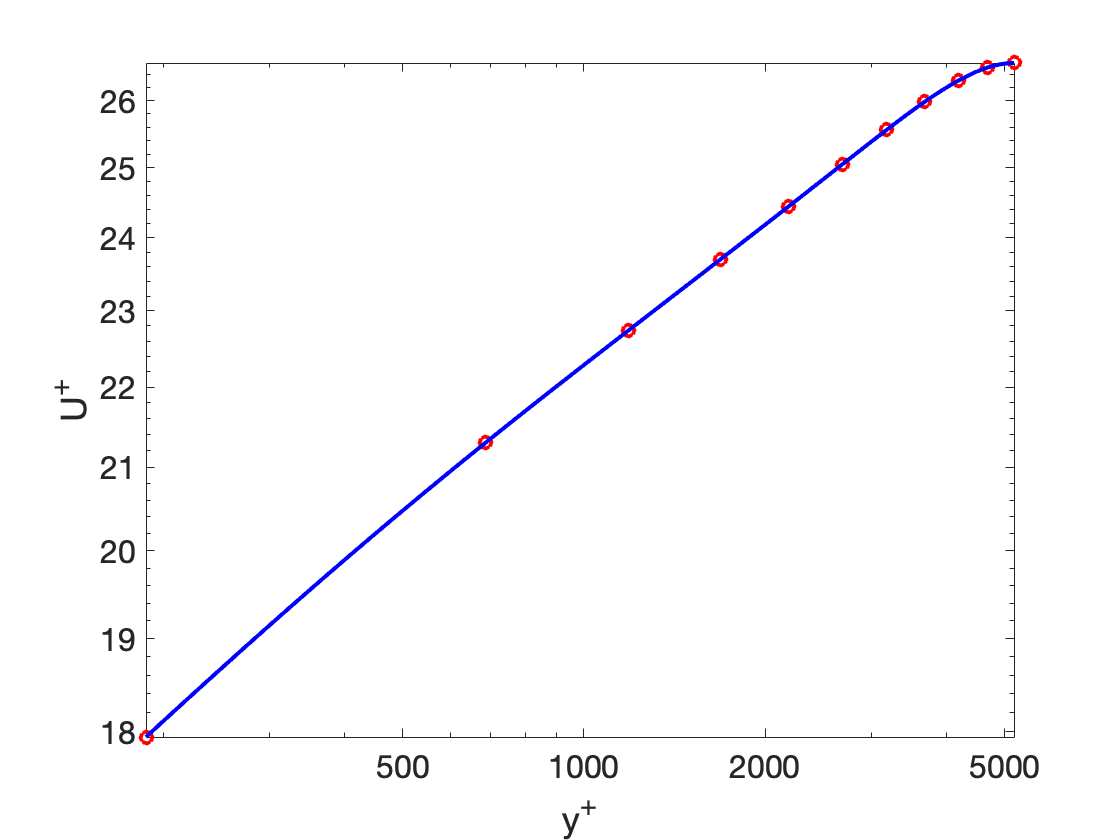}
        \includegraphics[width=0.4\textwidth]{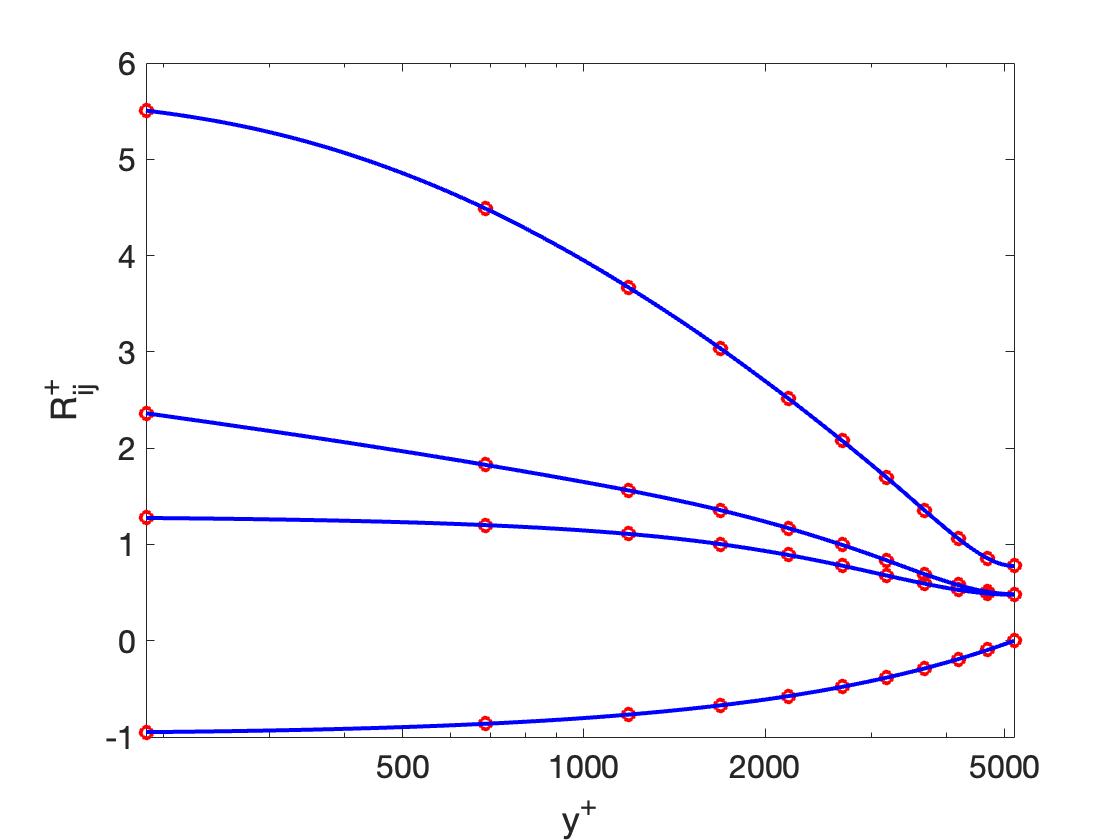}
    \caption{Reference (symbols) vs optimal attached eddy statistics for $Re_\tau \approx 5200$.  The moments are reconstructed from the inferred $I_1$ and $I_{ij}$ via Eqs.~\eqref{eq:U_of_y} and~\eqref{eq:rij_final}. The lowest $y^+$ shown corresponds approximately to $h_{\min}$, below which the AEM has no eddies and the model is not expected to apply.}
    \label{fig:Optimal2}
\end{figure}

Figure~\ref{fig:Optimal1} shows the optimal eddy contribution functions inferred from the channel flow data~\citep{lee2015direct} at friction Reynolds number $Re_{\tau}\approx 5200$. Figure~\ref{fig:Optimal2} confirms that the first and second velocity moments are perfectly reproduced across the entire channel. It is emphasized that the  perfect reproduction in Figure~\ref{fig:Optimal2} is expected by construction: the system is underdetermined ($2m$ unknowns, $m$ equations), and the minimum-norm solution exactly satisfies the constraints.  The sharp features near $\eta\approx 1$ visible in Figure~\ref{fig:Optimal1} are sensitive to the discretization and the minimum-norm selection criterion; by contrast, the qualitative structure (a near-plateau for $\eta \lesssim 1$ followed by rapid decay) is robust across different discretizations, regularization strategies, and data sources.  It is this robust qualitative structure, rather than the fine details, that motivates the simplified model in Section~\ref{sec:simplified_model}. The inversion uses moments across the full channel half-height ($y^+$ from $h_{\min}$ to $h_{\max}$), so the inferred $I_1(\eta)$ and $I_{ij}(\eta)$ are constrained by the entire wall-normal profile, not only the log layer.

\subsection{A simplified influence-function model and the log law}
\label{sec:simplified_model}

\begin{figure}
    \centering
    \includegraphics[width=0.5\textwidth]{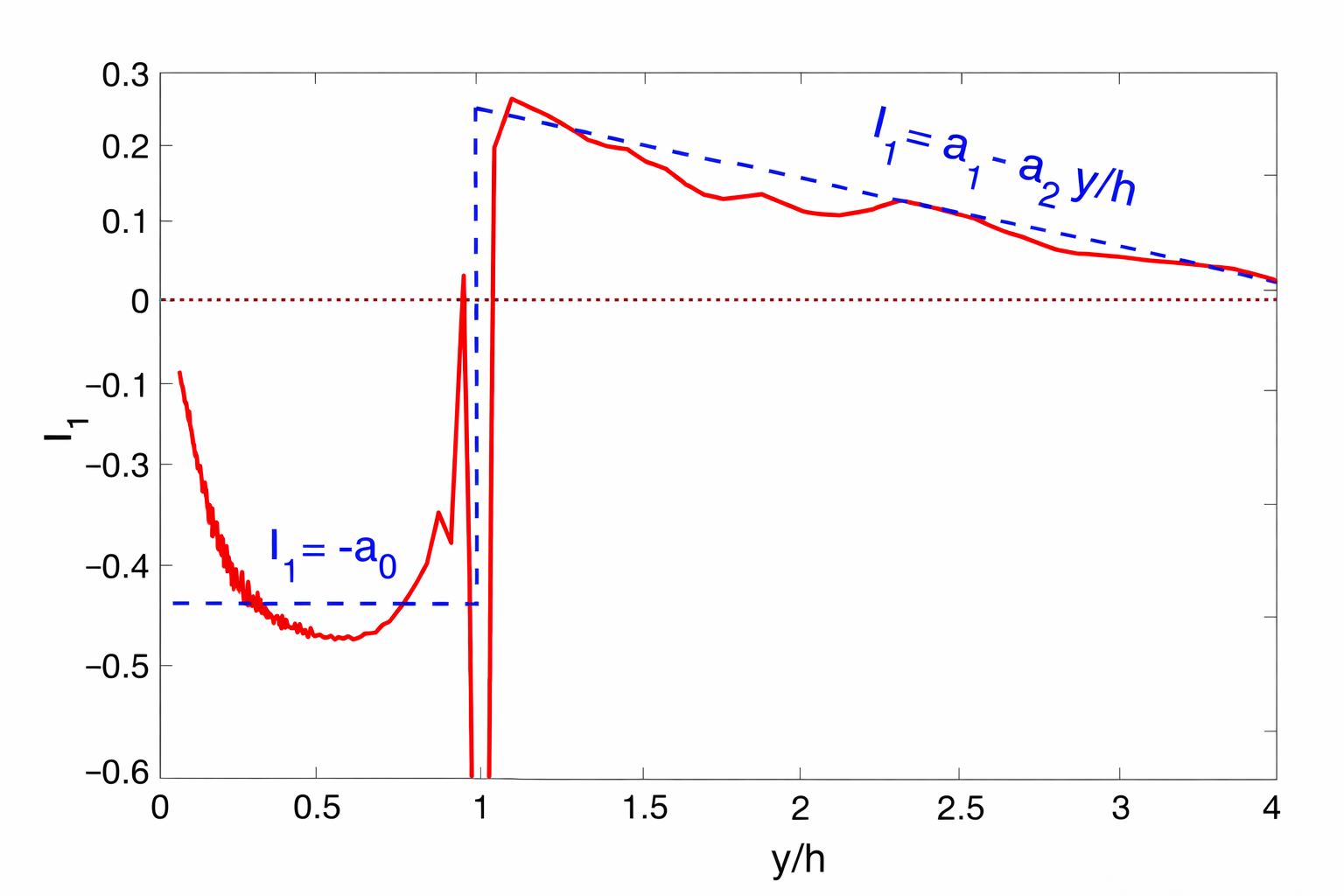}
    \caption{A hypothetical model of the eddy influence function corresponding to the mean streamwise velocity (blue dashed lines) compared to the optimal influence function.}
    \label{fig:model}
\end{figure}

We remark that the underlying physics is more complex than attached eddies can represent. The inverse problem, however, yields a function that, while smooth, has a visually interpretable structure: a near-plateau over $\eta=y/h\lesssim 1$ followed by a transition and decay for $\eta>1$. This motivates the simple piecewise model shown in Fig.~\ref{fig:model}.

Consider a collection of attached eddies that result in a simple function as shown in Fig.~\ref{fig:model}.
It will be shown below that it is indeed possible to construct an attached eddy that corresponds to such a function. Using this influence function, one can reconstruct the mean streamwise velocity as
\begin{align*}
U(y)&= \frac{2 \beta}{1/h_{\min}^2-1/h_{\max}^2} \int_{h_{\min}}^{h_{\max}}  \frac{I_1\left({y}/{h} \right)}{h} dh + U_{ref}\\
&\approx \frac{2 \beta}{1/h_{\min}^2-1/h_{\max}^2} \left[\int_{y/h_{\max}}^{1}  \frac{-a_0}{y/h} d(y/h) + \int_1^{a_1/a_2}  \frac{a_1-a_2 y/h}{y/h} d(y/h) \right] + U_{ref}\\
&= \frac{2 \beta}{1/h_{\min}^2-1/h_{\max}^2} \left[a_0 \log[y/h_{\max}]+a_1 \log[a_1/a_2]-a_1+a_2\right] + U_{ref}.
\end{align*}
It is thus clear that the K\'arm\'an constant can be constructed as
\begin{equation}
\kappa = \frac{1/h_{\min}^2-1/h_{\max}^2}{2 a_0 \beta}.
\label{eq:karman}
\end{equation}
Note that W\&M use a Taylor series approximation on a generic eddy to derive an alternate expression for $\kappa$. While that is insightful, it is perhaps a valid approximation only for certain types of eddy influence functions.

Eq.~\eqref{eq:karman} provides a direct route from measurable quantities ($\kappa$, $Re_\tau$) and the eddy population density $\beta$ to the plateau value $a_0$ of the influence function, or conversely, from the inferred $a_0$ and a prescribed $\beta$ to a prediction of $\kappa$. As will be shown in the following sections, employing a rectangular hairpin yields a precisely constant $I_1$ for $y/h \leq 1$. In sections 3 and 4, $\beta$ is chosen to fit the log law slope. In Section 6, a variable $\beta$ model is prescribed.


\section{Predictive Model and Insights}
\begin{figure}
    \centering
    \includegraphics[width=0.2\textwidth]{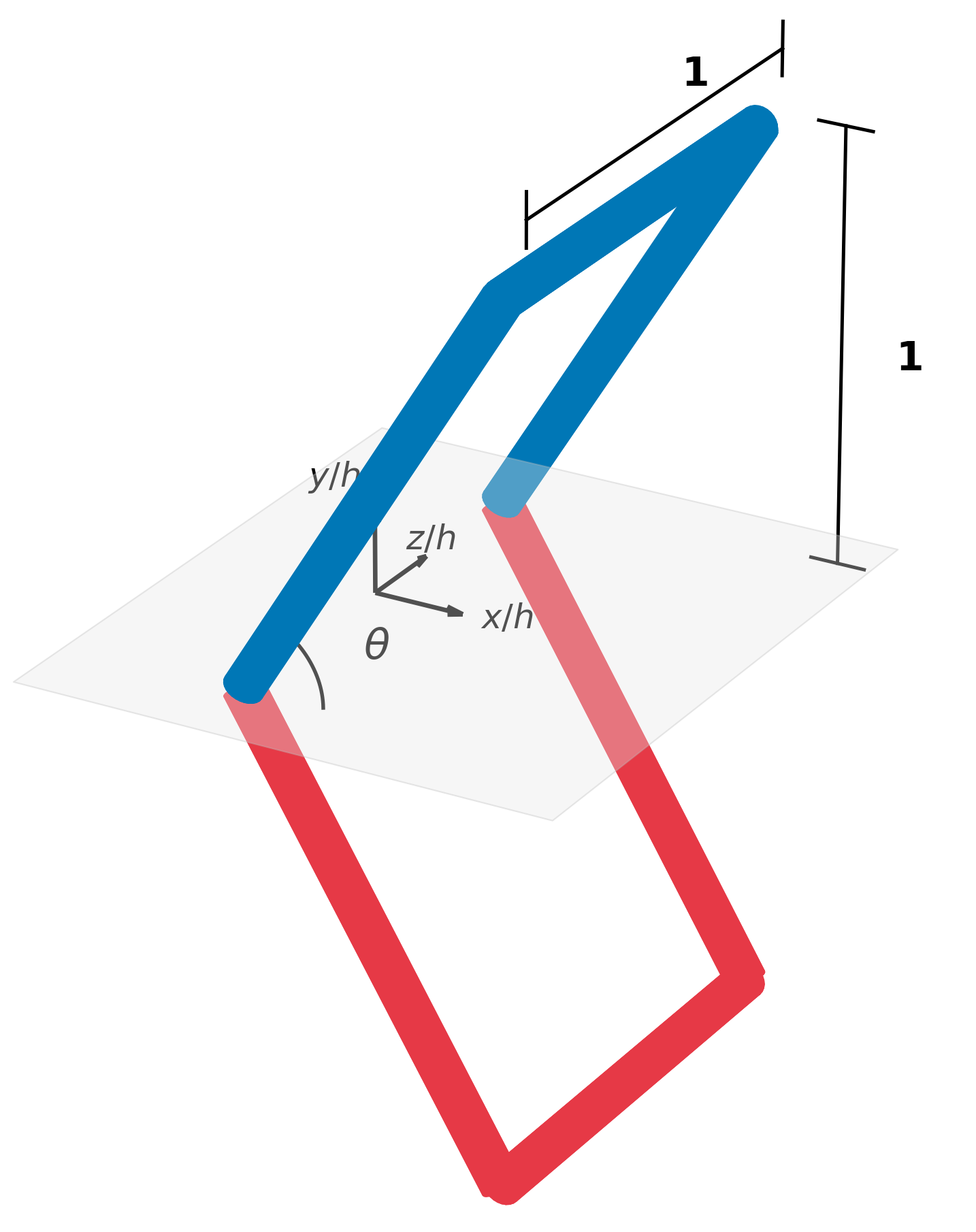}
    \caption{A prototypical hairpin-type eddy. }
    \label{fig:hairpin45}
\end{figure}

Now, we answer the question whether it is possible to recreate the above hypothetical influence function using an attached eddy. Consider a  hairpin (Figure~\ref{fig:hairpin45}) with unit circulation as the attached eddy, along with its image across the $y=0$ plane to enforce no-penetration. We consistently achieved good results using Rankine vortex-rods instead of vortex filaments. All the results henceforth correspond to vortex segments represented as Rankine vortex rods with core radius $\approx 0.025$ units relative to Fig.~\ref{fig:hairpin45}. 

\begin{figure}
    \centering
     \includegraphics[width=0.49\textwidth]{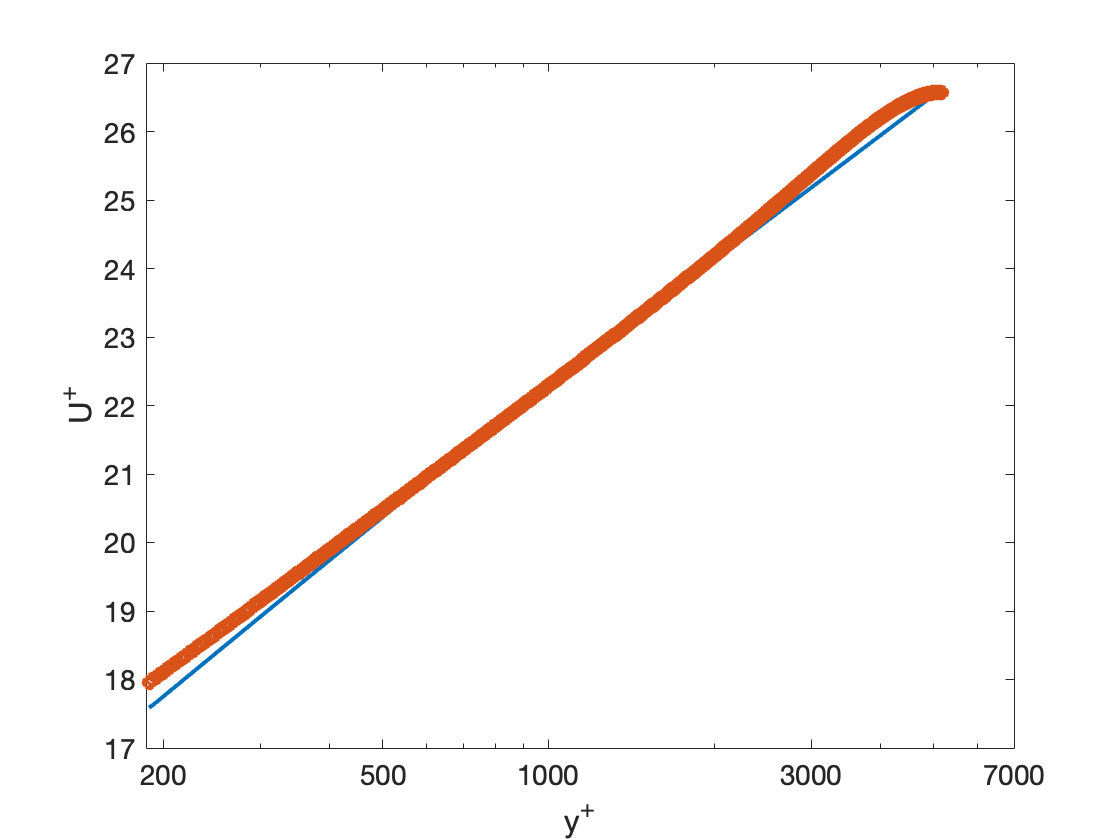}
        \includegraphics[width=0.49\textwidth]{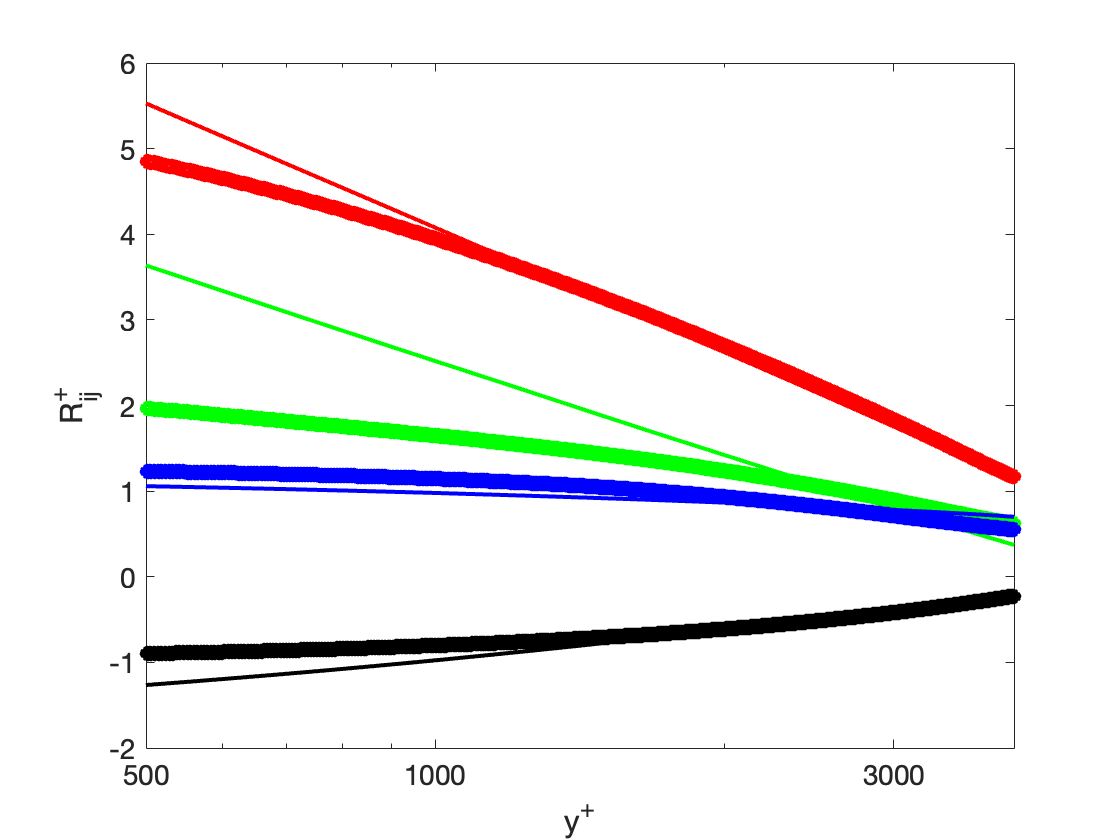}
    \includegraphics[width=0.49\textwidth]{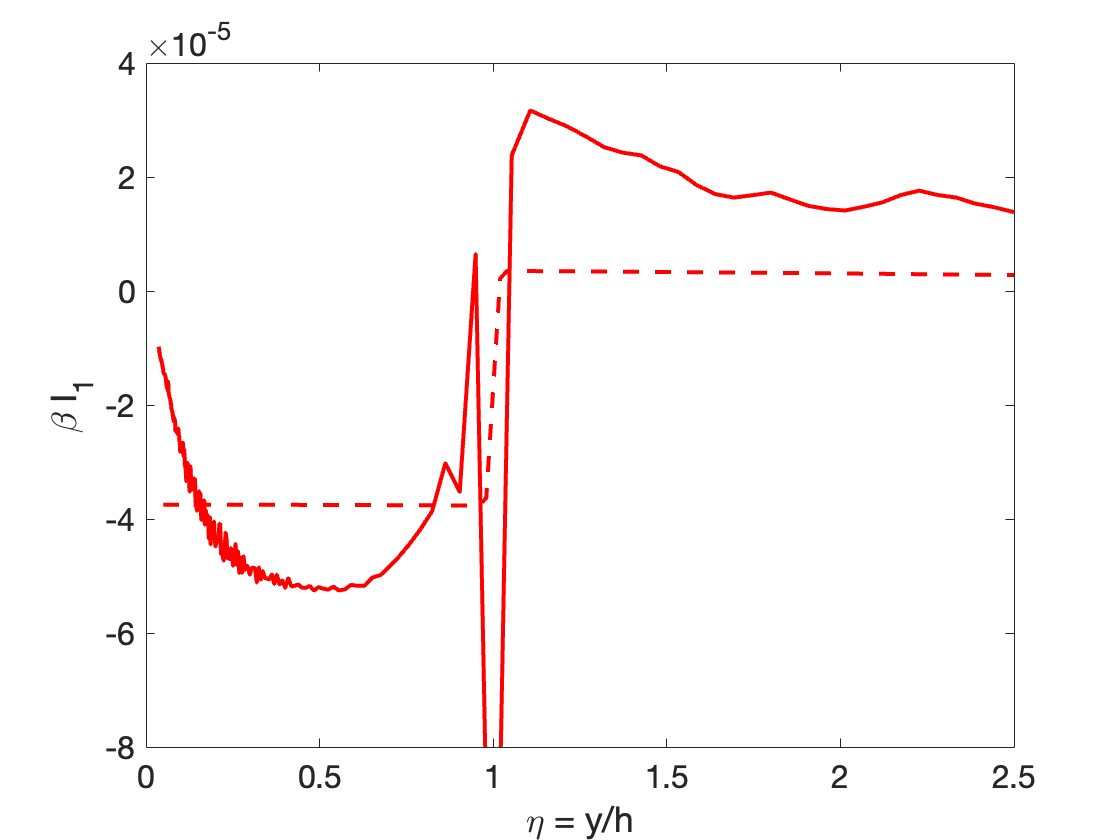}
        \includegraphics[width=0.49\textwidth]{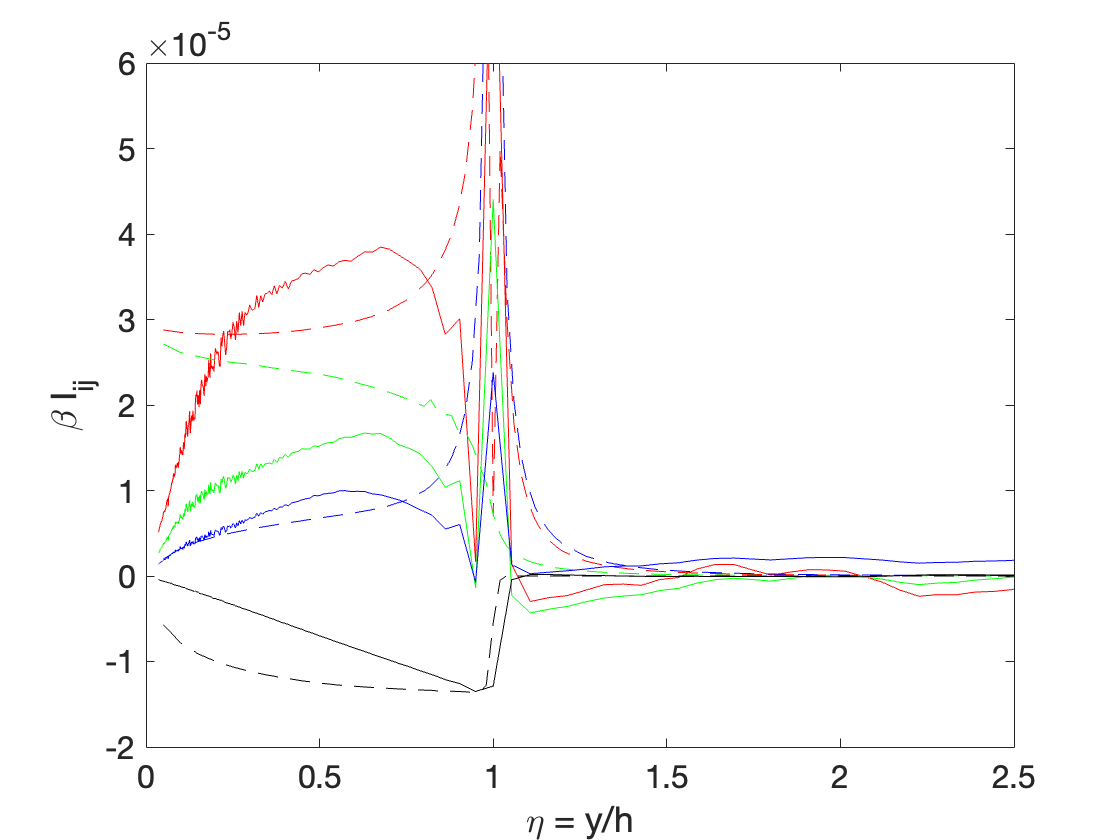}
    \caption{Top figures show velocity moments. Symbols: Data;  Lines: rectangular hairpin at $\theta= 60^\circ$..
    	Bottom figures show eddy influence functions. Solid lines: Optimal eddy influence function. Dashed lines: rectangular hairpin at $\theta= 60^\circ$. For the right figures, colors follow Figure~\ref{fig:Optimal1}.  Note that the $y^+ < 500$ region is not shown here because in the second moments, the log scaling takes hold in the outer part of the log layer. Refer Figure~\ref{fig:Optbeta} for a full view of this phenomenon.}
    \label{fig:hairpin45res}
\end{figure}

 It is remarked  that this eddy corresponds to a closed loop in the sense that the leg of the eddy aligned with the wall is canceled by an equal and opposite image vortex pair, and is thus consistent with the vorticity kinematics. Using the Biot--Savart law to compute the induced velocity $u_1$ in Eq.~\eqref{eq:zz}, Figure~\ref{fig:hairpin45res} shows that such a simple attached eddy can reproduce the mean streamwise velocity  accurately.  Further, the second moments are shown in Figure~\ref{fig:hairpin45res}, and the Reynolds stresses trends are well-predicted, though the  the spanwise stresses are over-estimated in the region  $y^+ < 2000$.

It is noted that an extensive optimization was performed on different eddy shapes (e.g. triangular hairpin) and  packet configurations. Yet, these did not yield improved predictions, particularly encountering difficulties in matching the log law slope while providing reasonable second moment and spectral behavior. The reasons behind this observation will be examined in detail in Section~\ref{sec:whysquare}. 

\begin{figure}
	\centering
	\includegraphics[width=0.7\textwidth]{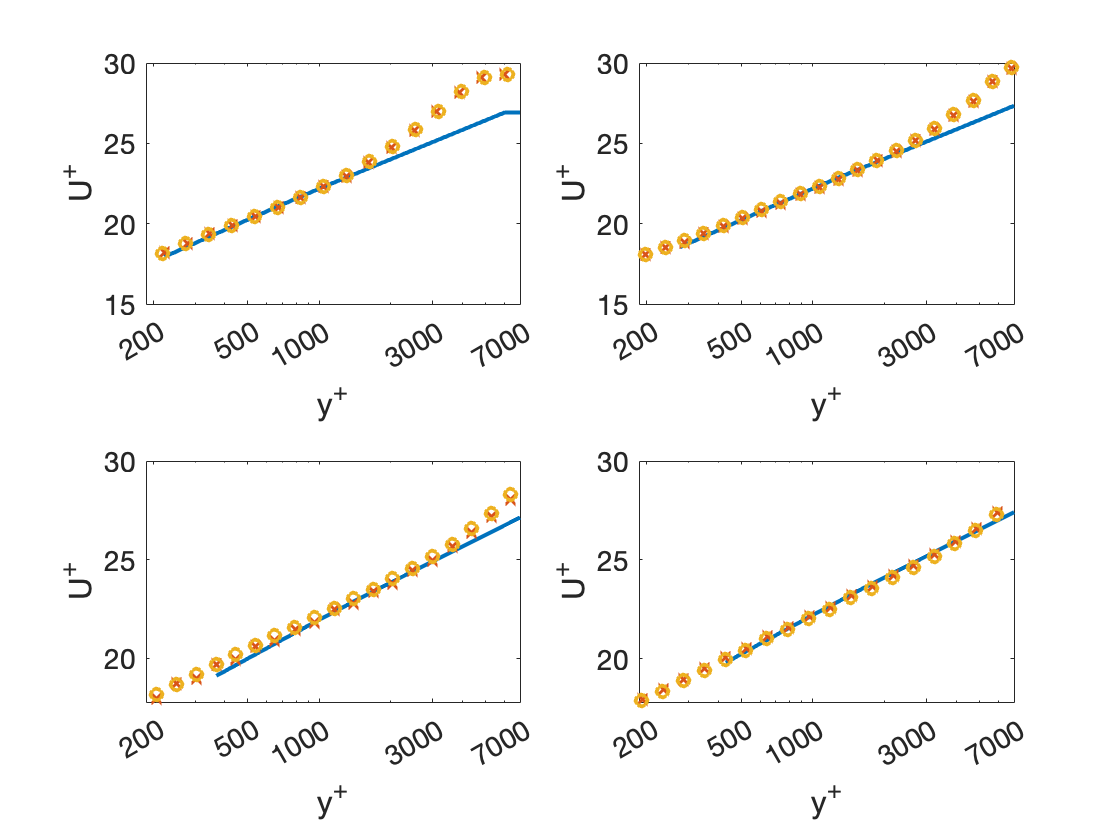}
	\caption{Mean velocity prediction using rectangular hairpin at $\theta= 60^\circ$ for $Re_\tau = [6000, 10000, 14500, 20000]$. Symbols: ~\cite{samie2018fully}.}
	\label{fig:ReU}
\end{figure}

\begin{figure}
	\centering
	\includegraphics[width=0.7\textwidth]{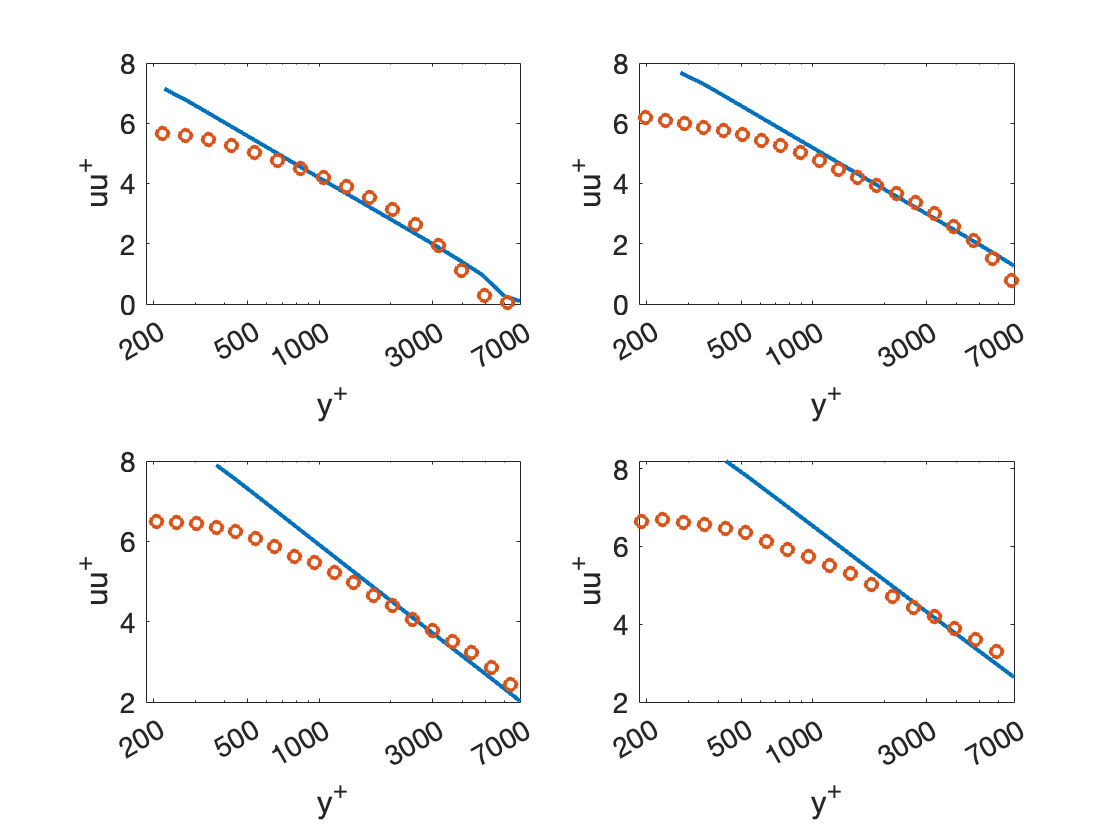}
	\caption{Streamwise velocity fluctuation predictions using rectangular hairpin at $\theta= 60^\circ$ for $Re_\tau = [6000, 10000, 14500, 20000]$. Symbols: ~\cite{samie2018fully}.}
	\label{fig:ReUU}
\end{figure}

\begin{figure}
	\centering
\includegraphics[width=0.35\textwidth]{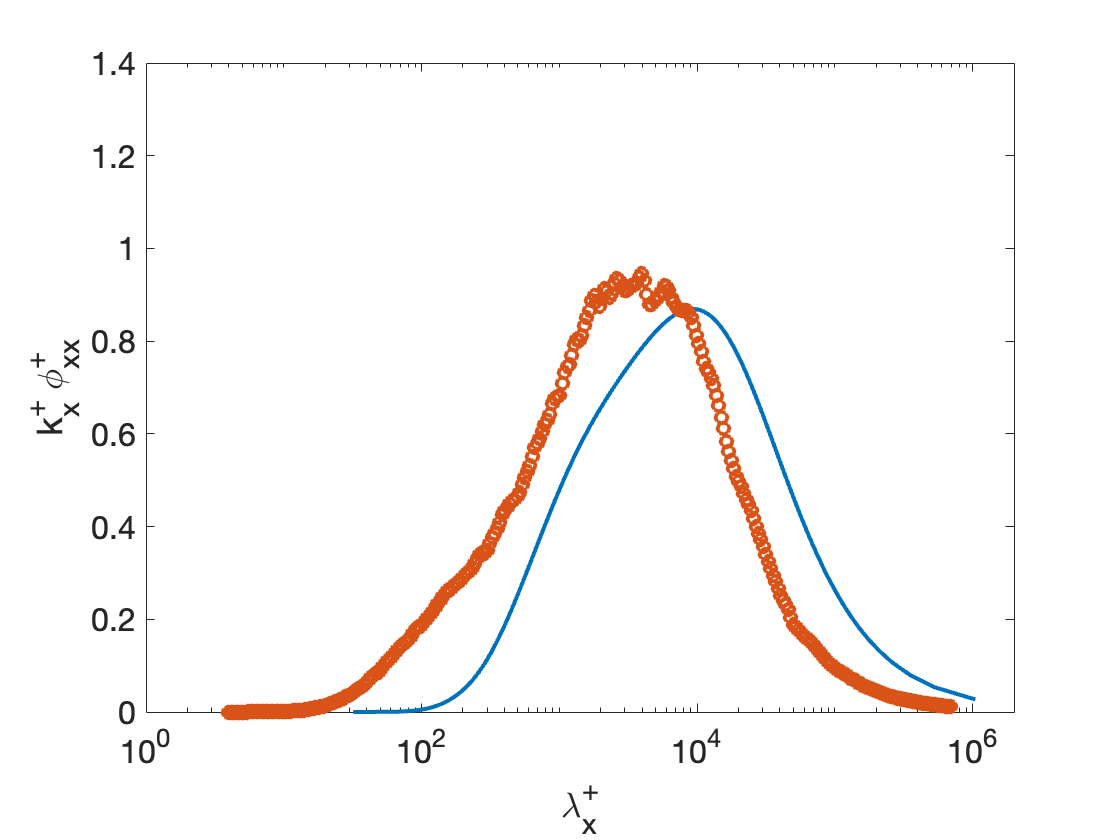}
\includegraphics[width=0.35\textwidth]{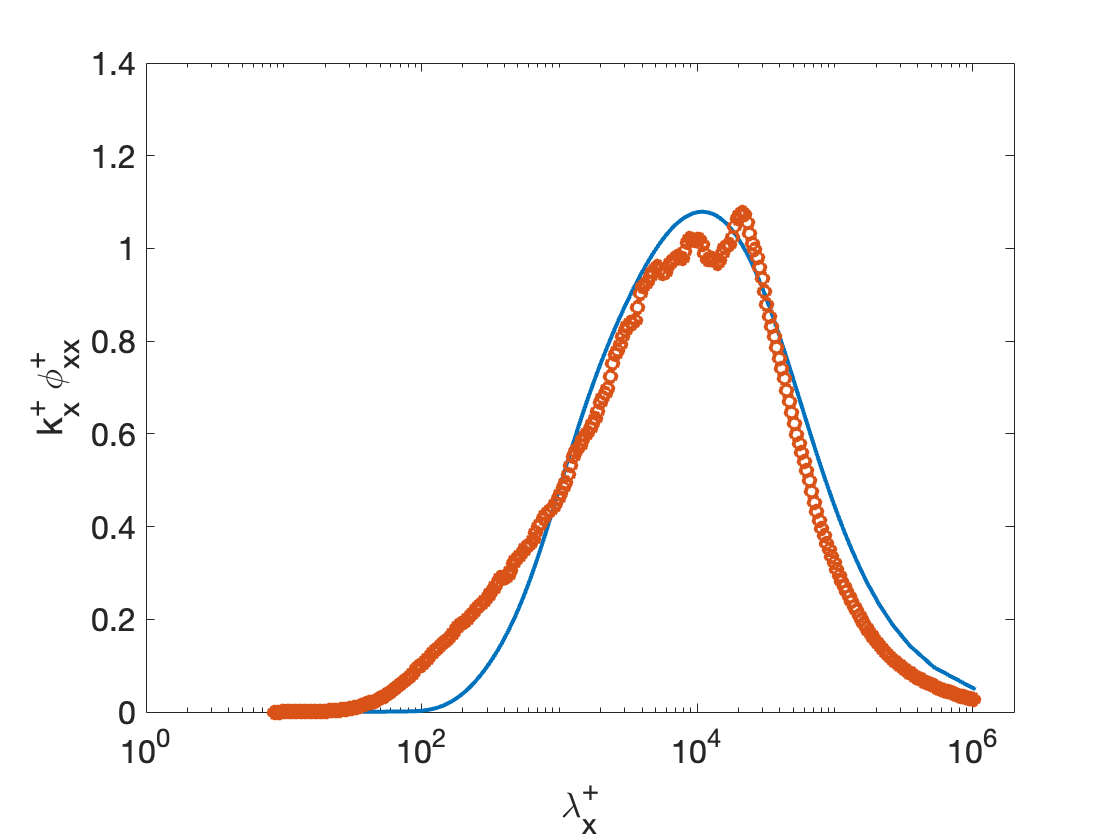}
\includegraphics[width=0.35\textwidth]{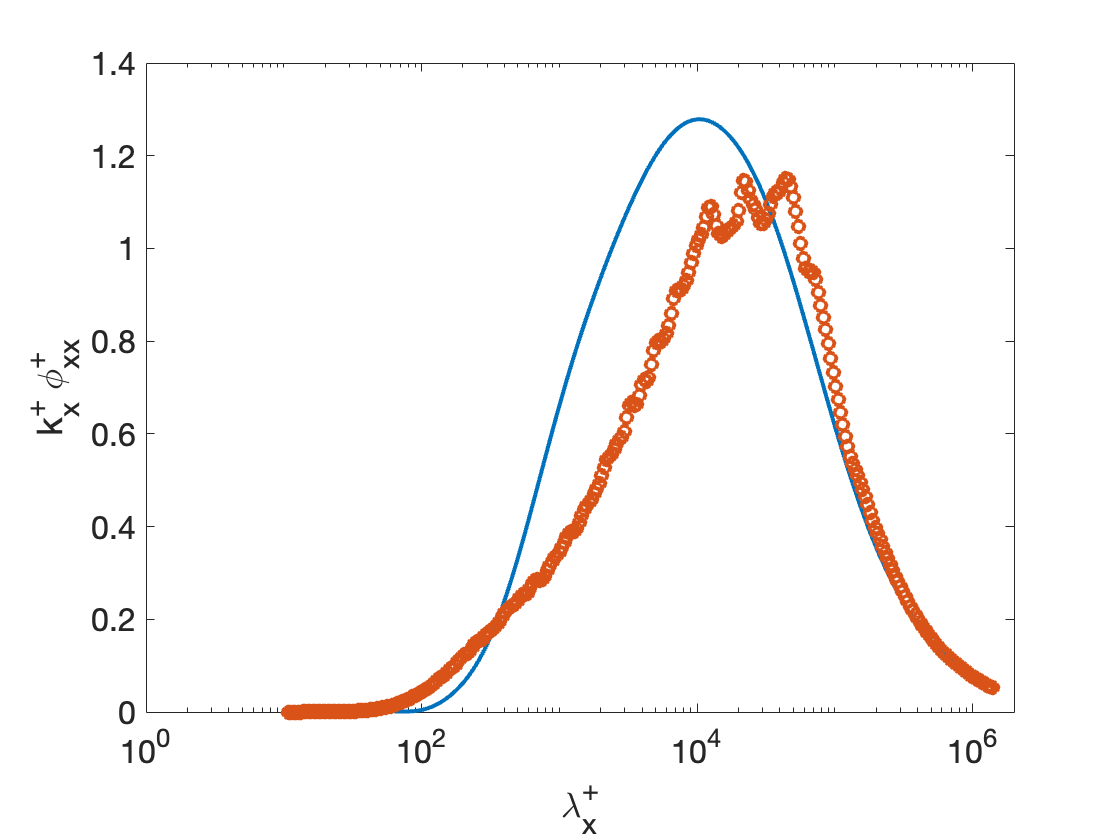}
\includegraphics[width=0.35\textwidth]{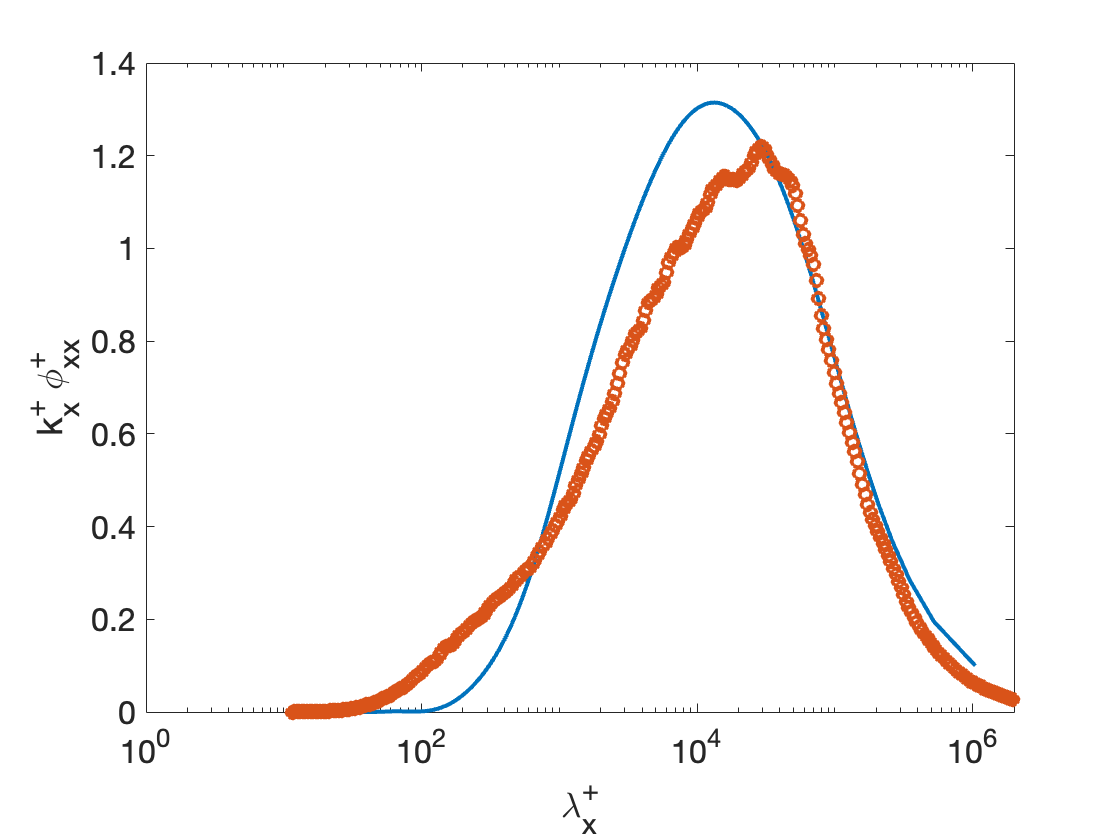}
	\caption{Pre-multiplied streamwise energy spectra predictions at $y^+  = 1184 $ using rectangular hairpin at $\theta= 60^\circ$ for $Re_\tau = [6000, 10000, 14500, 20000]$. Symbols: ~\cite{samie2018fully}.}
	\label{fig:Premul}
\end{figure}

The rectangular hairpin at an angle of $\theta = 60^\circ$ is close to the optimal solution across different metrics and shows comparisons to the experimental measurements from \cite{samie2018fully}. As shown in Figures~\ref{fig:ReU},\ref{fig:ReUU},\ref{fig:Premul}, the model remains remarkably accurate  in predicting mean velocity, streamwise Reynolds stress and spectra over a range of Reynolds numbers.

We note that the influence functions were inferred from channel flow DNS at $Re_\tau\approx 5200$~\citep{lee2015direct}, whereas the predictions in Figures~\ref{fig:ReU}--\ref{fig:Premul} are compared against zero-pressure-gradient boundary layer measurements at $Re_\tau=6000$--$20000$~\citep{samie2018fully}. The fact that the same eddy template transfers successfully across flow geometries supports the universality of the log-layer structure, but this correspondence deserves further scrutiny, particularly for the Reynolds stress components and for the outer region where channel and boundary layer statistics diverge.

We also note that the model predictions align better with the data above the log layer than within it. This is not contradictory despite $p(h) \propto h^{-3}$ being questionable in the outer region: the rapid decay of $I_1(\eta)$ for $\eta > 1$ ensures that the outer-region statistics are dominated by the largest eddies ($h \sim h_{\max}$), whose population amplitude is anchored by $\beta$ through the log-law calibration. The kernel structure naturally suppresses contributions from eddies smaller than the observation height, so the model works above the log layer primarily because of the influence-function shape rather than the population law.

The inverse problem  strongly constrains what any viable attached eddy must do in a wall-parallel plane. In particular, the inferred mean influence function $I_1(y/h)$ exhibits (a) a near-constant region for $y/h\lesssim 1$ and (b) rapid decay for $y/h\gtrsim 1$, implying that eddies larger than the observation height dominate the log-region population integral. The rectangular hairpin + image realizes these features because the two legs and head generate partially canceling far-field contributions, while the inviscid image enforces wall compatibility and suppresses the leading-order tail that would otherwise make the planar influence integrals marginal (Appendix~A). By contrast, a triangular hairpin produces a more monotone ``ramp'' in $I_1$ and does not reproduce the inferred plateau as cleanly (Figure~\ref{fig:hairpin45res}). Experimentation with a pack of triangular hairpins with variable circulation strengths did not improve results beyond the rectangular hairpin which appears optimal.

The inclined hairpin footprint produces elongated streamwise $u$ signatures, which is precisely the kind of wall-parallel signature that feeds the low-$k_x$ premultiplied hump. The log-layer statistics depend weakly on microscopic details once a few structural constraints are met.   This eddy structure appears to have some  key features that yield accurate streamwise energy spectra in the log layer. We  examine this below by providing an explicit similarity map from eddy morphology to the single-eddy spectrum as a function of relative height.

\section{The Influence kernel}
 The planar influence functions $I_1(y/h)$ and $I_{ij}(y/h)$ govern one-point moments via Eq.~\eqref{eq:U_of_y} and Eq.~\eqref{eq:rij_final}. For spectra, we introduce an analogous \emph{spectral} Influence kernel $I_\phi(\kappa_x,\eta)$ that compactly summarizes how a single self-similar eddy distributes streamwise energy across wavenumbers as a function of relative height $\eta=y/h$. To the author's knowledge, making this kernel explicit is not common in the attached-eddy literature; it is useful because it cleanly separates \emph{single-eddy geometry} (captured by $I_\phi$) from \emph{population effects} (captured by $p(h)$ and $\beta$).

Under our geometric  and circulation scaling assumptions,  the hairpin footprint is self-similar:
\[
\mathcal{U}\!\left(\frac{x}{h},\frac{z}{h},\frac{y}{h}\right)
=
\mathcal{U}(\xi,\zeta,\eta),
\]
where $\mathcal{U}$ is the unit-eddy footprint expressed in dimensionless coordinates. Taking the Fourier transform and changing variables yields the exact scaling
$
\widehat{u}_h(k_x,k_z;y)
\triangleq
\,h^2\,\widehat{\mathcal{U}}(\kappa_x,\kappa_z;\eta),
$
where
$
\widehat{\mathcal{U}}(\kappa_x,\kappa_z;\eta)
\triangleq
\int_{\mathbb{R}^2}\mathcal{U}(\xi,\zeta,\eta)\,e^{-i(\kappa_x\xi+\kappa_z\zeta)}\,d\xi\,d\zeta.$
Therefore
\[
\Phi_{xx}^{(h)}(k_x,k_z;y)
=
\frac{\,h^4}{(2\pi)^2}\,
\left|\widehat{\mathcal{U}}(\kappa_x,\kappa_z;\eta)\right|^2.
\]

Integrating over $k_z$ and using  $k_z=\kappa_z/h$ gives the scalewise spectrum
\[
\phi_{xx}^{(h)}(k_x;y)
\triangleq
\int_{-\infty}^{\infty}\Phi_{xx}^{(h)}(k_x,k_z;y)\,dk_z
=
\,h^3\,I_\phi(\kappa_x,\eta),
\]
where we define the \emph{Influence kernel}   as
\[
	I_\phi(\kappa_x,\eta)
	\triangleq
	\frac{1}{(2\pi)^2}\int_{-\infty}^{\infty}
	\left|\widehat{\mathcal{U}}(\kappa_x,\kappa_z;\eta)\right|^2\,d\kappa_z.
\]
This identity is a concrete mathematical object that will help decode the energy spectrum in greater detail.  In the following subsection, we will formulate an inverse problem to extract the shape of the ideal kernel.

It is noted that the spectral influence kernel is related to the one-point Reynolds stress kernel by $
I_{11}(\eta) = \int_0^\infty I_\phi(\kappa_x,\eta)\,d\kappa_x,$
 providing a consistency check between the spectral and one-point inverse problems.

\subsection{Inverse Problem to Extract Ideal Influence Kernel}
\label{sec:inverse_kernel}

The inversion uses the attached-eddy spectral superposition in the form
\begin{equation}
	\phi_{xx}(k_x;y)
	= \int_{h_{\min}}^{h_{\max}} \psi(k_x h, y/h)\,dh,
	\label{eq:forward_continuous}
\end{equation}
where
\[
\psi(\kappa,\eta) = \beta C I_{\phi}(\kappa_x,\eta),
\qquad
\kappa_x = k_x h,
\qquad
\eta = y/h.
\]
The numerical inversion solves for a gridded approximation to $\psi$.
The kernel is represented on a tensor-product grid in $(\kappa,\eta)$, and the unknown vector is the row-major flattening of the $n_\eta \times n_\kappa$ kernel array:
\[
\bm{x} = \operatorname{vec}(\Psi)
\in \mathbb{R}^{n_\eta n_\kappa},
\qquad
\Psi_{b,a} \approx \psi(\kappa_a,\eta_b),
\]
with the index map
$
\ell(b,a) = (b-1)n_\kappa + a,
\qquad
1 \le a \le n_\kappa,
\quad
1 \le b \le n_\eta.$
Hence $
\bm{x} = [\Psi_{1,1},\Psi_{1,2},\ldots,\Psi_{1,n_\kappa},\Psi_{2,1},\ldots,\Psi_{n_\eta,n_\kappa}]^T.
$ Equation~\eqref{eq:forward_continuous} is integrated over $h$ using $n_h$ geometrically spaced nodes $
h_1,\ldots,h_{n_h} \in [h_{\min},h_{\max}].$
The quadrature weights for a finite difference approximation are:
\begin{align*}
	w_1 &= h_2-h_1, \\
	w_q &= \tfrac12(h_{q+1}-h_{q-1}), \qquad 2\le q\le n_h-1, \\
	w_{n_h} &= h_{n_h}-h_{n_h-1}.
\end{align*}

We then solve the non-negative regularized least-squares problem
\begin{equation}
	\min_{\bm{x}\ge 0}
	\frac12\|\bm{A}\bm{x}-\bm{b}\|_2^2
	+ \frac{\lambda_\kappa^2}{2}\|\bm{L}_\kappa \bm{x}\|_2^2
	+ \frac{\lambda_\eta^2}{2}\|\bm{L}_\eta \bm{x}\|_2^2
	+ \frac{\lambda_0^2}{2}\|\bm{L}_0 \bm{x}\|_2^2,
	\label{eq:opt_problem}
\end{equation}
Equivalently, this is written as an augmented non-negative least-squares problem
\begin{equation}
	\min_{\bm{x}\ge 0}
	\frac12\|\widetilde{\bm A}\bm{x}-\widetilde{\bm{b}}\|_2^2,
	\label{eq:augmented_problem}
\end{equation}
with
\[
\widetilde{\bm A} =
\begin{bmatrix}
	\bm{A} \\
	\lambda_\kappa \bm{L}_\kappa \\
	\lambda_\eta \bm{L}_\eta \\
	\lambda_0 \bm{L}_0
\end{bmatrix},
\qquad
\widetilde{\bm{b}} =
\begin{bmatrix}
	\bm{b} \\
	\bm{0} \\
	\bm{0} \\
	\bm{0}
\end{bmatrix}.
\]
For the reported results,
$
\widetilde{\bm A} \in \mathbb{R}^{28159\times 5712},
\quad
\operatorname{nnz}(\widetilde{\bm A}) = 1029731.
$ and $
\lambda_\kappa = 10^{-1},
\quad
\lambda_\eta = 4\times 10^{-1},
\quad
\lambda_0 = 10^{-5}.$ While these numbers appear arbitrary, the goal is to obtain insight into the shape of the kernel, and thus the primary metric is to obtain a smooth solution that does not overfit to measurement noise.

\begin{figure}
	\centering
	\includegraphics[width=0.7\textwidth]{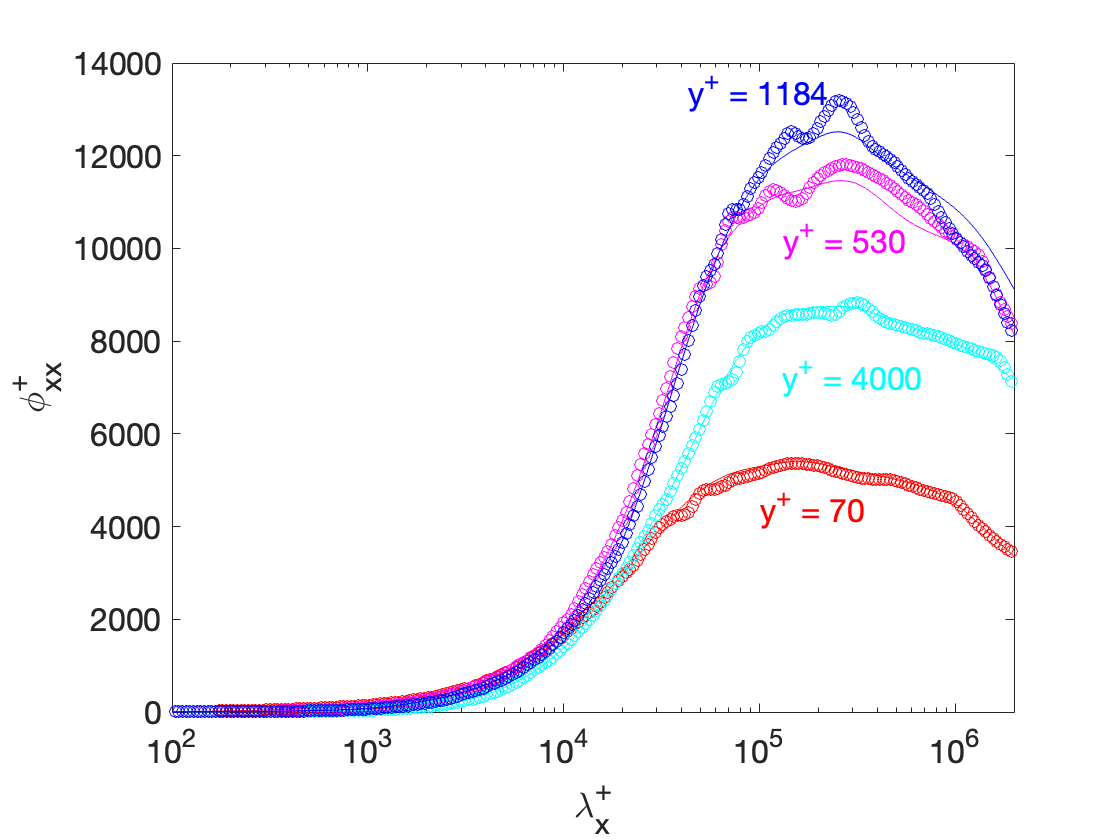}
	\caption{Data (symbols) and reconstructed (lines) streamwise energy spectra $\phi_{xx}^+(\kappa_x)$ for a few wall normal locations}
	\label{fig:reconstruct}
\end{figure}

 Appendix B provides more detail on the construction of these matrices. 
This is a more complicated inverse problem than the ones introduced earlier.
 Eq.~\eqref{eq:augmented_problem} is solved by projected FISTA~\citep{buccini2025krylov}. Figure~\ref{fig:reconstruct} shows the reconstructed streamwise energy spectra (from the inverse solution) and how it compares to the data. The agreement is good, and regularization is shown to smooth the solution to a small degree, but not enough to impact any conclusions.

The spectral data used for this inversion are from the experimental measurements of~\cite{samie2018fully} at $Re_\tau=20000$. A brief sensitivity study on the regularization weights $(\lambda_\kappa,\lambda_\eta,\lambda_0)$ confirmed that the qualitative shape of $I_\phi$, in particular the broad plateau at small $\eta$ and the localized hump at $\eta\sim O(1)$ is robust though finer features of the kernel near $\kappa_x\to 0$ are sensitive to the choice of $\lambda_\kappa$.

\begin{figure}
	\centering
	\includegraphics[width=0.7\textwidth]{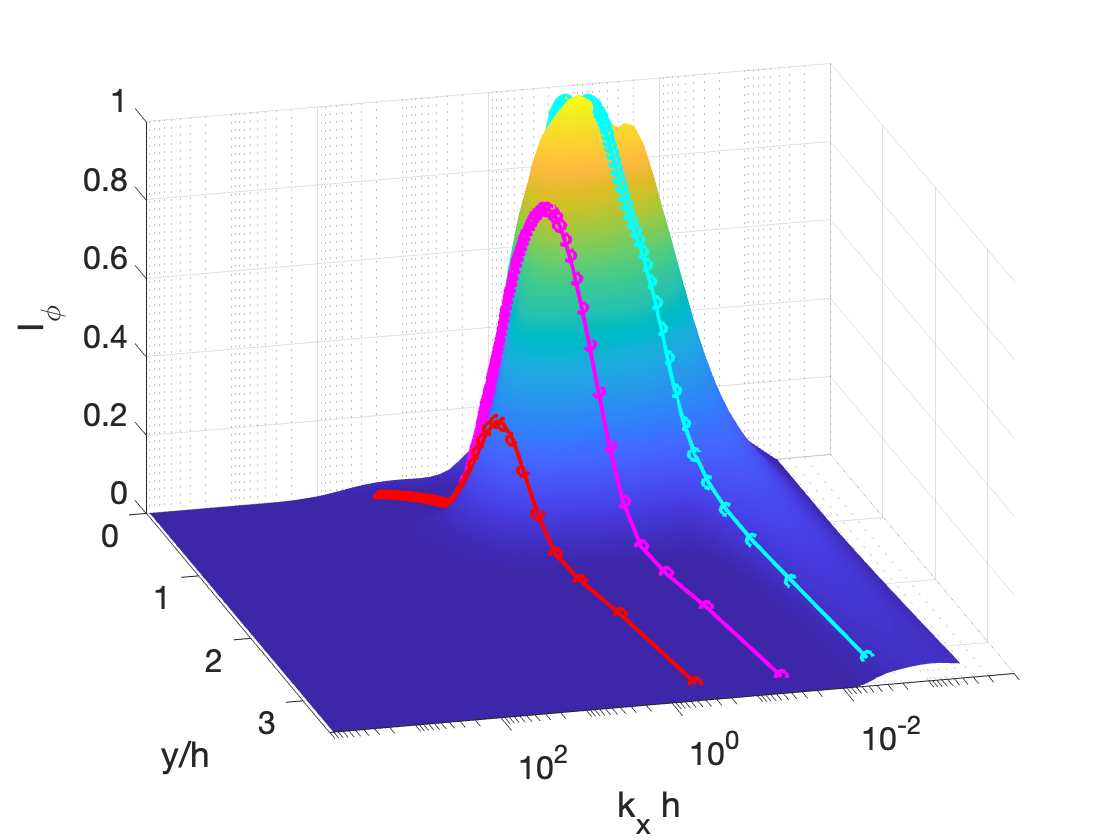}
	\caption{Inferred streamwise energy Influence kernel $I_\phi(\kappa_x,\eta)$ (normalized peak value). This figure highlights that the kernel has a broad, elevated region at small $\eta$ and small-to-moderate $\kappa$, and a more localized hump as $\eta$ increases.
		 Also shown are contributions to $\phi_{xx}^+(k_x^+)$ for $k_x^+ = 1 \times 10^{-5}$ (cyan) and $k_x^+ = 1 \times 10^{-4}$ (magenta) and $k_x^+ = 1 \times 10^{-3}$ (red)  at $y^+ = 1184$ for $Re_\tau=20000$.}
	\label{fig:Iphi}
\end{figure}

The extracted kernel $I_\phi(\kappa_x,\eta)$, which is a function of the dimensionless wavenumber $\kappa_x=k_x h$ at fixed relative height $\eta=y/h$ and is shown in Figure~\ref{fig:Iphi}.  It is immediately clear that eddies of scale $h < y $ have negligible influence on the statistics.

For any fixed $\eta$, typical scaling arguments yield a concentration of the influence \rev{$I_\phi(\kappa_x,\eta)$}  around $\kappa_x=O(1)$ because $\kappa_x$ is the only dimensionless streamwise wavenumber available to the unit eddy. Therefore, for a given size $h$ and fixed $y$, the dominant contribution of that eddy appears around
\[
k_x h = O(1)\quad \Longleftrightarrow\quad k_x = O(1/h),
\]
equivalently $\lambda_x=2\pi/k_x = O(h)$. This explains the peak of the premultiplied spectra in Fig.~\ref{fig:Premul}.  

\begin{figure}
	\centering
	\includegraphics[width=0.4\textwidth]{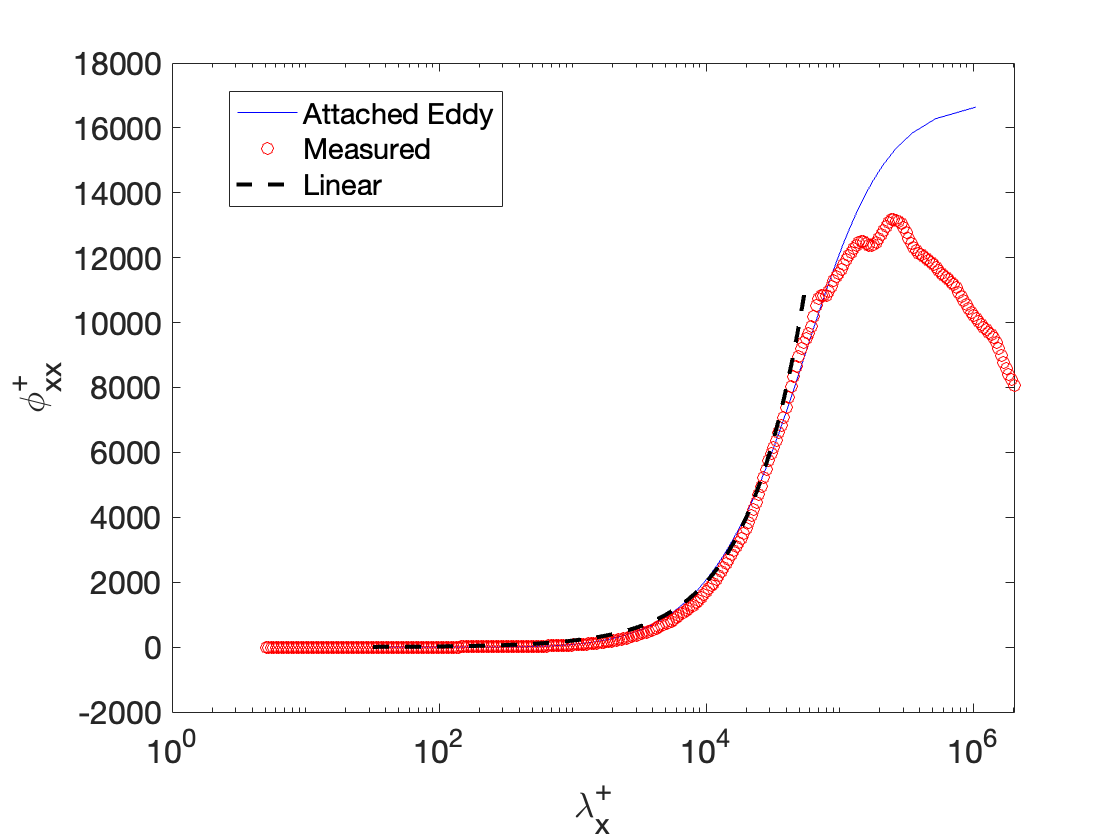}
		\includegraphics[width=0.4\textwidth]{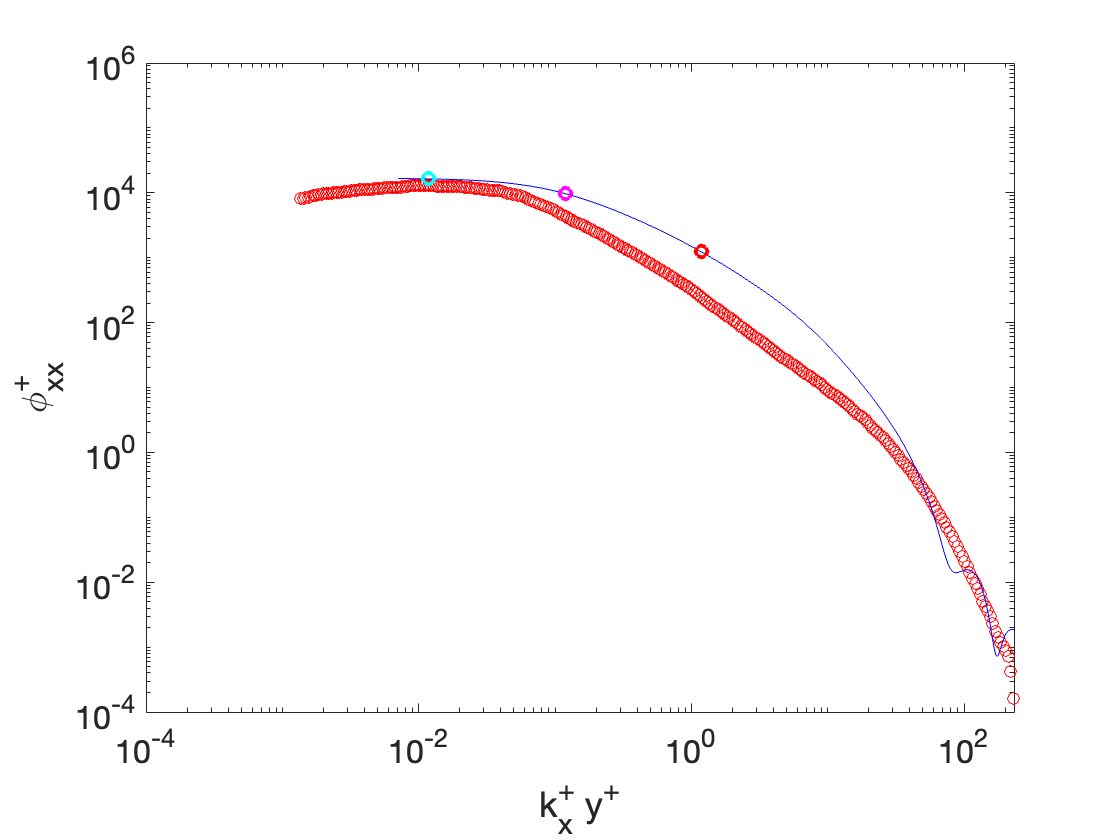}
	\caption{Streamwise energy spectra at $y^+ = 1184 $ for $Re_\tau=20000$.}
	\label{fig:linear}
\end{figure}

The rectangular hairpin Attached Eddy model is able to recover a near-perfect linear behavior of the streamwise energy spectrum in the intermediate scale range as shown in Fig.~\ref{fig:linear}. Below, we will give more concrete and quantitative explanations.

\subsection{From single-eddy spectra to the attached-eddy spectrum}

The attached-eddy prediction for the 1D spectrum is the population superposition
\[
\phi_{xx}(k_x;y)
=
\beta\int_{h_{\min}}^{h_{\max}} p(h)\,\phi_{xx}^{(h)}(k_x;y)\,dh,
\qquad p(h)=\frac{C}{h^3}.
\]
Insert the exact single-eddy factorization:
\[
\phi_{xx}(k_x;y)
=
\beta\int_{h_{\min}}^{h_{\max}} \frac{C}{h^3}\,
\big(\,h^3\,I_\phi(k_x h,y/h)\big)\,dh
=
\beta\,C\,\int_{h_{\min}}^{h_{\max}} I_\phi(k_x h,y/h)\,dh.
\]
Change variables $\kappa = k_x h$ so $dh = d\kappa/k_x$, and note that $y/h = yk_x/\kappa$:
\[
	\phi_{xx}(k_x;y)
	=
	\frac{\beta\,C\,}{k_x}
	\int_{k_x h_{\min}}^{k_x h_{\max}} I_\phi\!\left(\kappa,\frac{y k_x}{\kappa}\right)\,d\kappa.
\]
Define the integral factor
\[
F(k_x;y)
=
\int_{k_x h_{\min}}^{k_x h_{\max}} I_\phi \!\left(\kappa,\frac{y k_x}{\kappa}\right)\,d\kappa,
\]
so that
\[
\phi_{xx}(k_x;y)=\frac{\beta\,C\,}{k_x}\,F(k_x;y).
\]

\begin{figure}
	\centering
	\includegraphics[width=0.35\textwidth]{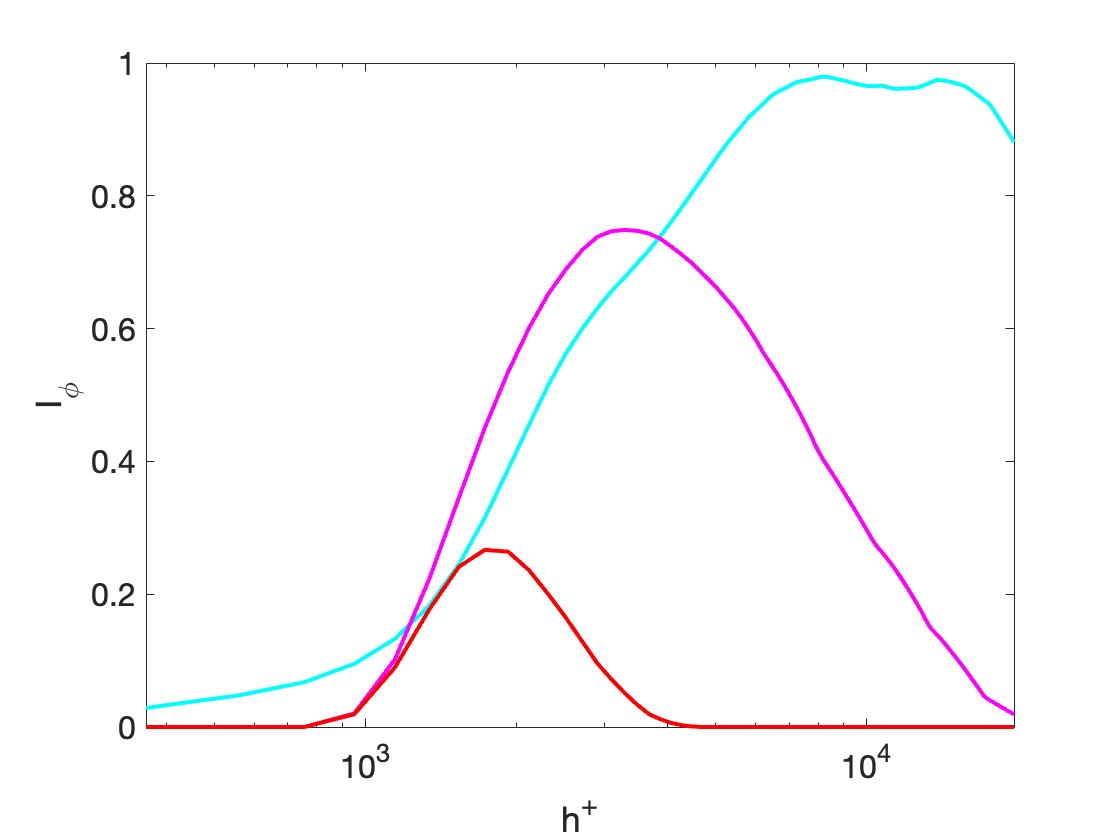}
		\includegraphics[width=0.35\textwidth]{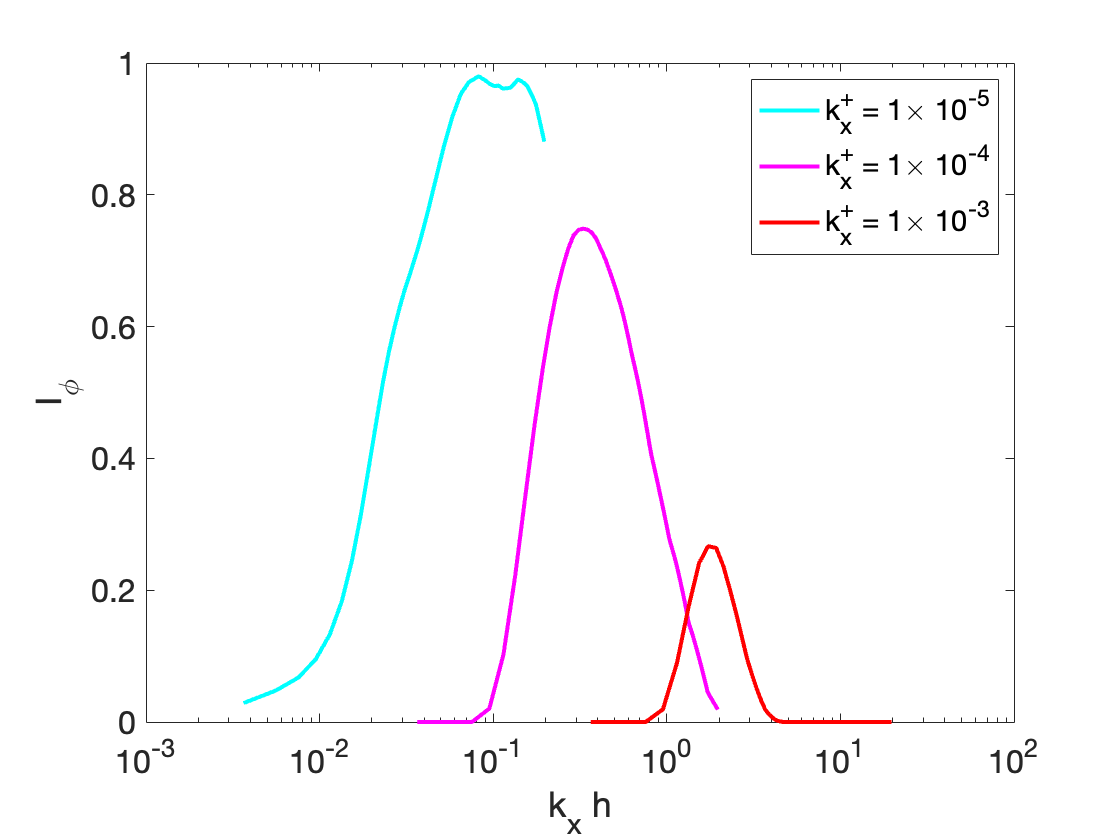}
	\caption{Contribution to $F$ integral from different scales at $y^+ = 1184 $ for $Re_\tau=20000$. The smallest-$k_x$ trajectory samples the left/low-$\kappa$ edge of the kernel and remains at small $\eta=y/h$.}
	\label{fig:cont}
\end{figure}

The $1/k_x$ scaling in the logarithmic region arises when $F(k_x;y)$ varies slowly with $k_x$ over an intermediate band. This occurs when the $\kappa$-window $[k_x h_{\min},k_x h_{\max}]$ covers the portion of the kernel $I_\phi(\kappa,\eta)$ that carries most of the contribution, and when the effective $\eta$ values encountered (through $\eta = y/h = yk_x/\kappa$) lie in the log-layer-relevant range. In that band,
\[
F(k_x;y)\approx \text{constant}\quad\Longrightarrow\quad \phi_{xx}(k_x;y)\propto \frac{1}{k_x} \propto \lambda_x.
\]
This is substantiated by the results in Fig.~\ref{fig:cont}.

	The above ``$1/k_x$'' argument corresponds to the  case that $F(k_x;y)$ is approximately constant over an intermediate $k_x$ range. However \emph{the low-$k_x$ behavior is entirely controlled by how $F(k_x;y)$ varies as $k_x\to 0$}. There is no universality guarantee there: when $k_x$ is small, the $\kappa$-window and the $\eta$-trajectory  can force the integral to sample the least robust part of the kernel.

	This is visible in the kernel visualization. Figure~\ref{fig:Iphi} shows the kernel surface, and Figure~\ref{fig:cont} shows three trajectories corresponding to fixed $k_x^+$ values. 
	The low-$\kappa$ region of the kernel is where $I_\phi(\kappa,\eta)$ typically changes most rapidly. In such a regime, the integral $F(k_x;y)$ cannot be expected to be constant: small shifts in $k_x$ move both the integration bounds and the trajectory through $(\kappa,\eta)$, producing non-negligible changes in $F$.

\begin{figure}
		\centering
		\includegraphics[width=0.7\textwidth]{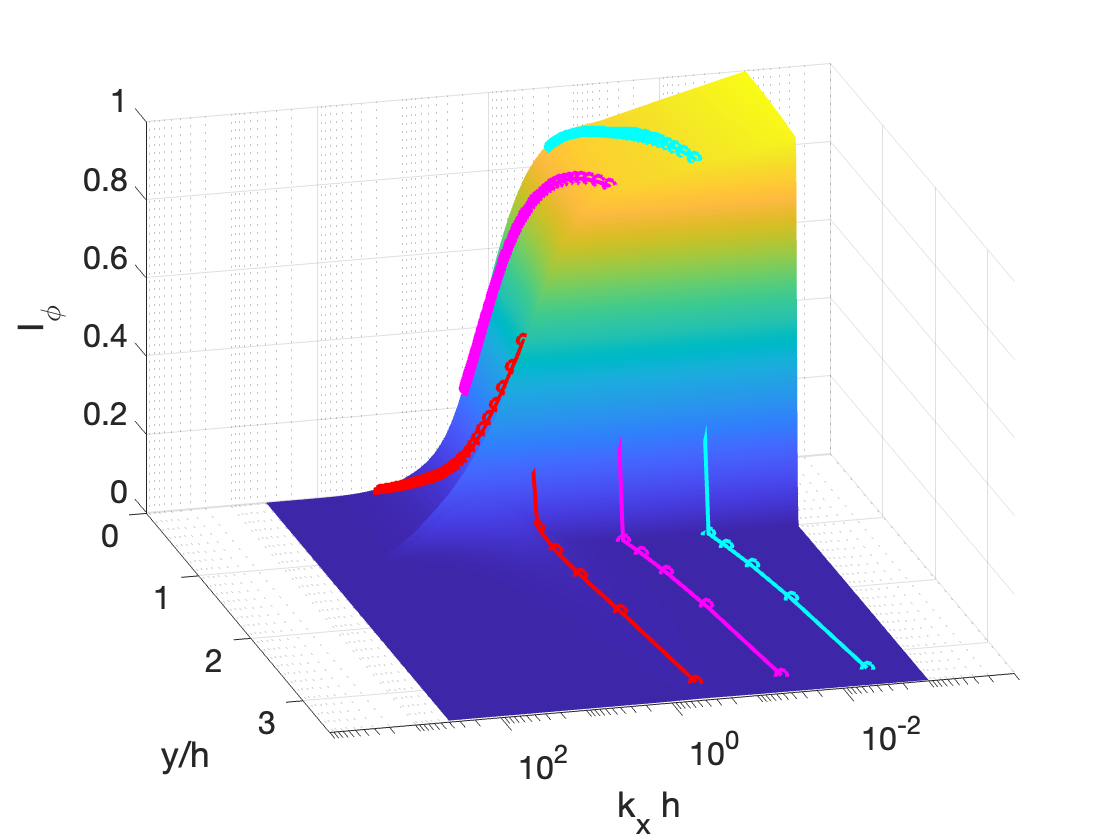}
		\caption{Streamwise energy Influence kernel $I_\phi(\kappa_x,\eta)$ (normalized peak value) for hairpin attached eddy at  $Re_\tau=20000$.}
		\label{fig:Iphi_AE}
	\end{figure}

\begin{figure}
		\centering
		\includegraphics[width=0.35\textwidth]{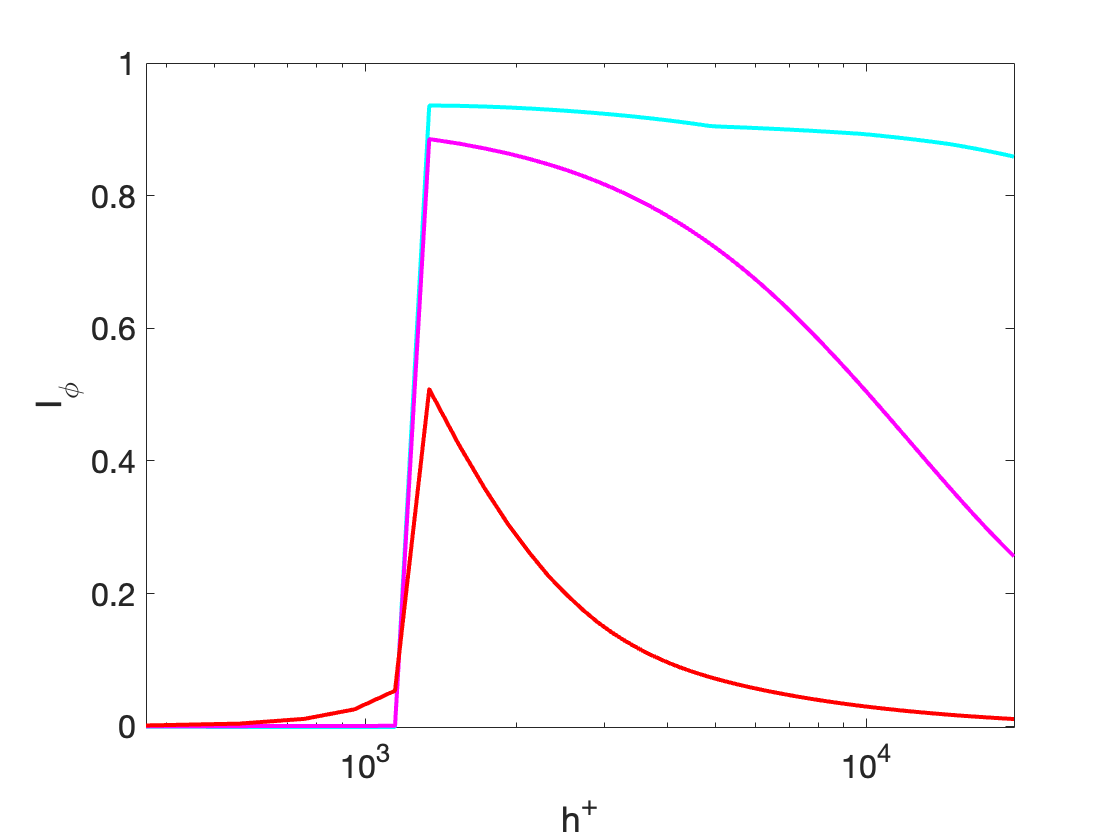}
		\includegraphics[width=0.35\textwidth]{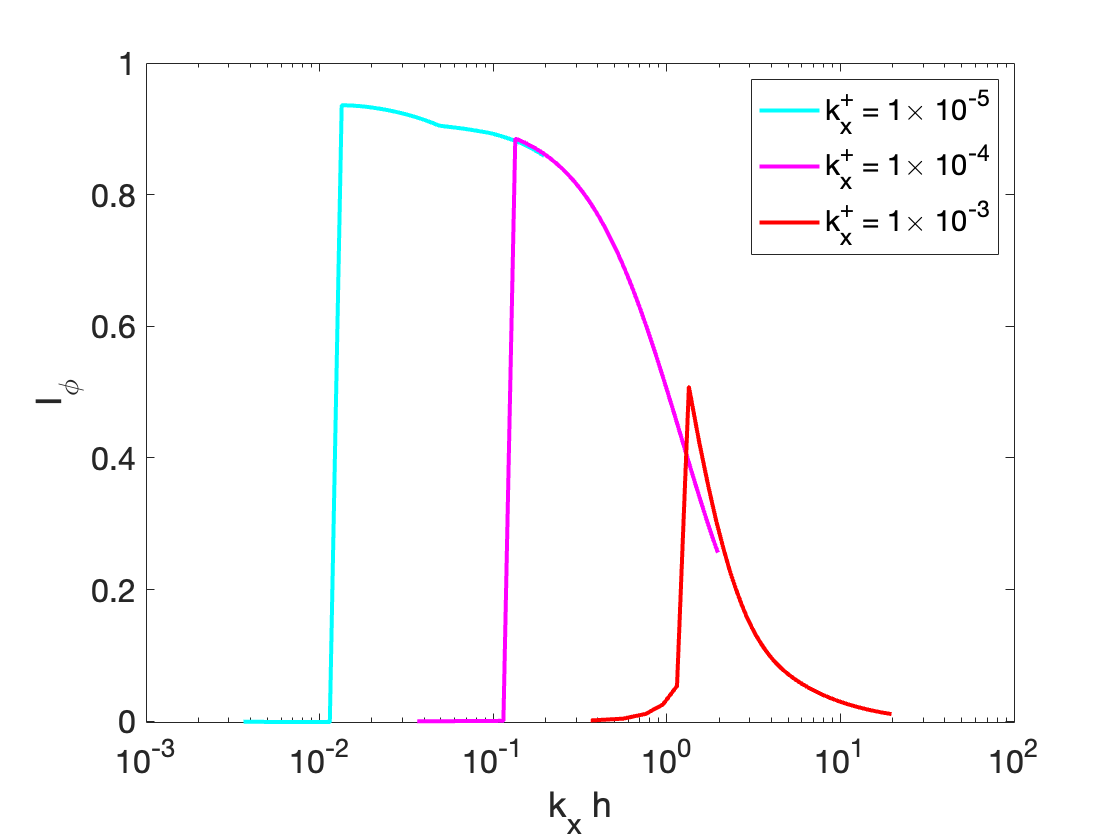}
		\caption{Contribution to $F$ integral from different scales at $y^+ = 1184 $ for $Re_\tau=20000$ for hairpin attached eddy.}
		\label{fig:cont_AE}
	\end{figure}

	The overprediction of streamwise energy spectra by the Attached Eddy model in the high wavelength limit is clearly explained by Figures~\ref{fig:Iphi_AE} and \ref{fig:cont_AE}. 	For small $\kappa_x$, this trajectory spends substantial ``arc length'' in the region of small $\kappa$ and small $\eta$ and $I_\phi$ does not reduce to zero for small $\kappa_x$ as in Figure~\ref{fig:Iphi}.  Nevertheless, the overall shape of the kernel and the  scale-by-scale contributions from the hairpin attached eddy appear impressive in relation to the inferred quantities.
	
	Physically, low $k_x$ corresponds to very long streamwise wavelengths. In the present attached-eddy framework with independent marks, the low-$k_x$ variance is built by incoherently summing contributions of the largest eddies.  In a Poisson superposition, long-wavelength modes are fed by many large eddies with essentially random phases, so variances add without cancellation. Real wall turbulence exhibits organization, intermittency, and correlations at the largest scales; those effects can reduce energy in the very lowest modes relative to an independent-eddy superposition. ~\cite{baars2020data1,baars2020data2} offer more quantitative insights in the form of 3 different eddy types, and that attached eddies represent one of these types and contribute a large fraction of the energy only at an intermediate range of scales in the log layer. The latter point has been  well argued by ~\cite{perry1995wall}.  The  independent-mark hypothesis therefore tends to overpopulate the lowest wavenumbers unless correlations are explicitly modeled or packaged into larger marks. It is possible that the influence kernel $I_\phi$ can be used in an inverse problem setting to `design' compositions of prototypical eddies to model desired spectral properties.

\section{Why the rectangular hairpin is unusually predictive, and  replacing it is difficult}
\label{sec:whysquare}
\begin{figure}
	\centering
	\includegraphics[width=0.5\textwidth]{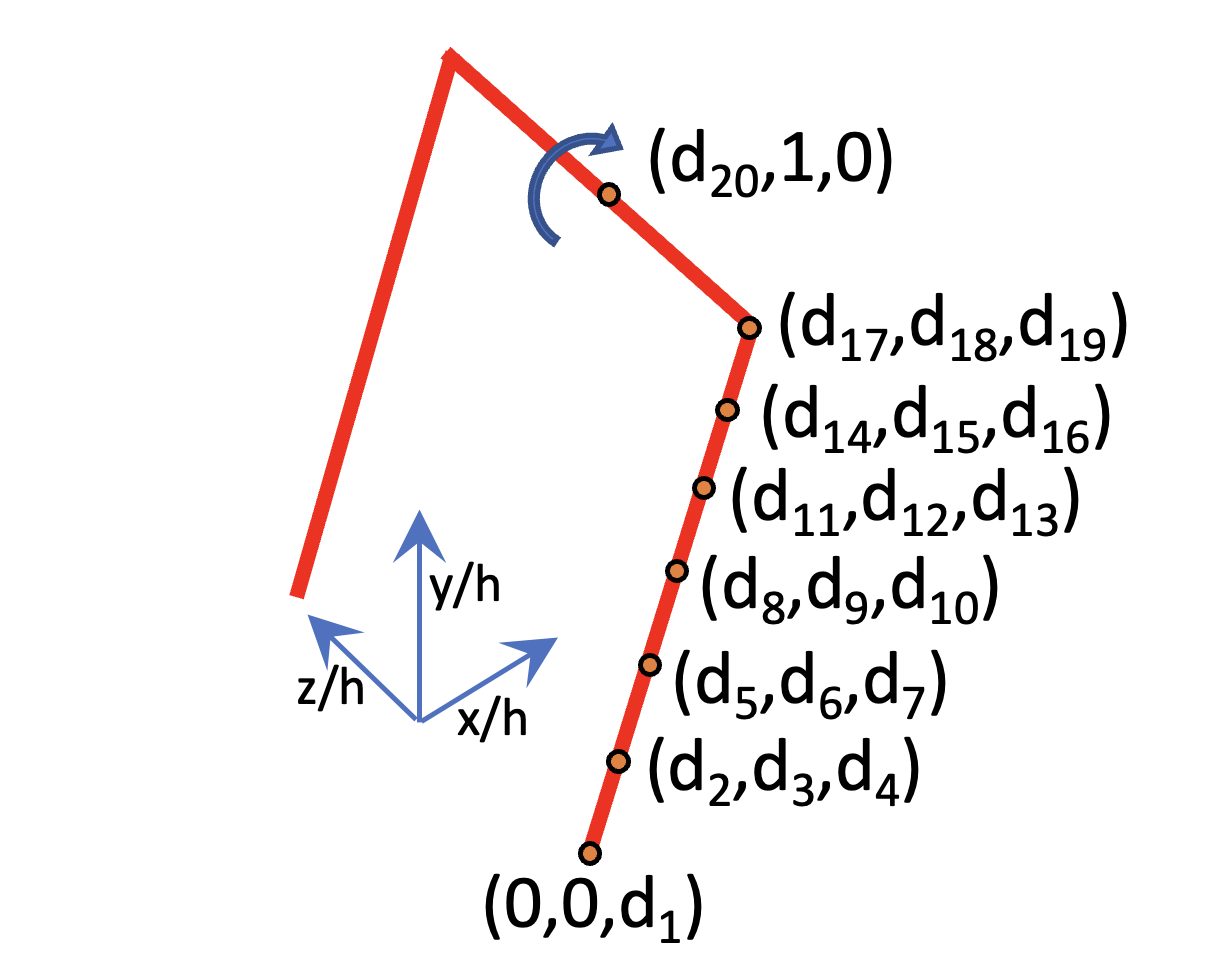}
	\caption{Setup of a sample optimization study conducted to find an optimal eddy shape. Design variables $d_i$ are coordinate locations which represent the eddy shape. No combination of design variables was found to result in a better prediction of the mean velocity and streamwise variance than the simple rectangular hairpin.}
	\label{fig:opt1}
\end{figure}

The rectangular hairpin presented above is not just another convenient template that happens to fit well. Rather, it concentrates the entire spanwise displacement at one height, which gives the exact step-shaped mean kernel required by the attached-eddy mean law, while its vertical legs remain spectrally broad and dominate the one-dimensional streamwise spectrum.
In other words, the rectangular hairpin is unusually predictive because it realizes an unusually clean separation of duties. The top segment concentrates the entire spanwise displacement at one height and therefore gives the exact $I_1$ plateau. The legs carry essentially none of $I_1$, but because they are strongly localized they dominate most of $I_\phi$. The angle $\theta$ then tunes the spectral peak largely through phase, without spoiling the mean. Replacing the rectangular hairpin is hard because most alternatives appear to spoil at least one of these properties, and many spoil two of them at once. This was a conclusion reached after a  number of optimization attempts on  configurations such as those shown in Figure~\ref{fig:opt1}. The optimization setup and code can be found at \url{https://github.com/CaslabUM/AttachedEddy}. 

\begin{figure}
	\centering
	\includegraphics[width=0.92\textwidth]{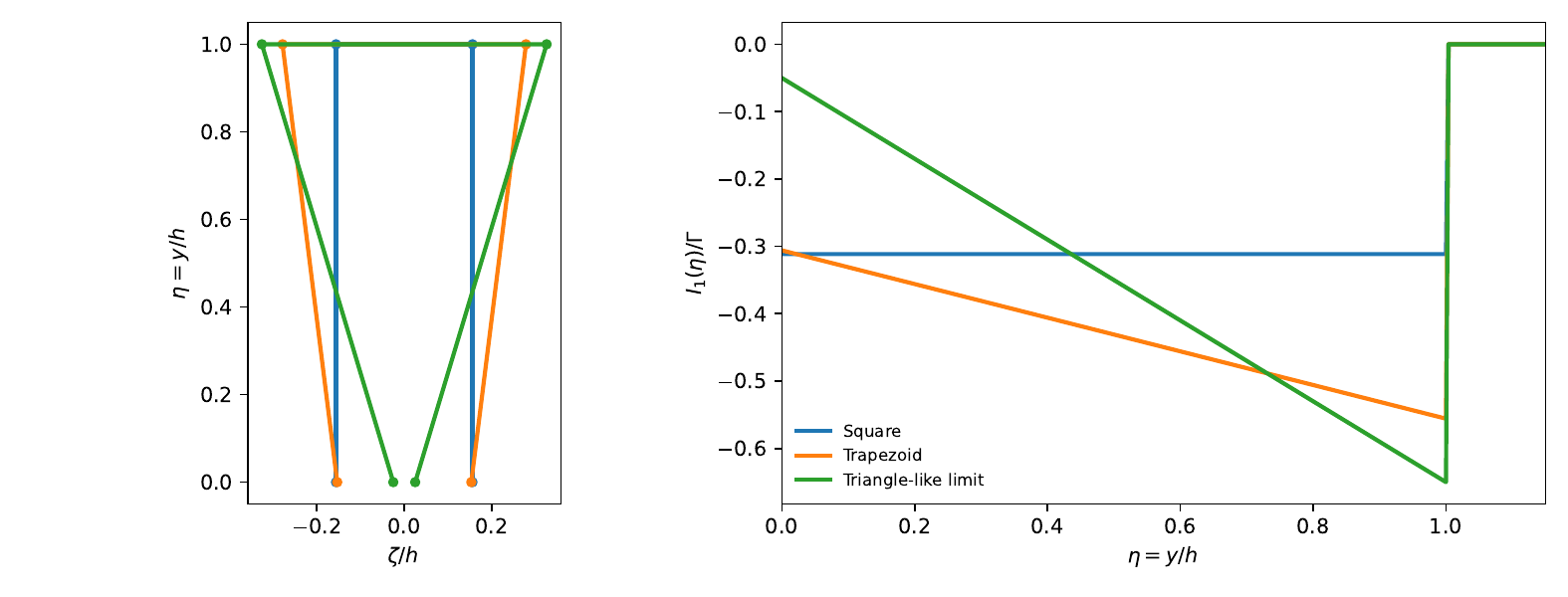}
	\caption{Straight-segment hairpin family and the corresponding exact $I_1(\eta)$ kernels.}
	\label{fig:geometry-family}
\end{figure}

The author is careful to note that the above statement  is not a proof, but a qualitative result. A more rigorous statement can be made in the case of a symmetric three-segment hairpin family whose edge nodes are given by
\[
(0,0,-b_0/2)
\rightarrow
(\lambda,1,-b_1/2)
\rightarrow
(\lambda,1,b_1/2)
\rightarrow
(0,0,b_0/2).
\]
 Figure~\ref{fig:geometry-family} shows the geometry in the $(y,z)$ projection.
Appendix C gives individual segment contributions.

\paragraph{Mean contribution}
Summing the three segment contributions from Eq.~\eqref{eq:I1-segment} gives the exact single-hairpin mean kernel
\begin{equation}
	I_1^{\mathrm{hp}}(\eta)
	=
	\Gamma\,[b_0+(b_1-b_0)\eta]H(1-\eta).
	\label{eq:I1-hairpin}
\end{equation}
This reduces to a plateau when $b_1=b_0$ and to a ramp when $b_1\ne b_0$.
For the attached-eddy population, the mean profile becomes
\begin{equation}
	U^+(y^+) = U_{\mathrm{ref}}^+ - \beta C\Gamma F_1(y^+),
	\label{eq:mean-model}
\end{equation}
with the exact population integral
\begin{equation}
	F_1(y^+) =b_0\log\!\left(\dfrac{h_{\max}^+}{y^+}\right)
		+(b_1-b_0)\left(1-\dfrac{y^+}{h_{\max}^+}\right), \ \  h_{\min}^+\le y^+\le h_{\max}^+.
	\label{eq:F1-piecewise}
\end{equation}
Thus, the rectangular hairpin $b_1=b_0=s$ collapses this to the exact logarithmic form
\[
F_1(y^+) = s\log(h_{\max}^+/y^+)
\qquad (h_{\min}^+\le y^+\le h_{\max}^+).
\]
This is the main reason the rectangular head remains so predictive: it gives the best possible morphology for the logarithmic mean.

\paragraph{Streamwise Kinetic Energy}

For the rectangular hairpin with $0<\eta<1$, Appendix C provides exact expressions for the {Fourier} transform of the streamwise {induced} velocities. It is instructive to split the contributions from the head of the hairpin and the legs as $\hat{u}_1 \triangleq \hat{u}_h + \hat{u}_\ell$. Then the spectral kernel can be reduced to an exact  quadrature in $\kappa_z$:
\begin{align}
	I_\phi^{hh}(\kappa_x,\eta)
	&=
	\frac{1}{2\pi^2}
	\int_0^\infty
	\left|
	\widehat{u}_h(\kappa_x,\kappa_z;\eta)
	\right|^2
	\, d\kappa_z,
	\label{eq:square-Iphi-hh-int}
	\\[0.3em]
	I_\phi^{\ell\ell}(\kappa_x,\eta)
	&=
	\frac{1}{2\pi^2}
	\int_0^\infty
	\left|
	\widehat{u}_{\ell}(\kappa_x,\kappa_z;\eta)
	\right|^2
	\, d\kappa_z,
	\label{eq:square-Iphi-ll-int}
	\\[0.3em]
	I_\phi^{h\ell}(\kappa_x,\eta)
	&=
	\frac{1}{\pi^2}
	\operatorname{Re}
	\int_0^\infty
	\widehat{u}_h(\kappa_x,\kappa_z;\eta)\,
	\widehat{u}_{\ell}(\kappa_x,\kappa_z;\eta)^*
	\, d\kappa_z.
	\label{eq:square-Iphi-hl-int}
\end{align}

For the representative rectangular hairpin at $\theta=60^\circ$, and $\eta=0.5$, these exact formulas give at $\kappa_x = 0$
\[
I_\phi^{hh}(0,0.5)=2.74\times 10^{-2},
\qquad
I_\phi^{\ell\ell}(0,0.5)=8.36\times 10^{-2},
\qquad
I_\phi^{h\ell}(0,0.5)=4.82\times 10^{-2}.
\]
This reveals two things: a) The fact that the kernels (and thus $\phi_{xx}$) do not decay to zero at  zero streamwise wavenumber; and b) even at $\kappa_x=0$, the head contributes only about $17\%$ of $I_\phi(0)$, the legs contribute about $53\%$, and the cross term contributes about $30\%$, even though the head contributes all of $I_1$. At the premultiplied peak ($\kappa_x \approx 1.55$) the split is approximately $8.5\%$, $71.4\%$, and $20.1\%$, and by $\kappa_x=5$ the legs carry about $98.1\%$ of the spectrum.

\begin{figure}[t]
	\centering
	\includegraphics[width=0.95\linewidth]{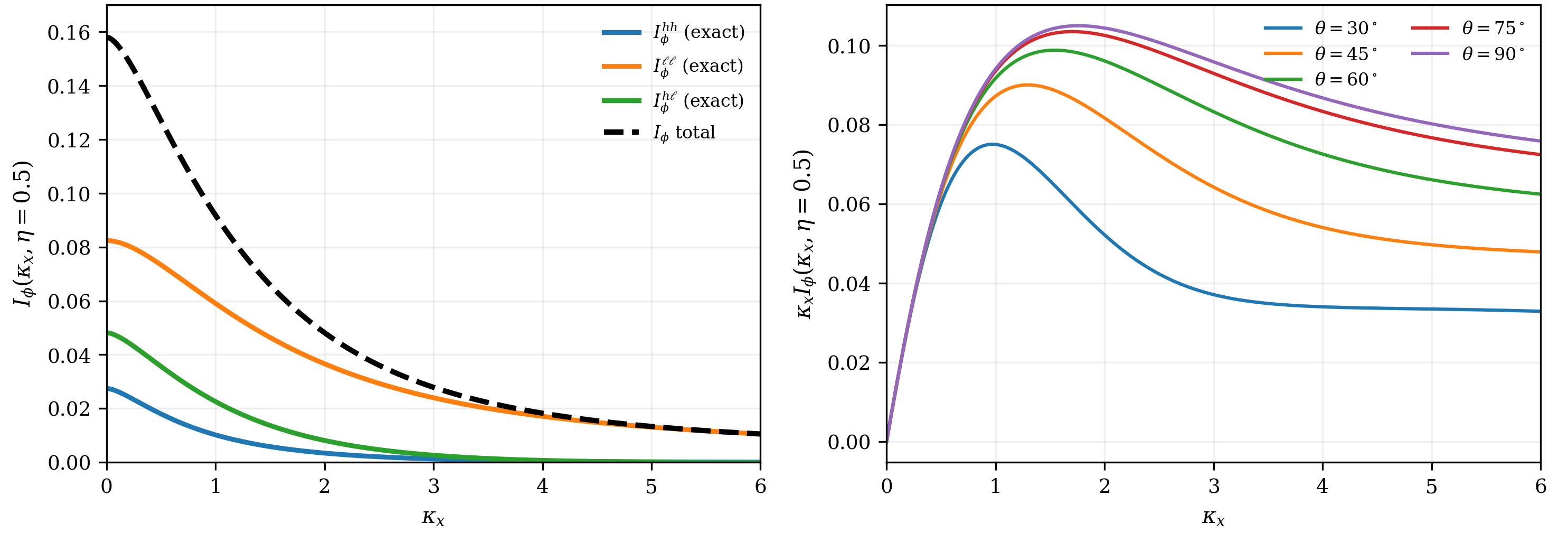}
	\caption{Left: Exact square-hairpin decomposition of the single-eddy spectrum into head, leg, and cross contributions at $\eta=0.5$. Right: premultiplied spectrum for the same square geometry at several inclination angles. The mean kernel is fixed by the top segment, while the angle mainly tunes the spectral peak through the leg phase $e^{-i\kappa_x s \cot\theta}$.}
	\label{fig:square-exact-decomp}
\end{figure}

This is the sharpest quantitative statement of the mean-variance duality: the head contributes  significantly to the mean kernel, but the legs determine most of the spectral energy. As mentioned earlier, the difficulty of replacing the rectangular head is precise rather than qualitative. Any replacement that spreads the top segment over a vertical band  immediately destroys the exact plateau in $I_1$. Triangles, rounded tops, tilted tops, and arch-like heads all behave the same way.

The same replacement also changes the head contribution to $I_\phi$. Experiments show that downward spreading of the head not only spoils the mean plateau; it also \emph{increases} the moderate-$\kappa_x$ head content because part of the head is brought closer to the observation plane. This is exactly the wrong direction if one wants to retain the clean head/leg role separation of the rectangular hairpin.

One might try to keep the top segment at a single height so that $I_1$ remains exact, and instead replace only the legs. This is a different failure mode. If the legs are broadened, bowed, packetized, or rounded in a way that gives them an effective spanwise width $w_\ell$, then the mean kernel is still unchanged to leading order because the leg contribution is small,  but the spectrum is no longer the same. Even when the mean is preserved exactly, broadening the legs narrows the Cauchy spectrum and shifts the premultiplied peak to lower $\kappa_x$. In other words, keeping the right $I_1$ is not enough: the rectangular hairpin also succeeds because its legs are highly localized and therefore spectrally broad.
  That is the main reason  the rectangular hairpin is unusually predictive: it is one of the very few simple filament templates in which the mean and spectral constraints are nearly decoupled.

  \section{Inferring and modeling the eddy density}

Once the eddy shape is chosen and $\beta$ is fixed, there are no degrees of freedom left.  Yet, the comparisons in the previous section suggest that  a minimal hairpin-type eddy can reproduce not only the one-point moments at $Re_\tau\approx 5200$ but also the evolution of $U^+$, $u'^2$ and the premultiplied spectrum across Reynolds numbers. This section provides a more substantial interpretation of why this occurs, which ingredients appear essential, and which limitations are inherent to an independent-eddy superposition.

It is useful to separate (i) the \emph{population law} $(\beta,p(h),h_{\min},h_{\max})$ from (ii) the \emph{single-eddy template} (geometry, inclination angle $\theta$, circulation scaling, and vortex-core regularization). Under these conventions, the only parameter used to anchor the data is the population amplitude $\beta$. 

\begin{figure}
	\includegraphics[width=0.33\textwidth]{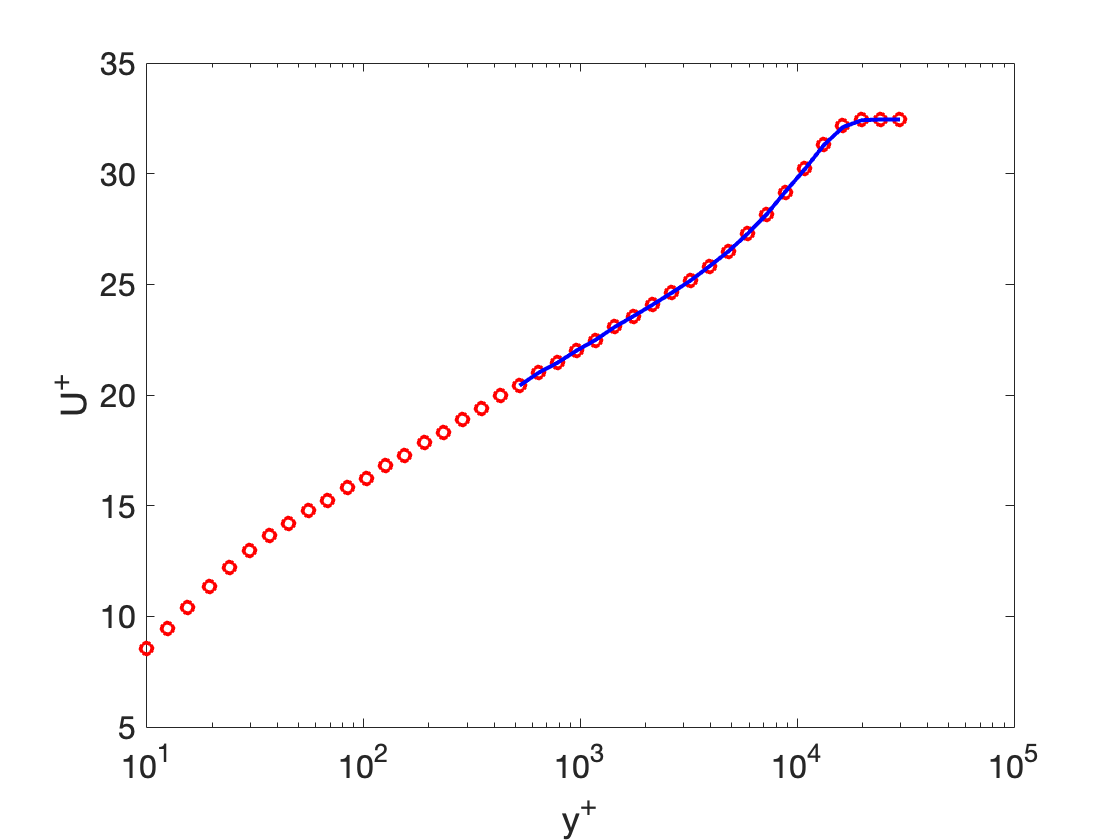}
	\includegraphics[width=0.33\textwidth]{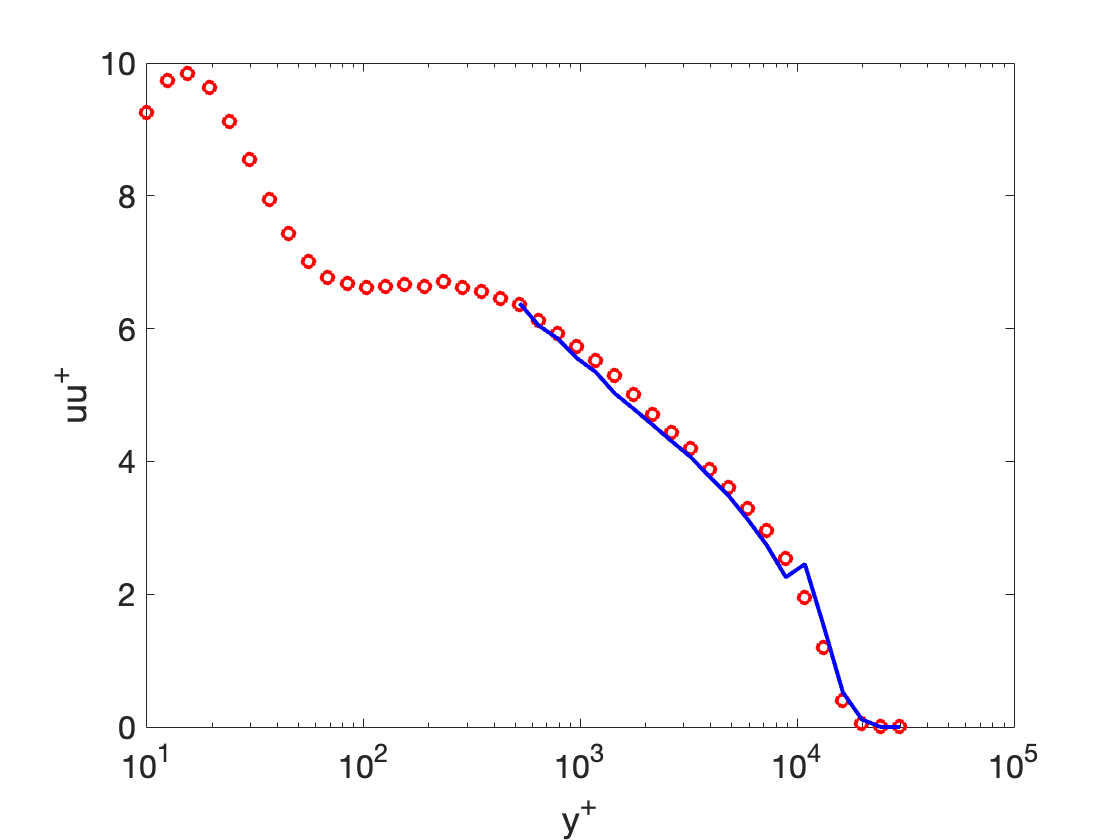}
	\includegraphics[width=0.33\textwidth]{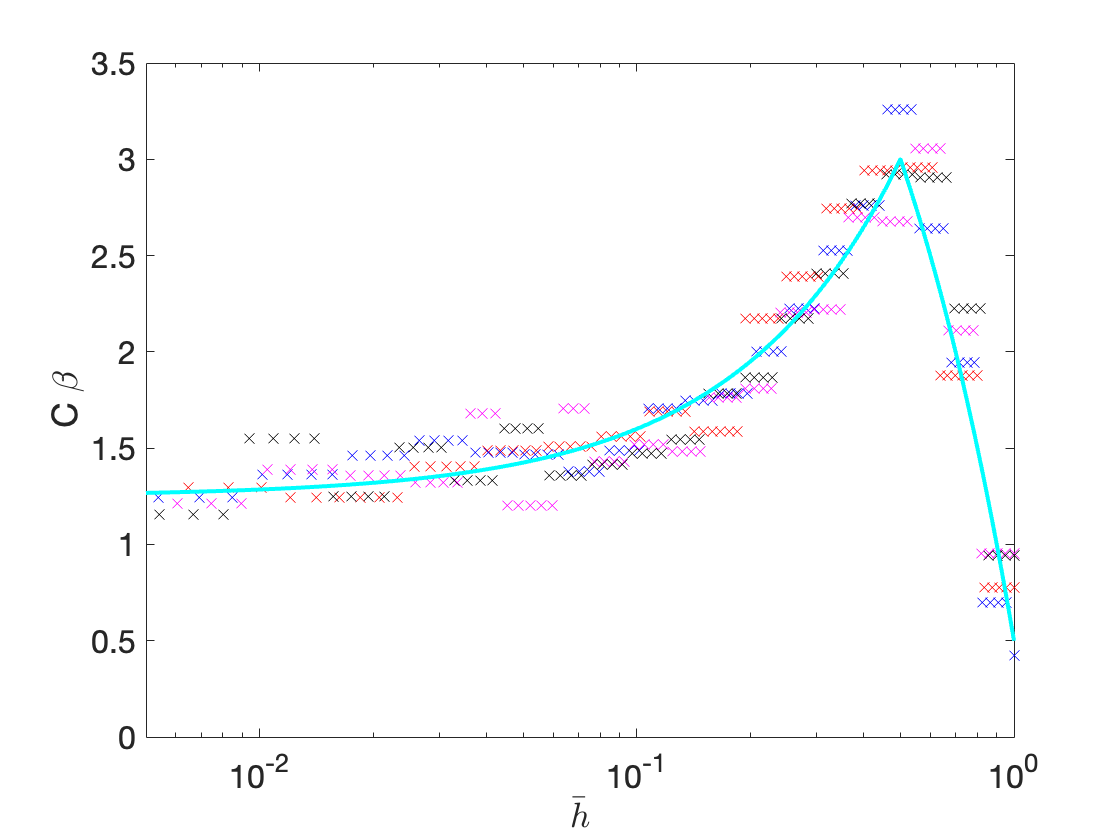}
	\caption{Left and middle: Joint optimal solution for eddy shape and density at $Re_\tau=20000$. Right: Inferred eddy density functions for $Re_\tau = [6000, 10000, 14500, 20000]$ and piecewise linear approximation. }
	\label{fig:Optbeta}
\end{figure}

It is, however, possible to improve the accuracy of the model by making the eddy density a function of the eddy size. A joint optimization run was performed for the eddy width, angle and $\beta(h^+)$. The optimization  for  $Re_\tau=20000$ converged to $\theta=75^\circ$ and an aspect ratio (ratio of hairpin width and height) of 1.8, and an eddy density as shown in Figure~\ref{fig:Optbeta}, which imply that it is possible to get a near-perfect prediction. Keeping the eddy shape and angle fixed, and examining the behavior of the eddy density function, the following  piecewise linear fit was found to be accurate
\begin{equation}
C \beta(\bar{h}) 
	=
	\begin{cases}
		1.25+3.5 \bar{h}& 0\le \bar{h} \le 0.5,\\[1.2em]
		 3.0-2.5 (\bar{h}-0.5)/0.5 &0.5 \le \bar{h} \le 1.
	\end{cases}
	\label{eq:hbar}
\end{equation}
where $\bar{h} = \frac{h^+ - h_{\min}^+}{h_{\max}^+ - h_{\min}^+}$.

Physically, the inferred $C\beta(\bar{h})$ peaks at mid-range eddy sizes ($\bar{h}\approx 0.5$) and decreases toward both the smallest and largest eddies. This is qualitatively consistent with the expectation that the largest eddies are geometrically constrained by the boundary layer thickness and may be less numerous than a pure $p(h)\propto h^{-3}$ law would predict, while the smallest eddies in the log layer interact with the buffer region where the attached-eddy hypothesis breaks down. However, we note that the piecewise linear form was chosen for simplicity and not derived from physical consistency requirements.

We also note that the jointly optimized angle ($\theta=75^\circ$) and aspect ratio (1.8) differ from the values used in Section~3 ($\theta=60^\circ$, aspect ratio 1); this suggests that when $\beta(h)$ absorbs additional modeling flexibility, the optimal eddy shape shifts accordingly, and the two parameterizations should not be directly compared.

\begin{figure}
	\centering
	\includegraphics[width=0.35\textwidth]{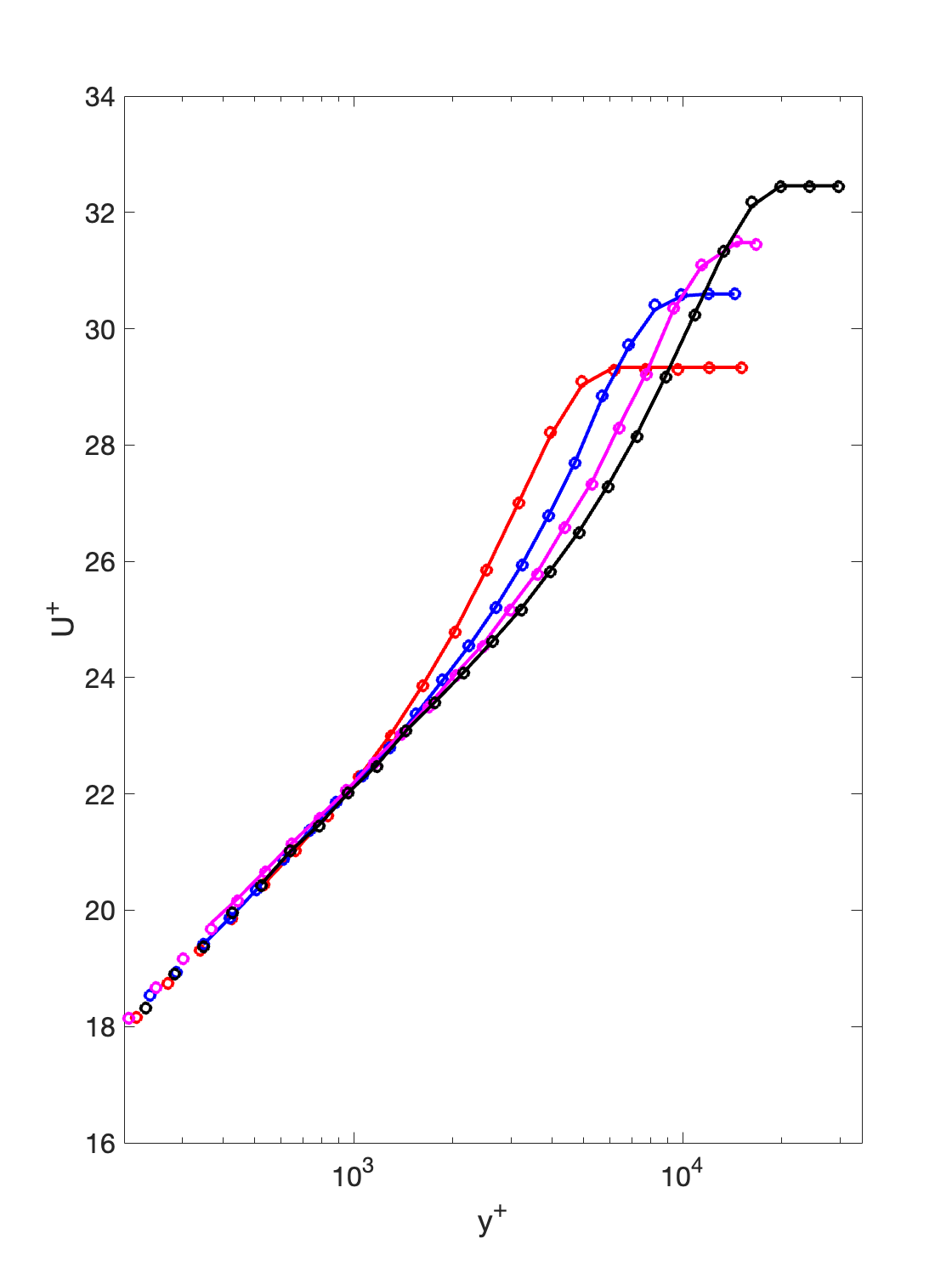}
	\includegraphics[width=0.35\textwidth]{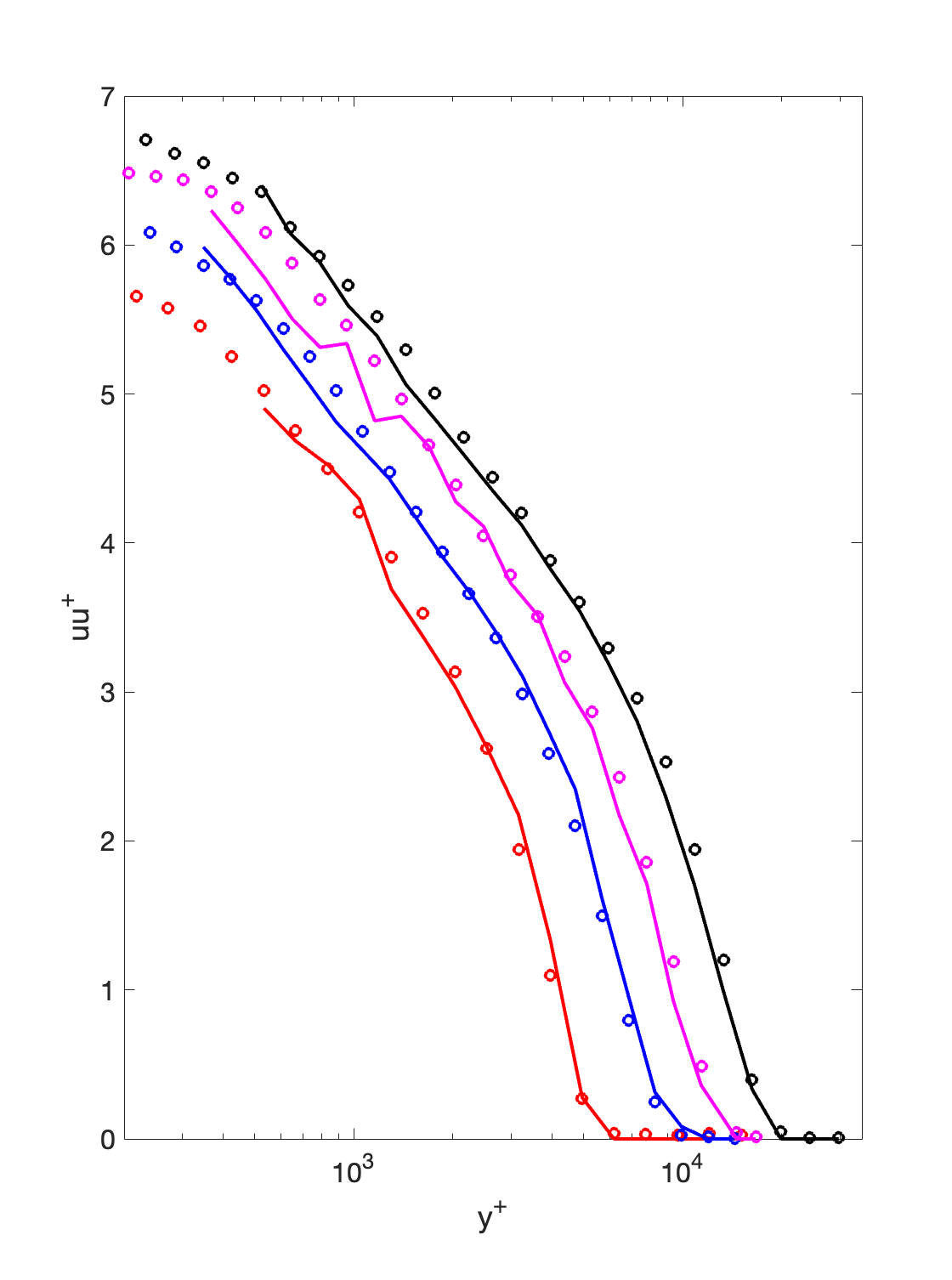}
	\caption{Improved predictions using analytical  eddy density  function $C \beta(h^+)$ for $Re_\tau = [6000, 10000, 14500, 20000]$.}
	\label{fig:Optbeta2}
\end{figure}

Using this analytical eddy density function, Figure~\ref{fig:Optbeta2} shows predictions at 4 different Reynolds numbers.

	\section{Related approaches and connections}
	
	The present work occupies a complementary position within the broader landscape of 
	attached-eddy theories and coherent-structure studies. An important parallel development 
	is the hierarchical random additive process (HRAP) model introduced by 
	\cite{YangMarusicMeneveau2016HRAP} which provides an alternative statistical 
	reinterpretation of Townsend's attached eddy hypothesis. This formulation elegantly recovers 
	logarithmic scaling laws for moments, structure functions, and generalized two-point 
	correlations without requiring explicit specification of eddy geometry. We note that the HRAP framework is a higher level abstraction than the present work, as it 
	encodes only the statistical consequence (additive contributions from a hierarchy of 
	scales) without committing to any particular morphology.
	
		~\cite{davidson2009simple} propose a deliberately simple log-layer model for streamwise fluctuations which reproduces the $k_x^{-1}$ behavior of the one-dimensional longitudinal spectrum in the log region and provides reasonable predictions for the variance $u'u'$ (up to an additive constant). The existence and criteria for a $k_x^{-1}$ spectral dependence have been investigated extensively, more recently by~\cite{nickels2005evidence}. Notably, the $k_x^{-1}$ trend is most apparent very close to the surface, whereas the present comparisons are shown at an intermediate wall-normal distance. The Davidson--Krogstad viewpoint supports the observation that once the mean and a small set of scaling assumptions are fixed, the streamwise spectrum in the log region has limited freedom.
	
	The above works  reinforce a key 
	message: different formulations can reproduce expected log-layer statistics. The present work attempts to provide a concrete mechanistic underpinning within a specific formulation of the attached eddy hypothesis, and thus the {\em conclusions of this study are limited purely to the specific formulation at hand}. 
	
	The inverse-problem analysis (Section~\ref{sec:inverse}) demonstrates 
	that the inferred influence functions $I_1(\eta)$ and $I_{ij}(\eta)$ represent the 
	essential kinematic constraints that any viable eddy must satisfy. As shown in Section~\ref{sec:whysquare}, the rectangular hairpin yields  a precise plateau and functions as a minimal template that satisfies 
	these influence-function requirements while remaining consistent with Biot--Savart 
	kinematics and wall boundary conditions. 
	
	It is important to distinguish the present statistical/kinematic AEM from the growing body 
	of work that seeks to identify wall-attached eddies as instantaneous, material features 
	in DNS or experimental data. Beginning with \cite{delAlamo2006Clusters}
	who identified self-similar vortex clusters in the logarithmic region of turbulent channels, 
	a series of studies have extracted three-dimensional coherent structures using various 
	detection criteria. \cite{DennisNickels2011} provided detailed 
	experimental measurements of vortex packets in turbulent boundary layers via tomographic 
	PIV. \cite{LozanoDuranFloresJimenez2012} characterized the three-dimensional 
	structure of momentum-transferring Q2/Q4 events, demonstrating that attached structures 
	become geometrically self-similar with sizes proportional to their distance from the wall. 
	More recently, \cite{HwangSung2018} developed a clustering methodology 
	to extract wall-attached structures of velocity fluctuations, and \cite{Cheng2019Identity,Cheng2020Uncovering} employed bidimensional empirical mode decomposition to 
	isolate attached eddies and quantify their contribution to skin-friction generation.
	Additionally, several studies have demonstrated that statistical properties can be recovered from stochastic representations of eddies with wall-distance scaling~\citep{bautista2019turbulent,ehsani2024attached,nath2026features}, and the clustering and larger-scale organization of attached eddies has been investigated by~\cite{hutchins2012towards} and~\cite{deshpande2023reynolds}, among others.
	
	These data-driven approaches validate the physical premise of wall-attached, self-similar 
	hierarchies, providing essential structures that models such as 
	the AEM seek to represent statistically. However, they also reveal complexity that the 
	minimal AEM intentionally abstracts away: instantaneous eddies exhibit significant 
	variability in shape, orientation, and internal structure; they interact nonlinearly; and 
	their dynamics involve processes (generation, merging, cascade) not captured by easily interpretable models.
 The present work thus occupies a middle ground: it provides more geometric specificity 
	than purely statistical models (HRAP) while remaining deliberately simpler than the 
	full complexity revealed by structure-extraction studies. The spectral influence kernel 
	$I_\phi(\kappa_x, \eta)$ introduced in Section~\rev{\ref{sec:inverse_kernel}} offers a bridge between these 
	perspectives and encodes the spectral signature of a single eddy template and can, in 
	principle, be compared against conditionally averaged spectra from identified structures. 
	Such comparisons, along with extensions to two-dimensional $(k_x, k_z)$ spectra and 
	cross-component correlations, represent promising directions for strengthening the 
	connection between kinematic models and data-driven structure identification.
	
	We also note a connection to the resolvent framework~\citep{mckeon2019self}: the self-similar hierarchy of response modes identified by resolvent analysis shares the same wall-attached scaling as the AEM, and the resolvent modes can be viewed as a dynamically informed basis for the eddies that the AEM treats kinematically. Integrating the spectral influence kernel $I_\phi$ with resolvent-based predictions of the two-dimensional spectral tensor represents a promising path toward a dynamically grounded attached-eddy model.

	\section{Conclusions and Perspectives}

This work extended the statistical foundations of the attached eddy modeling framework by clarifying the formulation and offering proofs of convergence.  Additional contributions of this work can be summarized as follows:
\begin{itemize}
	\item {\em Inverse problem for attached-eddy kernels.} We formulate the recovery of the single-eddy influence functions $I_1(y/h)$ and $I_{ij}(y/h)$ from reference one-point moments as an explicit inverse problem (Section~\ref{sec:inverse}). This makes the ``eddy contribution function'' a quantitative, inferable object rather than a qualitative sketch.
	\item {\em Ideal-kernel structure.} For the $Re_\tau\approx 5200$ DNS considered, the inferred mean-flow kernel exhibits a near-plateau over $y/h\lesssim 1$ and a transition/decay for $y/h>1$ (Figure~\ref{fig:Optimal1}). This is precisely the structural feature required for logarithmic scaling in Eq.~\eqref{eq:inv_eta_form}.
	\item {\em Minimal Biot--Savart-consistent eddy template.} A simple rectangular hairpin-type vortex-loop template (implemented with Rankine vortex rods) together with an inviscid image system across the wall can reproduce both the inferred kernels and the corresponding mean and Reynolds-stress profiles with good accuracy (Figure~\ref{fig:hairpin45res}).
	\item {\em A spectral Influence kernel.} We introduce the streamwise-energy Influence kernel $I_\phi(\kappa_x,\eta)$ (Section~\rev{\ref{sec:inverse_kernel}}), which provides a compact representation of how a self-similar eddy footprint populates the one-dimensional spectrum as a function of scale.  The spectral content of a single self-similar eddy as a function of dimensionless wavenumber and relative height is implicit in prior work, possibly dating back to ~\cite{perry1982mechanism}. In this work,   the kernel makes the emergence of the $k_x^{-1}$ scaling band and its limitations  explicit, and more transparent than population-level arguments alone and offers insight into constructing more sophisticated compositional eddy models. It is notable that $\int_0^\infty I_\phi(\kappa_x,\eta) d \kappa_x = I_{11}(\eta).$
	\item {\em Closed-form  results for the straight hairpin family.} For the three-segment straight-filament hairpin family with wall image, we derive exact expressions for the mean kernel $I_1^{\mathrm{hp}}(\eta)=\Gamma[b_0+(b_1-b_0)\eta]H(1-\eta)$ (Appendix~C, Eq.~\eqref{eq:I1-hairpin}), demonstrating that only the rectangular hairpin ($b_1=b_0$) produces an exact plateau and hence a purely logarithmic mean profile and serves as a minimal eddy that satisfies kinematic constraints and yields acceptable first and second streamwise moment behavior.
 Fourier-space expressions for the streamwise velocity (Eq.~\eqref{eq:uhat-segment-compact}) enable an exact head/leg/cross decomposition of the spectral kernel, quantifying the mean-variance duality: For a rectangular hairpin, the head contributes all of $I_1$ but only $\sim 17\%$ of $I_\phi$ at $\kappa_x=0$ for $\eta=0.5$, while the legs carry $\sim 53\%$ of the spectral energy.
	\item {\em Scale-dependent eddy density.} Allowing $\beta$ to vary with eddy size $h$  provides an additional degree of freedom that yields near-perfect predictions of both mean velocity and streamwise variance across $Re_\tau=6000$--$20000$ (Figure~\ref{fig:Optbeta2}), with a simple piecewise-linear analytical fit (Eq.~\eqref{eq:hbar}).
\end{itemize}

The original goal of this work was to use the inferred influence functions to uncover an optimal eddy or packet configuration. One of the conclusions of this work is that a simple Rankine rod-based rectangular hairpin is unusually predictive once the mean is anchored, and is able to represent attached eddy physics. The reasons are as follows:
\begin{itemize}
	\item The single-eddy spectrum is determined by $I_\phi(\kappa_x,\eta)$, a two-variable kernel. Population integration introduces only the scale window and one amplitude constant ($\beta$).
	\item $p(h)\propto h^{-3}$ cancels the $h^3$ prefactor exactly. This produces the $1/k_x$ factor explicitly and shifts the remaining complexity into the slowly varying integral $F(k_x;y)$.
	\item The tilted hairpin footprint produces streamwise-extended $u$ signatures on a wall-parallel plane. Through the $k_x h$ mapping, the dominant wavelengths at $y$ are in the decade $\lambda_x = O(10y)$, which at $y^+\sim 10^3$ sits near $\lambda_x^+\sim 10^4$.
	\item Reynolds-number variation enters mainly through the window endpoints. For fixed $y^+$ in the log region, changing $Re_\tau$ changes $h_{\max}$ and therefore the low-$k_x$ limit, while leaving much of the intermediate band governed by the same kernel shape.
\end{itemize}

At this juncture, we remark that the AEM is a statistical and non-dynamic model, and thus prototypical eddies cannot be interpreted as literal material structures, even if they provide useful building blocks for kinematic prediction. Nevertheless, the inverse-kernel viewpoint provides a bridge between \emph{statistical requirements} (what kernels are needed to match moments) and \emph{geometric hypotheses} (what minimal eddy templates can realize those kernels under Biot--Savart induction).  A good fit of a one-dimensional premultiplied spectrum does not uniquely identify a specific structure. Many different eddy templates can generate similar $k^{-1}$-like behavior once integrated over a scale-invariant population. Therefore, the fact that the Rankine-rod hairpin AEM matches experiments across many quantities of interest and Reynolds numbers is encouraging, but it is not, by itself, a proof that the flow is literally composed of those hairpins. Stronger discrimination can come from tests that 1D spectra smear out: two-dimensional spectra, cross-spectra, coherence, wall-normal and spanwise components, vorticity-related spectra, and conditional structure signatures. 	It is noted that while statistical results are presented only in the logarithmic region, where attached eddies are predominant, the integration limits span the bottom of the log layer all the way to the boundary layer height, where detached eddies persist. Future work may examine the sensitivity of the results to the limits of integration, and additional models for detached eddies.

\paragraph{Inverse problems as a pathway forward.} The inverse formulation suggests several immediate extensions: (i) simultaneous inference of multiple kernels using multiple moments and multiple Reynolds numbers with shared regularization and physically motivated constraints, (ii) inference of \emph{2D} spectral kernels (in $(k_x,k_z)$) rather than only 1D kernels, and (iii) incorporation of non-Poisson population effects (packets, clustering, exclusion) through pair-correlation corrections to Eq.~\eqref{eq:rij_final} and their spectral counterparts. A key limitation of the present validation is that it is restricted to the streamwise velocity component: the mean $U^+$, the streamwise variance $\overline{u'^2}$, and the 1D streamwise energy spectrum. Extending the comparisons to all components of the Reynolds stress tensor (wall-normal and spanwise variances and the Reynolds shear stress), two-point correlations, and two-dimensional $(k_x,k_z)$ spectra would provide a much more stringent test of the minimal eddy hypothesis and could distinguish between eddy templates that appear equivalent in 1D metrics.

\vspace{1cm}
\noindent {\bf Acknowledgement} This work was supported by the Aeronautics Research Mission Directorate at NASA
through the Transformative Aeronautical Concepts Program (TACP) and the D.2 Transformational
Tools and Technologies Project (TTT) under contract \# 80NSSC23M0215.
Early part of the research was supported by NASA grant \# 80NSSC18M0149.  The analytical expressions in Equations 36-39 were derived with the help of GPT5.2-Pro from OpenAI, Inc.

\section*{Appendix A:  Numerical convergence}

Let \(\rho \triangleq \sqrt{\xi^2+\zeta^2}\). A standard sufficient condition for convergence is that
\begin{equation}
	\int_{\mathbb{R}^2} |u_1(\xi,\zeta,\eta)|\,d\xi\,d\zeta < \infty.
	\label{eq:L1_condition}
\end{equation}
A convenient and interpretable sufficient condition is a power-law far-field bound
\(
|u_1(\xi,\zeta,\eta)| \le C(\eta)\,(1+\rho)^{-p}
\)
for some \(p>2\). Indeed, the tail can be bounded (up to an angular factor) by
\begin{equation}
	\big|I_1(\eta)-I_1(\eta;R)\big|
	\;\le\;
	\int_{\rho>R} |u_1|\,dA
	\;\lesssim\;
	2\pi C(\eta)\int_R^\infty \rho^{1-p}\,d\rho
	\;=\;
	\frac{2\pi C(\eta)}{p-2}\,R^{2-p}.
	\label{eq:I1_tail_bound}
\end{equation}
Hence, \(p>2\) guarantees convergence as \(R\to\infty\). The borderline case \(p=2\) produces a
logarithmic divergence \(\sim \log R\); for \(p=3\) the tail decays as \(\mathcal{O}(1/R)\).

\paragraph{Reynolds-stress kernels are less restrictive.}
Similarly, the second-moment influence functions are
\begin{equation}
	I_{ij}(\eta;R)
	\;\triangleq\;
	\int_{-R}^{R}\int_{-R}^{R} u_i(\xi,\zeta,\eta)\,u_j(\xi,\zeta,\eta)\,d\xi\,d\zeta,
	\qquad
	I_{ij}(\eta)\;\triangleq\;\lim_{R\to\infty} I_{ij}(\eta;R),
	\label{eq:Iij_def_R}
\end{equation}
whenever the limit exists.
A sufficient condition for \(I_{ij}\) to be well-defined is \(u_i(\cdot,\cdot,\eta)\,u_j(\cdot,\cdot,\eta)\in L^1(\mathbb{R}^2)\).
If each velocity component satisfies the far-field decay \(|u_k|\lesssim \rho^{-p}\), then
\(u_i u_j \lesssim \rho^{-2p}\), and the planar tail behaves like
\(
\int_R^\infty \rho\,\rho^{-2p}\,d\rho = \int_R^\infty \rho^{1-2p}\,d\rho
\),
which converges whenever \(p>1\).
Thus, the convergence requirement for the \emph{mean-flow} influence \(I_1\) (which requires \(p>2\))
is typically the more restrictive requirement; once \(I_1\) is well-defined, the second-moment kernels
\(I_{ij}\) are generally well-defined for the same eddy field.

\paragraph{Implications for Biot--Savart line-vortex eddies and the role of the image system.}
For line-vortex Biot--Savart fields, a generic open filament can generate a \(\rho^{-2}\) contribution to
certain velocity components in the far field, which is precisely the borderline case for \(I_1\) and would
lead to a logarithmic growth of \(I_1(\eta;R)\) with \(R\).
The specific square-hairpin eddy used herein is paired with an opposite-sign mirror image across
\(y=0\) (to enforce inviscid wall boundary conditions), and the wall-parallel segment cancels with its
image. This construction removes leading-order contributions responsible for a \(\rho^{-2}\) tail in the
streamwise component and yields a faster far-field decay. In other words, the image construction is not
a technical convenience: it is central to ensuring that Eq.~\eqref{eq:zz} is well posed for a line-vortex eddy model.

\paragraph{Numerical evidence.}
To verify the  decay/convergence directly, we performed a Biot--Savart calculation for a simple
rectangular hairpin (three straight segments with unit circulation) together with an opposite-sign
mirror image across \(y=0\), consistent with the construction described around Fig.~\ref{fig:hairpin45}. A small vortex-core
regularization was used to smooth the near-filament singularity; this does not affect the far-field decay
that controls convergence of the planar integrals.

\emph{Far-field decay check.} At \(\eta=y/h=0.5\), evaluating along \((\xi,0,\eta)\) with \(\xi\gg 1\) gives
\begin{align*}
	u_1(10,0,0.5) &\approx -3.92\times 10^{-4},\\
	u_1(20,0,0.5) &\approx -4.44\times 10^{-5},\\
	u_1(40,0,0.5) &\approx -5.26\times 10^{-6},\\
	u_1(80,0,0.5) &\approx -6.39\times 10^{-7}.
\end{align*}
Moreover, \(u_1(\xi,0,0.5)\,\xi^3\) is approximately constant (between \(\approx-0.33\) and \(\approx-0.39\)),
indicating the scaling
\begin{equation}
	u_1(\rho,\eta)\sim \frac{C(\eta)}{\rho^{3}}
	\qquad (\rho\to\infty),
	\label{eq:u1_r3}
\end{equation}
so that \(\big|I_1(\eta)-I_1(\eta;R)\big|=\mathcal{O}(1/R)\) as predicted by \eqref{eq:I1_tail_bound} with \(p=3\).

\emph{Direct growth of $I_1$ with domain size.}
For the same hairpin at \(\eta=0.5\), integrating over \([{-}R,R]^2\) yields the values shown in Table~\ref{tab:I1_convergence}.
The observed approach is consistent with the \(\mathcal{O}(1/R)\) tail implied by \eqref{eq:u1_r3}. In particular,
extrapolating with the model \(I_1(\eta;R)=I_{1,\infty}(\eta)+C/R\) gives
\begin{equation}
	I_{1,\infty}(\eta=0.5)\approx -0.997.
\end{equation}
Above the eddy in this toy geometry, the convergence is rapid; e.g. at \(\eta=1.2\),
\(I_1(1.2;R)\) approaches \(0\) quickly (with \(I_1\approx 0.005\) by \(R=24\)), again consistent with a
well-defined infinite-domain limit.

\paragraph{Practical implication for the homogenization argument.}
In the attached-eddy homogenization step, taking \(L/h\) ``large enough'' should be interpreted as
choosing \(R=L/h\) such that the residual tail \(\big|I_1(\eta)-I_1(\eta;R)\big|\) (and similarly for \(I_{ij}\))
is negligible at the level of accuracy desired. For the line-vortex eddies considered here, this condition
is satisfied because the image construction yields a sufficiently rapid far-field decay of the induced
streamwise velocity, ensuring that the planar influence integrals are well-defined in the limit \(R\to\infty\).

\begin{table}[t]
	\centering
	\caption{Convergence of the planar influence integral \(I_1(\eta;R)\) for the rectangular hairpin + image at \(\eta=0.5\).}
	\label{tab:I1_convergence}
	\begin{tabular}{c c}
		\hline
		$R=L/h$ & $I_1(\eta=0.5;R)$ \\
		\hline
		4  & $-0.885$ \\
		8  & $-0.941$ \\
		12 & $-0.960$ \\
		16 & $-0.969$ \\
		20 & $-0.975$ \\
		24 & $-0.979$ \\
		\hline
	\end{tabular}
\end{table}


\section*{Appendix B: Components of the Inverse Kernel Optimization}
More details are provided on the construction of the matrices involved in Eq.~\eqref{eq:opt_problem}.

\subsection*{Construction of the forward matrix A}
For each measurement pair $(i,j)$ and each quadrature node $h_q$, define
\[
\eta_{iq} = \frac{y_i}{h_q},
\qquad
\kappa_{ijq} = k_{x,j}h_q.
\]
If $(\kappa_{ijq},\eta_{iq})$ lies outside the inversion grid, that contribution is skipped. Otherwise, the kernel value is obtained by bilinear interpolation on the \emph{logarithmic} coordinate grid
$
\xi = \log \kappa,
\qquad
\zeta = \log \eta.$ Given
$
\kappa_a \le \kappa_{ijq} \le \kappa_{a+1},
\qquad
\eta_b \le \eta_{iq} \le \eta_{b+1}. $
The local interpolation coordinates are
\begin{align*}
	t_\kappa = \frac{\log \kappa_{ijq} - \log \kappa_a}{\log \kappa_{a+1}-\log \kappa_a}, \ \
	t_\eta = \frac{\log \eta_{iq} - \log \eta_b}{\log \eta_{b+1}-\log \eta_b}.
\end{align*}
The four bilinear weights are then
\begin{align*}
	\omega^{00}_{ijq} = w_q(1-t_\kappa)(1-t_\eta), \ \
	\omega^{10}_{ijq} = w_q t_\kappa(1-t_\eta), \ \
	\omega^{01}_{ijq} = w_q(1-t_\kappa)t_\eta, \ \
	\omega^{11}_{ijq} = w_q t_\kappa t_\eta.
\end{align*}

If row $m$ corresponds to $(i,j)$, these weights are accumulated into the sparse matrix as
\begin{align*}
	A_{m,\ell(b,a)} \mathrel{+}= \omega^{00}_{ijq}, \ \
	A_{m,\ell(b,a+1)} \mathrel{+}= \omega^{10}_{ijq}, \ \
	A_{m,\ell(b+1,a)} \mathrel{+}= \omega^{01}_{ijq}, \ \
	A_{m,\ell(b+1,a+1)} \mathrel{+}= \omega^{11}_{ijq}.
\end{align*}
Therefore each matrix row represents the quadrature approximation
\[
(A\bm{x})_m
\approx
\sum_{q=1}^{n_h} w_q\,\mathcal{I}_{\log}[\psi](\kappa_{ijq},\eta_{iq}),
\]
where $\mathcal{I}_{\log}$ denotes bilinear interpolation on the $(\log\kappa,\log\eta)$ grid.
For the present inversion,
$
\bm{A} \in \mathbb{R}^{11327\times 5712},
\qquad
\operatorname{nnz}(\bm{A}) = 990659.
$

\subsection*{Regularization matrices}
The inversion uses second-difference smoothness penalties in both coordinate directions plus a very small minimum-norm term. First define the standard second-difference matrix
\[
\bm{D}_n =
\begin{bmatrix}
	1 & -2 & 1 & 0 & \cdots & 0 \\
	0 & 1 & -2 & 1 & \cdots & 0 \\
	\vdots & & \ddots & \ddots & \ddots & \vdots \\
	0 & \cdots & 0 & 1 & -2 & 1
\end{bmatrix}
\in \mathbb{R}^{(n-2)\times n}.
\]
Then the two directional penalties are built by Kronecker products:
\begin{align*}
	\bm{L}_\kappa = I_{n_\eta} \otimes D_{n_\kappa}, \ \
	\bm{L}_\eta = D_{n_\eta} \otimes I_{n_\kappa}, \ \
	\bm{L}_0 = I_{n_\eta n_\kappa}.
\end{align*}
Their dimensions used were
\begin{align*}
	\bm{L}_\kappa \in \mathbb{R}^{5544\times 5712},  \ \
	\bm{L}_\eta \in \mathbb{R}^{5576\times 5712}, \ \
	\bm{L}_0 \in \mathbb{R}^{5712\times 5712}.
\end{align*}

\subsection*{Construction of vector b}
Each usable spectrum sample becomes one entry of the right-hand side. If measurement index $m$ corresponds to wall-normal location $y_i$ and streamwise wavenumber $k_{x,j}$, then $
b_m = \widehat{\Phi}_{ij}. $
The implementation only retains points satisfying
\[
\widehat{\Phi}_{ij} > 0,
\qquad
k_{x,j} > 0,
\qquad
\text{all quantities finite}.
\]

\section*{Appendix C: Analytical Expressions for Single Straight Filament with Image}
Let a straight segment run from 
\[
A=(\xi_a,\eta_a,\zeta_a), \qquad B=(\xi_b,\eta_b,\zeta_b),\]
with circulation $\Gamma$, and let the wall image be the reflected segment \[
A'=(\xi_a,-\eta_a,\zeta_a), \qquad B'=(\xi_b,-\eta_b,\zeta_b)\]
of circulation $-\Gamma$.
Define $
\Delta \xi = \xi_b-\xi_a,  \Delta \eta = \eta_b-\eta_a,  \Delta \zeta = \zeta_b-\zeta_a.$ The segment is parameterized as $
\mathbf r(s)=A+s(B-A), \quad 0\le s\le 1.$

\subsection*{Exact formula for $I_1(\eta)$}
The planar mean kernel is $
I_1(\eta)=\int_{\mathbb R^2} u_1(\xi,\eta,\zeta)\,d\xi\,d\zeta.$
Using the vorticity $\omega_z = \partial_x u_2 - \partial_y u_1$ and integrating over the wall-parallel plane gives
\[
\frac{dI_1}{d\eta} = -\int_{\mathbb R^2} \omega_z(\xi,\eta,\zeta)\,d\xi\,d\zeta.
\]
For one filament,
\[
\omega_z = \Gamma\int_0^1 \delta(\mathbf x-\mathbf r(s))\,\zeta'(s)\,ds,
\]
so
\[
I_1(\eta)=\Gamma\int_0^1 H(\eta(s)-\eta)\,\zeta'(s)\,ds.
\]
For the straight segment, $\zeta'(s)=\Delta\zeta$ and $\eta(s)=\eta_a+s\Delta\eta$, hence
\[
I_1^{AB}(\eta)=\Gamma\,\Delta\zeta\int_0^1 H(\eta_a+s\Delta\eta-\eta)\,ds.
\]
If $\Delta\eta\neq 0$, this evaluates to
\begin{equation}
	I_1^{AB}(\eta)
	=
	\Gamma\,\Delta\zeta\,
	\frac{(\eta_+-\eta)_+-(\eta_--\eta)_+}{\eta_+-\eta_-},
	\label{eq:I1-segment}
\end{equation}
where
\[
\eta_+=\max(\eta_a,\eta_b), \qquad \eta_-=\min(\eta_a,\eta_b), \qquad (q)_+=\max(q,0).
\]
If $\Delta\eta=0$, the segment is horizontal in the $(y,z)$ projection and
\begin{equation}
	I_1^{AB}(\eta)=\Gamma\,\Delta\zeta\,H(\eta_a-\eta).
	\label{eq:I1-horizontal}
\end{equation}
For a physical segment entirely in $y>0$, the image contribution vanishes identically for $\eta>0$, so the wall-inclusive $I_1$ is exactly the same as~\eqref{eq:I1-segment}--\eqref{eq:I1-horizontal}. This is the key simplification for optimization: $I_1$ depends only on the projected trace in the $(y,z)$ plane.

\subsection*{Closed form for $\widehat{u}_1(k_x,\eta,k_z)$}
\begin{figure}
	\centering
	\includegraphics[width=0.98\textwidth]{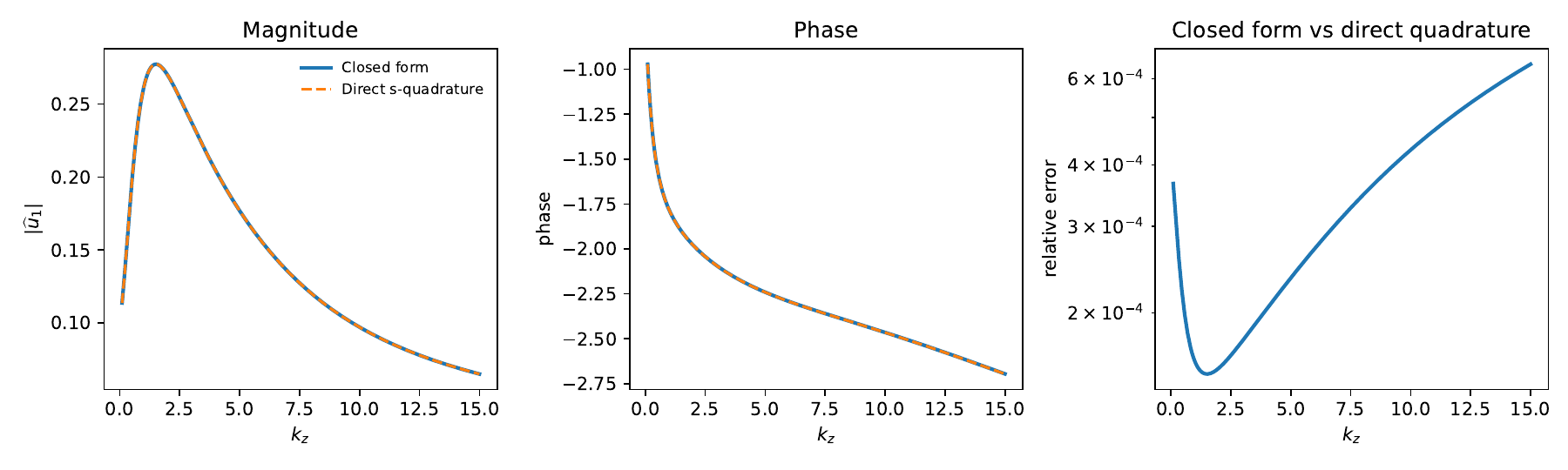}
	\caption{Validation of the closed-form straight-segment transform against direct quadrature in the segment parameter $s$.}
	\label{fig:segment-validation}
\end{figure}

Define
\[
\widehat{u}_1(k_x,\eta,k_z) \triangleq \int_{\mathbb R^2} u_1(\xi,\eta,\zeta)e^{-i(k_x\xi+k_z\zeta)}\,d\xi\,d\zeta,
\qquad q=\sqrt{k_x^2+k_z^2}.
\]
The partial Fourier transform of Biot--Savart for a filament segment plus image gives
\begin{align}
	\widehat{u}_1
	&= \frac{\Gamma}{2}\int_0^1 e^{-i(k_x\xi(s)+k_z\zeta(s))}
	\Big[(\zeta'-i\eta'k_z/q)e^{-q(\eta(s)-\eta)}H(\eta(s)-\eta)
	\notag\\
	&\qquad\qquad -(\zeta'+i\eta'k_z/q)e^{-q(\eta-\eta(s))}H(\eta-\eta(s))
	+ (\zeta'-i\eta'k_z/q)e^{-q(\eta+\eta(s))}\Big] ds.
	\label{eq:uhat-master}
\end{align}
For a straight segment, the phase is linear,
\[
\alpha = k_x\Delta\xi + k_z\Delta\zeta,
\qquad
k_x\xi(s)+k_z\zeta(s) = k_x\xi_a + k_z\zeta_a + \alpha s.
\]
It is convenient to define
\begin{align}
	J_+(s_0,s_1)
	&\triangleq \int_{s_0}^{s_1} e^{-i\alpha s}e^{-q(\eta_a+s\Delta\eta-\eta)} ds
	= e^{-q(\eta_a-\eta)}
	\frac{e^{-(q\Delta\eta+i\alpha)s_0}-e^{-(q\Delta\eta+i\alpha)s_1}}{q\Delta\eta+i\alpha},
	\label{eq:Jplus}
	\\
	J_-(s_0,s_1)
	&\triangleq \int_{s_0}^{s_1} e^{-i\alpha s}e^{-q(\eta-\eta_a-s\Delta\eta)} ds
	= e^{-q(\eta-\eta_a)}
	\frac{e^{(q\Delta\eta-i\alpha)s_1}-e^{(q\Delta\eta-i\alpha)s_0}}{q\Delta\eta-i\alpha},
	\label{eq:Jminus}
	\\
	J_{\mathrm{img}}
	&\triangleq \int_0^1 e^{-i\alpha s}e^{-q(\eta+\eta_a+s\Delta\eta)} ds
	= e^{-q(\eta+\eta_a)}
	\frac{1-e^{-(q\Delta\eta+i\alpha)}}{q\Delta\eta+i\alpha}.
	\label{eq:Jimg}
\end{align}
Then
\begin{equation}
	\widehat{u}_1
	=
	\frac{\Gamma}{2}e^{-i(k_x\xi_a+k_z\zeta_a)}
	\Big[(\Delta\zeta-i\Delta\eta\,k_z/q)J_+
	-(\Delta\zeta+i\Delta\eta\,k_z/q)J_-
	+(\Delta\zeta-i\Delta\eta\,k_z/q)J_{\mathrm{img}}\Big],
	\label{eq:uhat-segment-compact}
\end{equation}
where the interval used in $J_+$ and $J_-$ depends on whether the segment is above, below, or cut by the plane $y=\eta$.
If $s_\eta=(\eta-\eta_a)/\Delta\eta$ lies inside $(0,1)$, the physical segment is split into $[0,s_\eta]$ and $[s_\eta,1]$.
Figure~\ref{fig:segment-validation} validates the closed form against direct quadrature of~\eqref{eq:uhat-master} for a generic segment. 

\section*{Appendix D: Why $I_{11}$, and hence $uu^+$, are logarithmic}
	
	Let $I_{\phi}(\kappa,\eta)$ denote the single-eddy streamwise spectral kernel, with
	$\kappa = k_x h$ and $\eta = y/h$. The corresponding single-eddy second-moment kernel is
	\[
	I_{11}(\eta) = \int_0^\infty I_{\phi}(\kappa,\eta)\, d\kappa .
	\]
	In the attached-eddy hierarchy,
	\[
	R_{11}(y) = \overline{u'^2}(y)
	= \beta \int_{h_{\min}}^{h_{\max}} p(h)\, h^2\, I_{11}(y/h)\, dh .
	\]
	For the scale-invariant population law $
	p(h) = C h^{-3},$
	this becomes
	$
	R_{11}(y)
	= \beta C \int_{y/h_{\max}}^{y/h_{\min}} \frac{I_{11}(\eta)}{\eta}\, d\eta .
	$
	Therefore the criterion for a logarithm in $uu^+ = R_{11}/u_\tau^2$ is simple: if $I_{11}(\eta)$ is approximately constant for $0<\eta \lesssim 1$, then
	\[
	R_{11}(y) \sim \log\!\left(\frac{h_{\max}}{y}\right),
	\qquad
	uu^+(y) = B_1 - A_1 \log(y/\delta),
	\]
	with $h_{\max} \sim \delta$ and
	$
	A_1 = \frac{\beta C b_0}{u_\tau^2},
	\qquad
	b_0 \approx I_{11}(\eta<1).
	$ The issue is therefore reduced to the behavior of $I_{11}(\eta)$.
	
	For the validated rectangular hairpin surrogate at $\eta=0.5$,
	$
	I_{\phi}(\kappa,0.5)
	\approx
	I_{\phi}^{(hh)}(\kappa,0.5)
	+
	I_{\phi}^{(\ell\ell)}(\kappa,0.5)
	+
	I_{\phi}^{(h\ell)}(\kappa,0.5),
	$
	with
	\[
	I_{\phi}^{(hh)}(\kappa,0.5)
	\approx
	0.0296 e^{-1.058\kappa},
	\]
	\[
	I_{\phi}^{(\ell\ell)}(\kappa,0.5)
	\approx
	\frac{0.0812}{\left[1+(\kappa/1.276)^2\right]^{0.651}},
	\]
	\[
	I_{\phi}^{(h\ell)}(\kappa,0.5)
	\approx
	\frac{0.0495 e^{-0.525\kappa}}{\left[1+(\kappa/2.096)^2\right]^{1.208}}.
	\]
	
	Because this validation is explicit on the resolved band $0 \le \kappa \le 4.5$, define
	\[
	I_{11,4.5}(0.5) \triangleq \int_0^{4.5} I_{\phi}(\kappa,0.5)\, d\kappa .
	\]
	Then
	\[
	I_{11,4.5}(0.5)
	=
	I_{11,4.5}^{(hh)}(0.5)
	+
	I_{11,4.5}^{(\ell\ell)}(0.5)
	+
	I_{11,4.5}^{(h\ell)}(0.5).
	\]
	
	The head contribution:
	\[
	I_{11,4.5}^{(hh)}(0.5)
	=
	0.0296 \int_0^{4.5} e^{-1.058\kappa}\, d\kappa
	=
	0.0296 \frac{1-e^{-1.058\cdot 4.5}}{1.058}
	\approx
	0.0277.
	\]
	
	The other two surrogate integrals give
	\[
	I_{11,4.5}^{(\ell\ell)}(0.5) \approx 0.177,
	\qquad
	I_{11,4.5}^{(h\ell)}(0.5) \approx 0.0569.
	\]
	Hence
	$
	I_{11,4.5}(0.5) \approx 0.262.
	$

	So the dominant contribution to $I_{11}$ comes from the broad leg spectrum, not from the head.
	To see how this persists with $\eta$, use the componentwise continuation
	\[
	I_{\phi}^{(hh)}(\kappa,\eta)
	\approx
	A_h(\eta) e^{-2\kappa d_{\rm eff}(\eta)},
	\qquad
	d_{\rm eff}(\eta) \approx 1-\eta,
	\]
	and
	\[
	I_{\phi}^{(\ell\ell)}(\kappa,\eta)
	\approx
	A_\ell(\eta)
	\frac{\kappa^{p_\ell(\eta)}}{\left[1+(\kappa/\kappa_{0,\ell}(\eta))^2\right]^{q_\ell(\eta)}}.
	\]
	
	The leg contribution to $I_{11}$ is then explicit:
	\[
	I_{11}^{(\ell\ell)}(\eta)
	\approx
	A_\ell(\eta)
	\int_0^\infty
	\frac{\kappa^{p_\ell(\eta)}}{\left[1+(\kappa/\kappa_{0,\ell}(\eta))^2\right]^{q_\ell(\eta)}}\, d\kappa.
	\]
	With the change of variable $t = (\kappa/\kappa_{0,\ell})^2$,
	\[
	I_{11}^{(\ell\ell)}(\eta)
	\approx
	\frac{A_\ell(\eta)\, \kappa_{0,\ell}(\eta)^{1+p_\ell(\eta)}}{2}
	\int_0^\infty
	t^{(p_\ell(\eta)-1)/2} (1+t)^{-q_\ell(\eta)}\, dt.
	\]
	Therefore
	\[
	I_{11}^{(\ell\ell)}(\eta)
	\approx
	\frac{A_\ell(\eta)\, \kappa_{0,\ell}(\eta)^{1+p_\ell(\eta)}}{2}
	B\!\left(
	\frac{1+p_\ell(\eta)}{2},
	q_\ell(\eta)-\frac{1+p_\ell(\eta)}{2}
	\right),
	\]
	provided
	\[
	q_\ell(\eta) > \frac{1+p_\ell(\eta)}{2},
	\]
	which is satisfied by the fitted parameters.	Using the fitted values
	\[
	(\eta,A_\ell,p_\ell,\kappa_{0,\ell},q_\ell)
	=
	(0.2,0.082,-0.03,1.64,0.88),
	\]
	\[
	(0.5,0.072,-0.04,1.70,0.87),
	\]
	\[
	(0.7,0.059,-0.05,1.74,0.82),
	\]
	\[
	(0.9,0.040,-0.06,1.54,0.70),
	\]
	one obtains
	$
	I_{11}^{(\ell\ell)}(0.2) \approx 0.251,
	\qquad
	I_{11}^{(\ell\ell)}(0.5) \approx 0.230,
	\qquad
	I_{11}^{(\ell\ell)}(0.7) \approx 0.211,
	\qquad
	I_{11}^{(\ell\ell)}(0.9) \approx 0.173.
	$
	
	Thus the dominant part of $I_{11}(\eta)$ stays roughly constant for
	$0<\eta<1$. This is  the structure needed in the attached-eddy integral:
	\[
	I_{11}(\eta) \approx b_0
	\qquad
	(0<\eta \lesssim 1),
	\]
	so that
	\[
	R_{11}(y)
	=
	\beta C \int_{y/h_{\max}}^{y/h_{\min}} \frac{I_{11}(\eta)}{\eta}\, d\eta
	\approx
	\beta C b_0 \int_{y/h_{\max}}^{\eta_*} \frac{d\eta}{\eta}
	+ \text{const},
	\]
	and therefore
	\[
	R_{11}(y)
	\approx
	\beta C b_0 \log\!\left(\frac{\eta_* h_{\max}}{y}\right)
	+ \text{const}.
	\]
Thus,
	\[
	uu^+(y)
	=
	B_1 - A_1 \log(y/\delta),
	\qquad
	A_1 = \frac{\beta C b_0}{u_\tau^2}.
	\]
This asymmetry in the tolerance applied to $I_1$ and $I_{11}$ reflects a corresponding
asymmetry in the evidence each is required to reproduce. A clear logarithmic region in the mean velocity is established
at Reynolds numbers for which the streamwise variance exhibits no comparable region: the
variance instead displays an extended plateau (e.g.
Figure~\ref{fig:Optbeta}), and a logarithmic decay emerges only at
substantially higher $Re_\tau$ with a weaker slope (see, for instance, Figure 9 in \cite{samie2018fully}).The mean profile therefore constrains the kernel considerably
more tightly than the second moment does.
	
The logarithm in the mean comes from the exact step in $I_1$,
	whereas the logarithm in $uu^+$ comes from the approximately constant, leg-dominated
	second-moment kernel $I_{11}$. In the rectangular hairpin, the head carries the mean, but the legs carry
	most of the variance. It is noted that for $0 \le \eta \le 1$, $I_1$ is precisely constant, whereas $I_{11}$ is only approximately constant.
	
	Using the fitted leg parameters and the population constants from the $\theta = 60^\circ$ rectangular hairpin model, one obtains $A_1 \approx 1.2$, consistent with the experimentally observed slope reported by Marusic et al.\ (2013). This provides an independent consistency check linking the spectral structure of the minimal eddy to the observed log-linear decay of the streamwise variance.

\bibliographystyle{jfm}
\bibliography{jfm}

\begin{thebibliography}{32}
\expandafter\ifx\csname natexlab\endcsname\relax\def\natexlab#1{#1}\fi
\def\au#1{#1} \def\ed#1{#1} \def\yr#1{#1}\def\at#1{#1}\def\jt#1{\textit{#1}}
  \def\bt#1{#1}\def\bvol#1{\textbf{#1}} \def\vol#1{#1} \def\pg#1{#1}
  \def\publ#1{#1}\def\arxiv#1{#1}\def\org#1{#1}\def\st#1{\textit{#1}}

\bibitem[del {\'A}lamo {\em et~al.\/}(2006)del {\'A}lamo, Jim{\'e}nez,
  Zandonade \& Moser]{delAlamo2006Clusters}
{\sc \au{del {\'A}lamo, Juan~C.}, \au{Jim{\'e}nez, Javier}, \au{Zandonade,
  Paulo} \& \au{Moser, Robert~D.}} \yr{2006}  \at{Self-similar vortex clusters
  in the turbulent logarithmic region}.  \jt{Journal of Fluid Mechanics}
  \bvol{561},  \pg{329--358}.

\bibitem[Baars \& Marusic(2020{\natexlab{{\em a\/}}})]{baars2020data1}
{\sc \au{Baars, Woutijn~J} \& \au{Marusic, Ivan}} \yr{2020{\natexlab{{\em
  a\/}}}}  \at{Data-driven decomposition of the streamwise turbulence kinetic
  energy in boundary layers. part 1. energy spectra}.  \jt{Journal of Fluid
  Mechanics}  \bvol{882},  \pg{A25}.

\bibitem[Baars \& Marusic(2020{\natexlab{{\em b\/}}})]{baars2020data2}
{\sc \au{Baars, Woutijn~J} \& \au{Marusic, Ivan}} \yr{2020{\natexlab{{\em
  b\/}}}}  \at{Data-driven decomposition of the streamwise turbulence kinetic
  energy in boundary layers. part 2. integrated energy and}.  \jt{Journal of
  Fluid Mechanics}  \bvol{882},  \pg{A26}.

\bibitem[Buccini {\em et~al.\/}(2025)Buccini, Chen, Pasha \&
  Reichel]{buccini2025krylov}
{\sc \au{Buccini, Alessandro}, \au{Chen, Fei}, \au{Pasha, Mirjeta} \&
  \au{Reichel, Lothar}} \yr{2025}  \at{Krylov subspace based fista-type methods
  for linear discrete ill-posed problems}.  \jt{Numerical Linear Algebra with
  Applications}  \bvol{32}~(1),  \pg{e2610}.

\bibitem[Cheng {\em et~al.\/}(2019)Cheng, Li, Lozano-Dur{\'a}n \&
  Liu]{Cheng2019Identity}
{\sc \au{Cheng, Cheng}, \au{Li, Weipeng}, \au{Lozano-Dur{\'a}n, Adri{\'a}n} \&
  \au{Liu, Hong}} \yr{2019}  \at{Identity of attached eddies in turbulent
  channel flows with bidimensional empirical mode decomposition}.  \jt{Journal
  of Fluid Mechanics}  \bvol{870},  \pg{1037--1071}.

\bibitem[Cheng {\em et~al.\/}(2020)Cheng, Li, Lozano-Dur{\'a}n \&
  Liu]{Cheng2020Uncovering}
{\sc \au{Cheng, Cheng}, \au{Li, Weipeng}, \au{Lozano-Dur{\'a}n, Adri{\'a}n} \&
  \au{Liu, Hong}} \yr{2020}  \at{Uncovering {T}ownsend's wall-attached eddies
  in low-{R}eynolds-number wall turbulence}.  \jt{Journal of Fluid Mechanics}
  \bvol{889},  \pg{A29}.

\bibitem[Cuevas~Bautista {\em et~al.\/}(2019)Cuevas~Bautista, Ebadi, White,
  Chini \& Klewicki]{bautista2019turbulent}
{\sc \au{Cuevas~Bautista, Juan~Carlos}, \au{Ebadi, Alireza}, \au{White,
  Christopher~M.}, \au{Chini, Gregory~P.} \& \au{Klewicki, Joseph~C.}}
  \yr{2019}  \at{A uniform momentum zone--vortical fissure model of the
  turbulent boundary layer}.  \jt{Journal of Fluid Mechanics}  \bvol{858},
  \pg{609--633}.

\bibitem[Davidson \& Krogstad(2009)]{davidson2009simple}
{\sc \au{Davidson, PA} \& \au{Krogstad, P-{\AA}}} \yr{2009}  \at{A simple model
  for the streamwise fluctuations in the log-law region of a boundary layer}.
  \jt{Physics of Fluids}  \bvol{21}~(5),  \pg{055105}.

\bibitem[Davidson {\em et~al.\/}(2006)Davidson, Nickels \&
  Krogstad]{davidson2006logarithmic}
{\sc \au{Davidson, PA}, \au{Nickels, TB} \& \au{Krogstad, P-{\AA}}} \yr{2006}
  \at{The logarithmic structure function law in wall-layer turbulence}.
  \jt{Journal of Fluid Mechanics}  \bvol{550},  \pg{51--60}.

\bibitem[Dennis \& Nickels(2011)]{DennisNickels2011}
{\sc \au{Dennis, David J.~C.} \& \au{Nickels, Timothy~B.}} \yr{2011}
  \at{Experimental measurement of large-scale three-dimensional structures in a
  turbulent boundary layer. part 1. vortex packets}.  \jt{Journal of Fluid
  Mechanics}  \bvol{673},  \pg{180--217}.

\bibitem[Deshpande {\em et~al.\/}(2023)Deshpande, Van Den~Bogaard, Vinuesa,
  Lindi{\'c} \& Marusic]{deshpande2023reynolds}
{\sc \au{Deshpande, Rahul}, \au{Van Den~Bogaard, Aron}, \au{Vinuesa, Ricardo},
  \au{Lindi{\'c}, Luka} \& \au{Marusic, Ivan}} \yr{2023}  \at{Reynolds-number
  effects on the outer region of adverse-pressure-gradient turbulent boundary
  layers}.  \jt{Physical Review Fluids}  \bvol{8}~(12),  \pg{124604}.

\bibitem[Ehsani {\em et~al.\/}(2024)Ehsani, Heisel, Li, Voller, Hong \&
  Guala]{ehsani2024attached}
{\sc \au{Ehsani, Roozbeh}, \au{Heisel, Michael}, \au{Li, Jiarong}, \au{Voller,
  Vaughan}, \au{Hong, Jiarong} \& \au{Guala, Michele}} \yr{2024}
  \at{Stochastic modelling of the instantaneous velocity profile in rough-wall
  turbulent boundary layers}.  \jt{Journal of Fluid Mechanics}  \bvol{979},
  \pg{A12}.

\bibitem[Hutchins {\em et~al.\/}(2012)Hutchins, Chauhan, Marusic, Monty \&
  Klewicki]{hutchins2012towards}
{\sc \au{Hutchins, Nicholas}, \au{Chauhan, Kapil}, \au{Marusic, Ivan},
  \au{Monty, Jason} \& \au{Klewicki, Joseph}} \yr{2012}  \at{Towards
  reconciling the large-scale structure of turbulent boundary layers in the
  atmosphere and laboratory}.  \jt{Boundary-layer meteorology}  \bvol{145}~(2),
   \pg{273--306}.

\bibitem[Hwang \& Sung(2018)]{HwangSung2018}
{\sc \au{Hwang, Jinyul} \& \au{Sung, Hyung~Jin}} \yr{2018}  \at{Wall-attached
  structures of velocity fluctuations in a turbulent boundary layer}.
  \jt{Journal of Fluid Mechanics}  \bvol{856},  \pg{958--983}.

\bibitem[Hwang \& Eckhardt(2020)]{hwang2020attached}
{\sc \au{Hwang, Yongyun} \& \au{Eckhardt, Bruno}} \yr{2020}  \at{Attached eddy
  model revisited using a minimal quasi-linear approximation}.  \jt{Journal of
  Fluid Mechanics}  \bvol{894}.

\bibitem[Lee \& Moser(2015)]{lee2015direct}
{\sc \au{Lee, Myoungkyu} \& \au{Moser, Robert~D}} \yr{2015}  \at{Direct
  numerical simulation of turbulent channel flow up to}.  \jt{Journal of fluid
  mechanics}  \bvol{774},  \pg{395--415}.

\bibitem[Lozano-Dur{\'a}n \& Bae(2019)]{lozano2019characteristic}
{\sc \au{Lozano-Dur{\'a}n, Adri{\'a}n} \& \au{Bae, Hyunji~Jane}} \yr{2019}
  \at{Characteristic scales of townsend’s wall-attached eddies}.  \jt{Journal
  of fluid mechanics}  \bvol{868},  \pg{698--725}.

\bibitem[Lozano-Dur{\'a}n {\em et~al.\/}(2012)Lozano-Dur{\'a}n, Flores \&
  Jim{\'e}nez]{LozanoDuranFloresJimenez2012}
{\sc \au{Lozano-Dur{\'a}n, Adri{\'a}n}, \au{Flores, Oscar} \& \au{Jim{\'e}nez,
  Javier}} \yr{2012}  \at{The three-dimensional structure of momentum transfer
  in turbulent channels}.  \jt{Journal of Fluid Mechanics}  \bvol{694},
  \pg{100--130}.

\bibitem[Marusic \& Monty(2019)]{marusic2019attached}
{\sc \au{Marusic, Ivan} \& \au{Monty, Jason~P}} \yr{2019}  \at{Attached eddy
  model of wall turbulence}.  \jt{Annual Review of Fluid Mechanics}  \bvol{51},
   \pg{49--74}.

\bibitem[Marusic {\em et~al.\/}(2013)Marusic, Monty, Hultmark \&
  Smits]{marusic2013logarithmic}
{\sc \au{Marusic, Ivan}, \au{Monty, Jason~P}, \au{Hultmark, Marcus} \&
  \au{Smits, Alexander~J}} \yr{2013}  \at{On the logarithmic region in wall
  turbulence}.  \jt{Journal of Fluid Mechanics}  \bvol{716}.

\bibitem[McKeon(2019)]{mckeon2019self}
{\sc \au{McKeon, Beverley~J}} \yr{2019}  \at{Self-similar hierarchies and
  attached eddies}.  \jt{Physical Review Fluids}  \bvol{4}~(8),  \pg{082601}.

\bibitem[Nath \& Hickey(2026)]{nath2026features}
{\sc \au{Nath, Pranav} \& \au{Hickey, Jean-Pierre}} \yr{2026}  \at{Features of
  the attached-eddy hypothesis in one-dimensional turbulence models of
  turbulent boundary layers}.  \jt{Physical Review Fluids}  \bvol{11}~(1),
  \pg{014604}.

\bibitem[Nickels {\em et~al.\/}(2005)Nickels, Marusic, Hafez \&
  Chong]{nickels2005evidence}
{\sc \au{Nickels, T.~B.}, \au{Marusic, Ivan}, \au{Hafez, S.} \& \au{Chong,
  M.~S.}} \yr{2005}  \at{Evidence of the $k_1^{-1}$ law in a
  high-{R}eynolds-number turbulent boundary layer}.  \jt{Physical Review
  Letters}  \bvol{95}~(7),  \pg{074501}.

\bibitem[Perry \& Chong(1982)]{perry1982mechanism}
{\sc \au{Perry, AE} \& \au{Chong, MS}} \yr{1982}  \at{On the mechanism of wall
  turbulence}.  \jt{Journal of Fluid Mechanics}  \bvol{119},  \pg{173--217}.

\bibitem[Perry \& Maru{\v{s}}i{\'c}(1995)]{perry1995wall}
{\sc \au{Perry, AE} \& \au{Maru{\v{s}}i{\'c}, Ivan}} \yr{1995}  \at{A wall-wake
  model for the turbulence structure of boundary layers. part 1. extension of
  the attached eddy hypothesis}.  \jt{Journal of Fluid Mechanics}  \bvol{298},
  \pg{361--388}.

\bibitem[Rice(1944)]{rice1944mathematical}
{\sc \au{Rice, Stephen~O}} \yr{1944}  \at{Mathematical analysis of random
  noise}.  \jt{The Bell System Technical Journal}  \bvol{23}~(3),
  \pg{282--332}.

\bibitem[Samie {\em et~al.\/}(2018)Samie, Marusic, Hutchins, Fu, Fan, Hultmark
  \& Smits]{samie2018fully}
{\sc \au{Samie, M}, \au{Marusic, I}, \au{Hutchins, N}, \au{Fu, MK}, \au{Fan,
  Y}, \au{Hultmark, M} \& \au{Smits, AJ}} \yr{2018}  \at{Fully resolved
  measurements of turbulent boundary layer flows up to}.  \jt{Journal of Fluid
  Mechanics}  \bvol{851},  \pg{391--415}.

\bibitem[de~Silva {\em et~al.\/}(2016{\natexlab{{\em a\/}}})de~Silva, Hutchins
  \& Marusic]{de2016uniform}
{\sc \au{de~Silva, Charitha~M}, \au{Hutchins, Nicholas} \& \au{Marusic, Ivan}}
  \yr{2016{\natexlab{{\em a\/}}}}  \at{Uniform momentum zones in turbulent
  boundary layers}.  \jt{Journal of Fluid Mechanics}  \bvol{786},
  \pg{309--331}.

\bibitem[de~Silva {\em et~al.\/}(2016{\natexlab{{\em b\/}}})de~Silva, Woodcock,
  Hutchins \& Marusic]{de2016influence}
{\sc \au{de~Silva, Charitha~M}, \au{Woodcock, James~D}, \au{Hutchins, Nicholas}
  \& \au{Marusic, Ivan}} \yr{2016{\natexlab{{\em b\/}}}}  \at{Influence of
  spatial exclusion on the statistical behavior of attached eddies}.
  \jt{Physical Review Fluids}  \bvol{1}~(2),  \pg{022401}.

\bibitem[Townsend(1976)]{townsend1976structure}
{\sc \au{Townsend, AAR}} \yr{1976} {\em The structure of turbulent shear
  flow\/}.  \publ{Cambridge university press}.

\bibitem[Woodcock \& Marusic(2015)]{woodcock2015statistical}
{\sc \au{Woodcock, JD} \& \au{Marusic, I}} \yr{2015}  \at{The statistical
  behaviour of attached eddies}.  \jt{Physics of Fluids}  \bvol{27}~(1),
  \pg{015104}.

\bibitem[Yang {\em et~al.\/}(2016)Yang, Marusic \&
  Meneveau]{YangMarusicMeneveau2016HRAP}
{\sc \au{Yang, Xiang I.~A.}, \au{Marusic, Ivan} \& \au{Meneveau, Charles}}
  \yr{2016}  \at{Hierarchical random additive process and logarithmic scaling
  of generalized high order, two-point correlations in turbulent boundary layer
  flow}.  \jt{Physical Review Fluids}  \bvol{1}~(2),  \pg{024402}.

\end{thebibliography}

\end{document}